\documentclass{elsarticle}
\usepackage{lineno,hyperref}
\modulolinenumbers[5]
\journal{International Journal of Thermal Sciences}

\biboptions{square,numbers,sort&compress}

\usepackage{marginnote}
\usepackage{bibentry}
\usepackage{amsmath}
\usepackage{amssymb}
\usepackage[english]{babel}
\usepackage{color}
\usepackage{epstopdf, epsfig}
\usepackage{graphicx}
\usepackage{float}
\usepackage{subfigure}

\usepackage{a4wide}

\newcommand{\red}[1]{\textcolor[rgb]{1,0,0}{#1}}
\newcommand{\RED}[1]{\textcolor[rgb]{1,0,0}{#1}}
\newcommand{\Ta}{\widetilde{T}}
\newcommand{\Tb}{\widetilde{\mathbf{T}}}
\newcommand{\Tc}{\widehat{\mathbf{T}}}
\newcommand{\xvec}[1]{\mathbf{#1}}

\renewcommand{\RED}[1]{\textcolor[rgb]{0,0,0}{#1}}

\newcommand{\Ja}{\RED{\widetilde{J}}}
\newcommand{\Jb}{\RED{\bar{J}}}
\newcommand{\Jc}{\RED{J'}}

\renewcommand{\Ja}{\RED{J_1}}
\renewcommand{\Jb}{\RED{J_2}}
\renewcommand{\Jc}{\RED{J_3}}

\newcommand{\Jax}{\RED{J_1^0}}
\newcommand{\Jbx}{\RED{J_2^0}}

\newlength\fwidth
\newlength\fheight

\begin{document}

\begin{frontmatter}

\title{Fast fluid heating by adaptive flow reorientation}

\author[mymainaddress,mysecondaryaddress]{R. Lensvelt}
\ead{r.lensvelt@tue.nl}

\author[mymainaddress]{M.F.M Speetjens\corref{cor1}}
\ead{m.f.m.speetjens@tue.nl}
\author[mysecondaryaddress]{H. Nijmeijer}
\ead{h.nijmeijer@tue.nl}
\address[mymainaddress]{Section Energy Technology \& Fluid Dynamics, Mechanical Engineering Department, Eindhoven University of Technology, P.O. Box 513, 5600 MB, Netherlands}
\address[mysecondaryaddress]{Section Dynamics \& Control, Mechanical Engineering Department, Eindhoven University of Technology, P.O. Box 513, 5600 MB, Netherlands}
\cortext[cor1]{Corresponding author}

\date{}

\begin{abstract}

Transport of scalar quantities such as e.g. heat or chemical species in laminar flows is key to many industrial activities and stirring of the fluid by flow reorientation is a common way to enhance this process. However, ``How best to stir?'' remains a major challenge. The present study aims to contribute to existing solutions by the development of a dedicated flow-control strategy for the fast heating of a cold fluid via a hot boundary in a representative case study. In-depth analysis of the dynamics of heating in fluid flows serves as foundation for the control strategy and exposes fluid deformation as the ``thermal actuator'' via which the flow affects the heat transfer. This link is non-trivial, though, in that fluid deformation may both enhance and diminish local heat exchange between fluid parcels and a fundamental ``conflict'' between local heat transfer and thermal homogenisation tends to restrict the beneficial impact of flow to short-lived episodes. Moreover,
the impact of fluid deformation on the global fluid heating is primarily confined to the direct proximity of the moving boundary that drives the flow. These insights imply that incorporation of the thermal behaviour is essential for effective flow-based enhancement strategies and efficient fluid mixing, the conventional approach adopted in industry for this purpose, is potentially sub-optimal. The notion that global heating encompasses two concurrent processes, i.e. increasing energy content (``energising'') and thermal homogenisation, yields the relevant metrics for the global dynamics and thus enables formulation of the control problem as the minimisation of a dedicated cost function. This facilitates step-wise determination of the ``best'' flow reorientation from predicted future evolutions of actual
intermediate states and thereby paves the way to (real-time) regulation of scalar transport by flow control in practical applications.
Performance analyses reveal that this ``adaptive flow reorientation'' significantly accelerates the fluid heating throughout the considered parameter space and thus is superior over conventional periodic schemes (designed for efficient fluid mixing) both in terms of consistency and effectiveness. The controller in fact breaks with conventions by, first, never selecting these periodic schemes and, second, achieving the same superior performance for all flow conditions irrespective of whether said mixing occurs. The controller typically achieves this superiority by thermal plumes that extend from the hot wall into the cold(er) interior and are driven by two alternating and counter-rotating circulations.

\end{abstract}
\end{frontmatter}

\section{Introduction}
\label{Intro}

Transport of scalar quantities such as e.g. heat or chemical species in laminar flows is key to many industrial activities ranging from viscous mixing of polymers and foodstuffs \cite{Todd2004,Erwin2011} via process intensification and micro-fluidic devices \cite{tabeling2005,gubanov2010,bayareh2020} to subsurface resource recovery \cite{Chen2015,Guo2016,speetjens2019} and groundwater remediation \cite{Piscopo2013,huo2020}. Such systems often employ reorientations of a laminar base flow via the flow forcing (e.g. moving impellers or alternating pumps) to enhance scalar transport and thus lean on the intuitive notion that stirring of a fluid benefits this transport. However, design and engineering of such reoriented flows faces a major challenge, namely ``How best to stir?'' and, intimately related to this, ``What defines efficient transport?'' in a given system.

The conventional approach towards tackling this challenge consists of assuming that enhanced scalar transport, regardless of the nature of the problem, is automatic with efficient fluid mixing and utilising periodic flow reorientations either in space or time to accomplish such mixing in laminar flows via so-called ``chaotic advection'' \cite{ottino1989,aref2017,Speetjens2021}. Numerous studies adopt this approach for a variety of configurations and design strategies ranging from parametric optimisation of activation times to maximising entropy \cite{dalessandro1999,lester2008a,elomari2010,gubanov2010,gubanov2012}. However, a reoriented flow so designed, even if effectively accomplishing chaotic advection, has important limitations for enhancing scalar transport. First, it substantially restricts permissible flow reorientations and thus potentially excludes more optimal scenarios. Second, it is non-dedicated by discounting the actual scalar transport relevant to the system. Third, it lacks robustness to unforeseen disturbances and changing process conditions. Fourth, it omits diffusive transport both internally and across non-adiabatic boundaries.

The above limitations motivated a host of efforts to design and create flows for enhancing scalar transport by way of optimal control \cite{vikhansky2002,noack2004,mathew2007,liu2008,cortelezzi2008,couchman2010,lin2011,lunasin2012,foures_caulfield_schmid_2014,hu2018,dolk2018}.
This overwhelmingly concerns scalar homogenisation in adiabatic systems and the control strategy essentially consists of {\it a priori} determining the flow (forcing) that maximally accelerates said homogenisation in terms of an optimality criterion. To this end various measures for scalar transport have successfully been employed, primarily the mix-norm \cite{mathew2007,cortelezzi2008,lin2011,gubanov2012,foures_caulfield_schmid_2014,hu2018,eggl_schmid_2020} yet also e.g. stretching rates of fluid interfaces \cite{vikhansky2002} or the intensity of segregation \cite{dolk2018}. Ref.~\cite{noack2004} thus designed a control law for a four-point vortex flow that maximizes the scalar flux across certain fluid interfaces via low-frequency modulation of the vortex motion. Ref.~\cite{eggl_schmid_2020} thus determined the motion of stirrers to accomplish maximum scalar homogenisation inside a circular domain using the mix-norm by \cite{mathew2005}.

However, these control strategies generally concern scalar transport only by (chaotic) advection and are often restricted to highly-idealised configurations and/or forcing mechanisms with limited practical relevance. Several studies address the effect of diffusion (characterised by the well-known P\'{e}clet number $Pe$) by demonstrating that control laws and measures designed for advective transport (i.e. limit $Pe\rightarrow\infty$) may remain effective for finite $Pe$. Flow forcing optimised for advective transport in Ref.~\cite{cortelezzi2008} e.g. proved effective for $Pe\gtrsim\mathcal{O}(10^4)$ and Ref.~\cite{foures_caulfield_schmid_2014} e.g. successfully applied the mix-norm down to $Pe\sim\mathcal{O}(10^3)$. Furthermore, Ref.~\cite{cortelezzi2008} introduces some degree of robustness by performing step-wise optimal control using the intermediate state. However, existing studies nonetheless primarily address the first and second limitations mentioned above; limited robustness and omission of diffusion remain two important shortcomings in most optimal-control approaches. Moreover, goals other than homogenisation such as e.g. enhanced scalar flux across non-adiabatic boundaries receive scant attention to date.

The above findings motivate the present study, which aims at contributing to the development of dedicated flow-control strategies for enhancement of advective-diffusive scalar transport in realistic flow systems with non-adiabatic boundaries. To this end the present study adopts heating of an initially cold fluid via an isothermal hot boundary in the 2D unsteady Rotated Arc Mixer (RAM) according to \cite{baskan2015} as representative case study. Control target is accomplishing ``fast'' fluid heating in a robust manner by a control strategy that step-wise determines the most effective reorientation of the RAM base flow from predicted future evolutions of the actual intermediate state. This expands the exploratory study in \cite{lensvelt2020} by, first, foundation of the control strategy on insights into the dynamics of heating in fluid flows and, second, an extensive performance analysis.

The study contributes to existing work in literature in several ways. First, it involves advective-diffusive scalar transport including flux across a non-adiabatic boundary in a realistic configuration.
\RED{The RAM namely is experimentally realisable and admits laboratory studies both on (chaotic) advection and thermal transport \cite{baskan2015,Baskan2015b}.} Second, the control problem is of great (practical) relevance yet, contrary to scalar homogenisation in adiabatic systems, scarcely investigated. Third, the control strategy paves the way to (real-time) regulation of scalar transport in practical applications. This sets the present study apart from related studies on realistic boundary-driven flows in \cite{vikhansky2002,dolk2018,eggl_schmid_2020}. The latter namely concern enhancement of scalar homogenisation. Moreover, Refs.~\cite{vikhansky2002,eggl_schmid_2020} again employ optimal control for this purpose; only Ref.~\cite{dolk2018} (to the best of our knowledge) adopts prediction-based state-feedback control akin to the present study.

The study is organised as follows. Sec.~\ref{sec:systems_dynamics} introduces the problem of interest and the general control strategy including the (to this end relevant) structure of the temperature field in reoriented flows. The dynamics of heating in fluid flows are investigated in Sec.~\ref{DynamicsOfHeating} and insights thus gained are employed to develop the control strategy in Sec.~\ref{ControlStrategy}. The performance of adaptive flow reorientation is examined in Sec.~\ref{sec:example_and_numerical_results} by a computational analysis and reconciled with flow and thermal physics. Conclusions and recommendations for future work are presented in Sec.~\ref{sec:conclusions_and_recommendations}.

\section{Scalar transport in reoriented flows}
\label{sec:systems_dynamics}

\subsection{Flow configuration}
\label{RAM}

\begin{figure}[h]
\centering

\subfigure[Geometry]{\includegraphics[width=.3\textwidth]{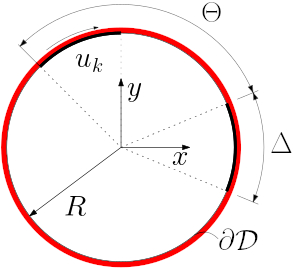}\label{RAM1a}}
\hspace*{6pt}
\subfigure[Base flow]{\includegraphics[width=.27\textwidth]{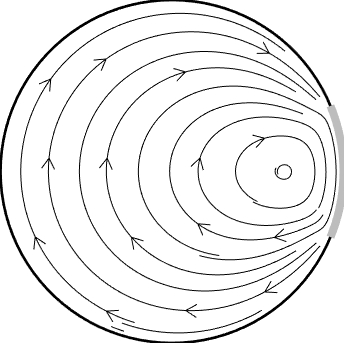}\label{RAM1b}}
\hspace*{6pt}
\subfigure[Reversed flow]{\includegraphics[width=.27\textwidth]{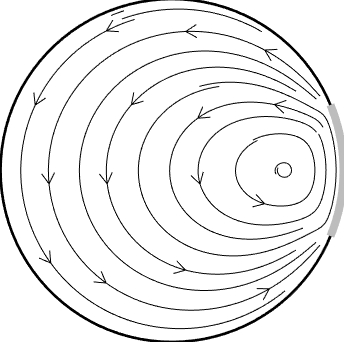}\label{RAM1c}}
\subfigure[Reoriented flows]{
\includegraphics[width=.27\textwidth]{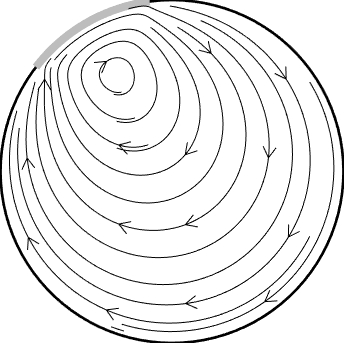}
\hspace*{6pt}
\includegraphics[width=.27\textwidth]{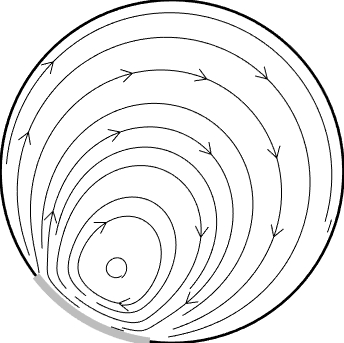}
\label{RAM1d}
}

\caption{Flow configuration of 2D unsteady RAM: (a) geometry bounded by circular wall of radius $R$ containing apertures of arc length $\Delta=\pi/4$ and offset by angle $\Theta$; (b) streamline pattern
of base flow driven by arc at $-\Delta/2\leq\theta\leq\Delta/2$; (c) reversed base flow; (d) reorientations of the base flow.}
\label{fig:ram_geometric_definition}
\end{figure}

The flow configuration of the $2D$ unsteady RAM is given in Fig.~\ref{RAM1a} and consists of a circular container $\mathcal{D}=\left\{\left.\left(r,\theta\right)\in\mathbb{R}^2\right|r\leq R,-\pi\leq\theta\leq\pi\right\}$  enclosed by a circular boundary $\Gamma = \partial \mathcal{D}$ of unit radius $R$ (red circle). The circumference of the RAM contains $N$ apertures (black arcs in Fig.~\ref{RAM1a}) with arc length $\Delta$ and angular offset $\Theta = 2\pi/N$ (i.e. the centerline of arc $1\leq k \leq N$ is located at angle $\theta_k = (k-1)\Theta$). Sliding wall segments along these apertures (practically realizable by belts \cite{baskan2015}) via viscous drag drive the flow inside the RAM. Activation of the first arc (i.e. centred on the $x$-axis) in clockwise direction at an angular velocity $\Omega$ thus sets up a steady flow $\mathbf{v}_1$ with streamline pattern following Fig.~\ref{RAM1b}. This constitutes the {\it base flow} of the RAM. Assumed are an instantaneous fluid response and negligible inertia, implying that the base flow is a steady Stokes flow symmetric about the $x$-axis and admitting an analytical solution following \cite{hwu1997}. These properties have the important consequence that other activations of the arcs result in flows that are direct transformations of base flow $\mathbf{v}_1$. Reversal of the motion of the first arc simply reverses the base flow, \RED{yielding a flow $\mathbf{v}(r,\theta)=-\mathbf{v}_1(r,\theta)$}, while maintaining the streamline portrait following Fig.~\ref{RAM1c}. Activation of arc $k>1$ rotates the base flow \RED{and yields a flow} following $\mathbf{v}\left(r,\theta\right) = \mathbf{v}_1\left(r,\theta+\left(k-1\right)\Theta\right)$\RED{, resulting} in reoriented flows as shown in Fig.~\ref{RAM1d} for $k=2$ (left) and $k=3$ (right) \RED{for} reorientation angle $\Theta = 2\pi/3$.

Systematic reorientation of a base flow by the flow forcing as exemplified in Fig.~\ref{fig:ram_geometric_definition} for the RAM can be accomplished in many industrial applications by similar wall activations
yet also via mechanic stirrers, electro-magnetic stirring and an array of (micro-fluidic) body forcings such as e.g. electro-osmosis, acoustic streaming or electro-wetting \cite{Speetjens2021}. Moreover, recent studies demonstrated that subsurface flows in e.g. enhanced geothermal systems, {\it in situ} mining or groundwater remediation admit similar flow reorientation via unsteady pumping schemes for injection and production wells \cite{Trefry2012,speetjens2019}. Thus the RAM indeed captures the essence of a wide range of systems.

\subsection{Thermal problem and general control strategy}
\label{System1}

Scalar transport in the RAM is investigated in terms of the heating of an initially cold fluid at uniform temperature $T_0$ inside $\mathcal{D}$
via the hot boundary $\Gamma$ with constant temperature $T_\infty>T_0$ (red circle in Fig.~\ref{RAM1a}). The evolution of the corresponding temperature field $T\left(\mathbf{x},t\right)$ is described by the advection-diffusion equation (ADE), which in non-dimensional form is given by
\begin{equation}
\frac{\partial T}{\partial t} = -\mathbf{v}\cdot\pmb{\nabla}T - \xvec{\nabla}\cdot\xvec{q} = -\mathbf{v}\cdot\pmb{\nabla}T + \frac{1}{Pe}\,\pmb{\nabla}^2T,\quad\quad
\xvec{q} = -\frac{1}{Pe}\,\pmb{\nabla}T,
\label{eq:ade_pde}
\end{equation}
with $\mathbf{v}\left(\mathbf{x},t\right)$ the unsteady flow, $Pe = UR/\nu$ the well-known P\`{e}clet number (determined by characteristic length and velocity scales $R$ and $U = \Omega R$, respectively, and thermal diffusion $\nu$) and $\xvec{q}$ the diffusive heat flux according to Fourier's law \cite{baskan2015,lensvelt2020}. (Geometry and flow forcing rescale to unit radius $R=1$ and unit
angular velocity $\Omega=1$, respectively).
The initial and boundary conditions corresponding with the above heating problem are $T\left(\mathbf{x},0\right)=T_0$ and $\left.T\left(\mathbf{x},t\right)\right|_{\Gamma}=T_\infty$ for all $\mathbf{x}\in\mathcal{D}$ and $t\geq 0$. Thermal problems within the scope of our study typically have $Pe\sim\mathcal{O}(10^2 - 10^4)$, implying advection-dominated heat transfer yet
with significant diffusion. Presence of diffusion has the important consequence that, regardless of the flow $\mathbf{v}$, any initial temperature field $T\left(\mathbf{x},0\right)$ {\it always} evolves towards
the uniform final state $\lim_{t\rightarrow\infty} T(\mathbf{x},t) = T_\infty$ (Sec.~\ref{GlobalHeatingDynamics1}). Thus the dynamic behaviour induced by a particular flow $\mathbf{v}$ is entirely incorporated in
the transient temperature
\begin{eqnarray}
\widetilde{T}(\mathbf{x},t) \equiv T(\mathbf{x},t) - T_\infty
\label{Temp2}
\end{eqnarray}
which is governed by the ADE
\begin{equation}
    \frac{\partial \widetilde{T}}{\partial t} = -\mathbf{v}\cdot\pmb{\nabla}\widetilde{T} + \frac{1}{Pe}\,\pmb{\nabla}^2\widetilde{T},\qquad \widetilde{T}\left(\mathbf{x},0\right)=T_0-T_\infty,\qquad \left.\widetilde{T}\left(\mathbf{x},t\right)\right|_{\Gamma}=0.
    \label{eq:ade_pde_discrete}
\end{equation}
as readily follows from substitution of \eqref{Temp2} into \eqref{eq:ade_pde}. Heating of the fluid then becomes equivalent to progression of the transient temperature towards the final state $\widetilde{T}_\infty = \lim_{t\rightarrow\infty}\widetilde{T}=0$. Initial and boundary conditions can be set to $T_0=0$ and $T_\infty=1$ hereafter without loss of generality and thus give $\widetilde{T}\left(\mathbf{x},0\right)=-1$ for the initial condition in \eqref{eq:ade_pde_discrete}.

The flow $\mathbf{v}(\mathbf{x},t)$ in \eqref{eq:ade_pde_discrete} plays a crucial role in the evolution of $\widetilde{T}$ towards $\widetilde{T}_\infty$. It is namely well-known that stirring of a fluid has a major impact on its heating and thereby on the duration of the transient. However, the central question is: ``How best to stir to obtain the fastest fluid heating?'' The RAM (and reoriented flows in general) aims at achieving optimal stirring in this sense via an unsteady flow $\mathbf{v}$ generated by switching between steady flows \red{$\mathbf{v}_{u}$} following
\begin{equation}
\mathbf{v}\left(\mathbf{x},t\right) = \RED{\mathbf{v}_{u_n}}\left(\mathbf{x}\right)
\label{eq:reorientated_flow}
\end{equation}
with $\mathcal{U} = \{u_0,u_1,\dots \RED{u_n},\dots\}$ the ``reorientation scheme'' that activates a selected flow \red{$\mathbf{v}_{u}$} with reorientation \RED{$u$} at time step $t_n = n\tau$ ($n\in\{0,1,\dots\}$) for a time
interval $t_n\leq t\leq t_n+\tau$ of duration $\tau$. The activated flow \RED{$\mathbf{v}_{u}$} can for a RAM with $N$ arcs at each $t_n$ be selected from the set of $2N+1$ flows $\mathbf{v}_k$ defined
by transformations of the base flow $\mathbf{v}_1$ according to
\begin{equation}
\mathbf{v}_{u}\left(\mathbf{x}\right) =
\left\{
\begin{array}{lc}
sign(u)\mathbf{v}_1\left(\mathcal{R}_{u}\left(\mathbf{x}\right)\right) & u\neq 0\\
\mathbf{0} & u=0,
\end{array}
\right.
\label{ReorientedFlow1}
\end{equation}
with $u\in \{-N,\dots,N\}$. Here operators $\mathcal{R}_{u}: (r,\theta)\rightarrow(r,\theta - (|u|-1)\Theta)$ and $sign(u)$ determine
rotation and flow direction, respectively, for a given $u\neq 0$. For example: $u=3$ rotates the base flow by angle $2\Theta$ by activation of aperture $u=3$ in clockwise direction; $u=-3$ in addition reverses
aperture motion and flow; $u=0$ deactivates the flow and heat is transferred by diffusion only. Reorientation scheme $\mathcal{U} = \{1,-3,2\}$ e.g. subsequently activates apertures $(1,3,2)$, each for a duration $\tau$ and with clockwise $(1,2)$ and counter-clockwise $(3)$ motion.

Determining the ``best'' reorientation scheme $\mathcal{U}$ \RED{to achieve the control target, i.e. accomplishing the ``fastest'' fluid heating,} is a major challenge. Existing approaches employ reorientation schemes consisting of {\it ad infinitum} periodic repetitions of a fixed sequence such as e.g. $\mathcal{U}=\{1,2,\dots,N\}$ or sequences determined {\it a priori} via optimal control. However, both approaches have important conceptual shortcomings and current optimal-control schemes are in general dedicated to homogenisation (Sec.~\ref{Intro}). This limits the suitability of these approaches for the present control problem. Sec.~\ref{ControlStrategy} resolves this by a dedicated control strategy that identifies the ``best'' $\mathcal{U}$ specifically for fast fluid heating by step-wise determining the most effective flow reorientation from predicted future evolutions of the actual intermediate state (denoted ``adaptive flow reorientation'' hereafter). \RED{Crucial to this end is an adequate definition and
quantification of what constitutes ``best'' and ``fastest'' in the present context. This is to be specified in Sec.~\ref{ControlStrategy2}.}

\subsection{Reorientation of the temperature field}
\label{System2}

Key for realizing a control strategy for advective-diffusive transport \RED{using reorientations of a base flow for actuation}
is the property that flow reorientations \eqref{eq:reorientated_flow} and \eqref{ReorientedFlow1} carry over to the temperature field. This hinges on the spectral decomposition of the Perron-Frobenius evolution operator $\mathcal{P}_t$ governing the temperature evolution for the base flow $\mathbf{v}_1$, given by
\begin{equation}
\widetilde{T}(\mathbf{x},t) = \mathcal{P}_t\widetilde{T}(\mathbf{x},0) = \sum_{m=0}^\infty \alpha_m \phi(\mathbf{x})e^{\lambda_m t},\quad \widetilde{T}(\mathbf{x},0) = \sum_{m=0}^\infty \alpha_m \phi(\mathbf{x}),
\label{Spectral1}
\end{equation}
with $\{\phi_m,\lambda_m\}$ the eigenfunction-eigenvalue pairs defined by the eigenvalue problem
\begin{equation}
Pe^{-1} \pmb{\nabla}^2\phi_m-\mathbf{v}_1\cdot\pmb{\nabla}\phi_m = \lambda_m\phi_m,\quad
\phi_m\left.(\mathbf{x})\right|_{\Gamma}=0,
\label{Spectral1b}
\end{equation}
corresponding with the advection-diffusion operator in \eqref{eq:ade_pde} and $\alpha_m$ the expansion coefficients determined by the initial condition \cite{lester2008a,baskan2015}. The terms in \eqref{Spectral1} constitute fundamental dynamic states (commonly denoted ``eigenmodes'') and are ordered by increasing decay rate according to $\dots < Re(\lambda_1)<Re(\lambda_0)<0$, where $m=0$ is the slowest-decaying (or ``dominant'') mode with characteristic decay time $\tau_0 = -1/Re(\lambda_0)$.

The spectral decomposition of reoriented flows is governed by eigenvalue problem \eqref{Spectral1b} upon substitution of $\mathbf{v}_1$ by $\mathbf{v}_u$ following \eqref{ReorientedFlow1} and directly relates to the base-flow decomposition \eqref{Spectral1} via
\begin{equation}
\widetilde{T}(\mathbf{x},t) = \mathcal{P}_t^{({u})}\widetilde{T}(\mathbf{x},0) = \sum_{m=0}^\infty \alpha_m^{({u})}\psi_m^{({u})}(\mathbf{x})e^{\lambda_m t},\quad
\widetilde{T}(\mathbf{x},0) = \sum_{m=0}^\infty \alpha_m^{({u})}\psi_m^{({u})}(\mathbf{x}),
\label{eq:spectral1}
\end{equation}
with $\mathcal{P}_t^{({u})}$ the corresponding Perron-Frobenius operator and
\begin{equation}
\mu_m =
\left\{
\begin{array}{lc}
\lambda_m & u\neq 0\\
\lambda_m^0 & u=0
\end{array}\right.
,\quad\quad
\psi_m^{({u})}(\mathbf{x}) =
\left\{
\begin{array}{lc}
\phi_m\left(\mathcal{G}_{u}\left(\mathbf{x}\right)\right) & u\neq 0\\
\phi_m^0(\mathbf{x}) & u=0
\end{array}\right.,
\label{eq:spectral1b}
\end{equation}
where $\{\phi_m^0,\lambda_m^0\}$ are the eigenfunction-eigenvalue pairs for the diffusion-only case $u=0$ governed by \eqref{Spectral1b} for deactivated base flow $\mathbf{v}_1=\mathbf{0}$, and
\begin{equation}
\mathcal{G}_{u}: (r,\theta)\rightarrow
\left\{
\begin{array}{lc}
\mathcal{R}_{u}(r,\theta)& u>0\\
\mathcal{S}\left(\mathcal{R}_{u}(r,\theta)\right)& u<0
\end{array}\right.,
\label{eq:spectral1c}
\end{equation}
the transformation operator for the eigenfunctions and $\mathcal{S}: (r,\theta)\rightarrow(r,-\theta)$ a reflection about the symmetry
axis $\theta=0$ of the base flow. Thus flow reorientation (and reversal) results in the same reorientation (and reflection) of the eigenfunction basis $\phi_m$ of the base flow while maintaining the eigenvalue spectrum $\lambda_m$ and, inherently, the decay rates. Flow deactivation results in a projection onto to the eigenfunction basis $\phi_m^0$ and
corresponding eigenvalues $\lambda_m^0$ of the diffusion-only case.

Spectral decomposition \eqref{eq:spectral1} admits direct expression of the evolution of the transient temperature $\Ta$ for any flow reorientation (including deactivation)
in terms of the base-flow decomposition \eqref{Spectral1} and its diffusion-only counterpart for $\mathbf{v}_1=\mathbf{0}$. This facilitates efficient prediction of $\Ta$ for arbitrary flow reorientations and is an essential element for the control strategy proposed in Sec.~\ref{ControlStrategy}.

\section{Dynamics of heating in fluid flows}
\label{DynamicsOfHeating}

A further essential element for the control strategy, besides the reorientation property of Sec.~\ref{System2}, is adequately capturing the system dynamics and in particular the impact of the (base) flow on the heating process. To this end the dynamics of this process are investigated below.

\subsection{Fluid deformation as ``thermal actuator''}
\label{LagrangianHeatTransfer}

The main reasoning behind the belief that fluid mixing and chaotic advection automatically enhance heat transfer is that this tends to expand fluid interfaces and increase scalar gradients via exponential stretching of fluid parcels and thereby yields faster scalar exchange over larger areas \cite{Speetjens2021}. Fluid deformation may thus indeed act as a ``thermal actuator'' yet the underlying mechanisms and resulting behaviour are rather delicate and, contrary to said belief, not necessarily conducive to efficient heat transfer. This is investigated below and leans on the so-called ``Lagrangian representation'' of fluid flow and heat transfer (i.e. relative to the fluid parcels).

\subsubsection{Lagrangian dynamics of fluid motion}

The Lagrangian representation of fluid motion expresses the Eulerian flow field $\xvec{v}$ in \eqref{eq:ade_pde} in terms of the evolution of the current positions $\xvec{x}(t)$ of fluid parcels released at initial position $\xvec{x}(0) = \xvec{x}_0$. This Lagrangian motion is governed by the kinematic equation
\begin{eqnarray}
\frac{d\xvec{x}(t)}{dt} = \xvec{u}\left(\xvec{x}(t),t\right),\quad \xvec{x}(0)=\xvec{x}_0
\quad\Rightarrow\quad
\xvec{x}(t) = \xvec{\Phi}_t(\xvec{x}_0),
\label{KinEq}
\end{eqnarray}
with flow $\xvec{\Phi}_t$ as its formal solution \cite{ottino1989}. Relevant in the present context of heat transfer is, besides the displacement of fluid parcels, in particular also their deformation due to viscous
stresses. Classical continuum mechanics describes this deformation in terms of the so-called ``Lagrangian coordinates''\RED{, which are} defined by the initial parcel positions $\xvec{x}_0$ and via
\RED{inversion of \eqref{KinEq}, i.e.}
%
\begin{eqnarray}
\xvec{x}_0 = \xvec{\Phi}_t^{-1}(\xvec{x}).
\label{LagrangianCoord}
\end{eqnarray}
\RED{relate} to the Eulerian coordinates $\xvec{x}$ \cite{Chadwick1999}. The Eulerian representation shows the momentary situation at time $t$ in fixed positions $\xvec{x}$ in physical space as actually seen by an observer; the Lagrangian representation gives this situation relative to the moving fluid parcels (labelled by initial positions $\xvec{x}_0$) and thus enables description of the corresponding material behaviour. Key to this are the deformation gradient tensor and right Cauchy-Green deformation tensor, given by
\begin{eqnarray}
F_0 = \partial\xvec{x}/\partial\xvec{x}_0 = (\xvec{\nabla}_0\xvec{x})^\dagger,\quad
C_0 = F_0^\dagger F_0 = \lambda_1\xvec{v}_1^0\xvec{v}_1^0 + \lambda_2\xvec{v}_2^0\xvec{v}_2^0 = \Lambda^{-1}\xvec{v}_1^0\xvec{v}_1^0 + \Lambda\xvec{v}_2^0\xvec{v}_2^0,
\label{RightCauchyTensor}
\end{eqnarray}
respectively, where $F_0$ describes the motion of initial material line segments $d\xvec{x}_0$ in the reference frame co-moving with a parcel released at $\xvec{x}_0$ via $d\xvec{x} = F_0d\xvec{x}_0$ and $C_0$ describes its corresponding material deformation via $|d\xvec{x}|^2 = d\xvec{x}^\dagger\cdot d\xvec{x} = d\xvec{x}_0^\dagger\cdot C_0\cdot d\xvec{x}_0$ \RED{($\dagger$ indicates transpose)}.
\RED{Relation \eqref{RightCauchyTensor} gives $C_0$ in terms of its eigenvalue-eigenvector pairs $(\lambda_i,\xvec{v}_i^0)$ and thus}
exposes the principal compression axis $\xvec{v}_1^0$ (factor $\lambda_1=\Lambda^{-1}<1$) and principal stretching axis $\xvec{v}_2^0\perp\xvec{v}_1^0$ (factor $\lambda_2=\Lambda>1$) of the deforming fluid parcel in question. ($|C_0| = \lambda_1\lambda_2 = 1$ for the 2D flow of an incompressible fluid.) The associated left Cauchy-Green deformation tensor
\begin{eqnarray}
B_0 = F_0 F_0^\dagger = R_0 C_0 R_0^\dagger = \Lambda^{-1}\xvec{v}_1\xvec{v}_1 + \Lambda\xvec{v}_2\xvec{v}_2,\quad\xvec{v}_i = R_0\xvec{v}_i^0,
\label{LeftCauchyTensor}
\end{eqnarray}
describes the material deformation in terms of the principal deformation axes $\xvec{v}_{1,2}$ in current position $\xvec{x}$ in the Eulerian frame. These axes are simply
rotations of their companions $\xvec{v}_{1,2}^0$ in the Lagrangian frame by $R_0$ from the well-known polar decomposition $F_0=R_0U_0$ \cite{Chadwick1999}. Further relevant tensors are the counterparts of \eqref{RightCauchyTensor} and \eqref{LeftCauchyTensor} in the reversed flow
$\xvec{x}_0 = \xvec{\Phi}_t^{-1}(\xvec{x})$, i.e.
\begin{eqnarray}
F = (\xvec{\nabla}\xvec{x}_0)^\dagger = F_0^{-1},\quad
C = B_0^{-1} = \Lambda\xvec{v}_1\xvec{v}_1 + \Lambda^{-1}\xvec{v}_2\xvec{v}_2,\quad
B = C_0^{-1} = \Lambda\xvec{v}_1^0\xvec{v}_1^0 + \Lambda^{-1}\xvec{v}_2^0\xvec{v}_2^0,
\label{TensorsReversedFlow}
\end{eqnarray}
constituting inverses of the tensors of the forward flow following \eqref{KinEq} and revealing that the principal contraction and stretching axes interchange. The stretching rate $\Lambda$ \RED{introduced above} admits alternative expression \RED{as} the so-called ``Finite-time Lyapunov exponent'' (FTLE), i.e.
\begin{eqnarray}
\sigma(\xvec{x}_0,t) \equiv \frac{\log\Lambda(\xvec{x}_0,t)}{2t},
\label{FTLE}
\end{eqnarray}
which sets the upper bound for the material stretching rate of parcel $\xvec{x}_0$ in the finite time span up to $t$: $|\xvec{x}(t)|/|\xvec{x}_0|\leq \exp(\sigma t)$. Positive FTLEs (i.e. $\sigma>0$) signify exponential stretching and, if persistent for all $t$, are regarded as ``fingerprints'' of chaotic advection \cite{ottino1989}.

\begin{figure}[htbp]
\centering

\begin{tabular}{cc}

\begin{tabular}{c}
\hfill\\

\subfigure[Fluid deformation]{\includegraphics[trim=6cm 1cm 5cm 1cm,clip,width=.4\textwidth]{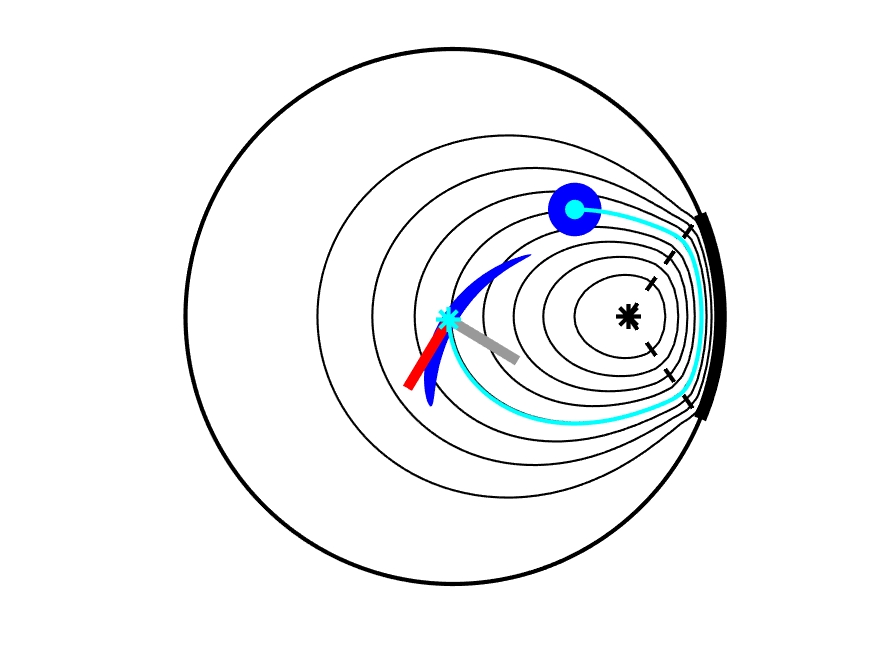}
\label{DeformationCharacteristics1a}}\\
\hfill
\end{tabular}

&

\begin{tabular}{c}
\subfigure[FTLE $\sigma(\xvec{x}_0,t)$]{
\includegraphics[trim=0cm 0cm 0cm 0cm,clip,width=.45\textwidth,height=.25\textwidth]{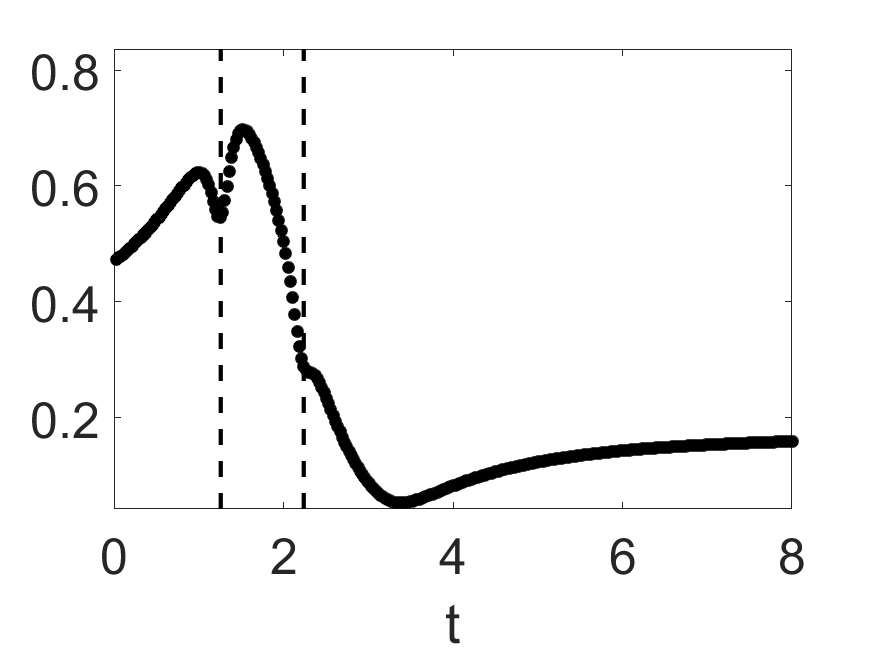}
\label{DeformationCharacteristics1b}}\\
\subfigure[Relative orientation $\rho(\xvec{x}_0,t)/\pi$]{
\includegraphics[trim=0cm 0cm 0cm 0cm,clip,width=.45\textwidth,height=.25\textwidth]{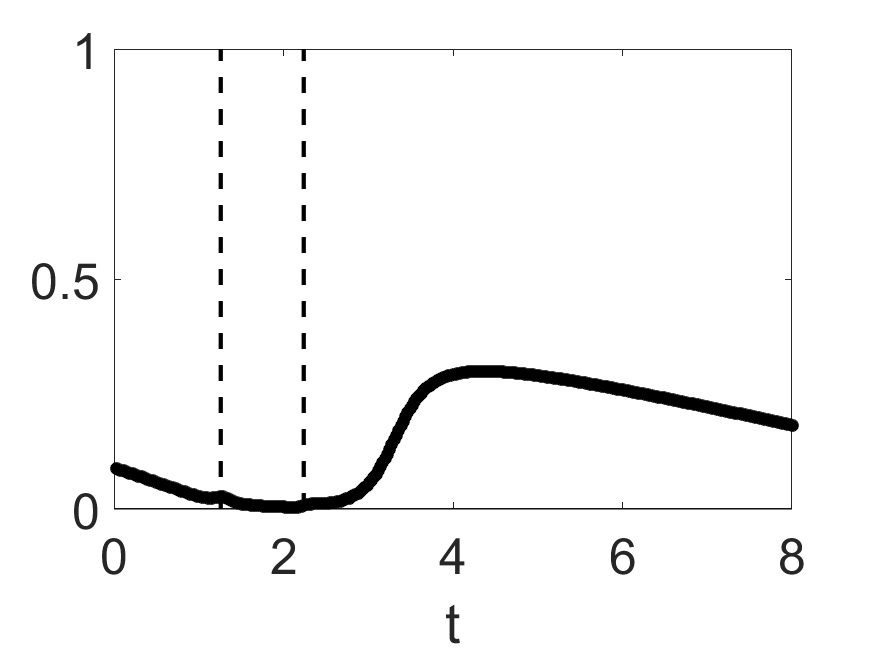}
\label{DeformationCharacteristics1c}}
\end{tabular}\\
\end{tabular}

\caption{Lagrangian dynamics in base flow: (a) deformation of fluid element (blue) along streamlines (black) versus principal stretching (red) and
compression (gray) axes; (b) quantitative deformation in terms of FTLE \RED{following \eqref{FTLE}}; (c) orientation of principal stretching axis relative to streamlines.}
\label{DeformationCharacteristics1}

\end{figure}

Fig.~\ref{DeformationCharacteristics1a} illustrates typical Lagrangian dynamics in the base flow $\mathbf{v}_1$ (Fig.~\ref{RAM1b}) by the evolution of a fluid element (blue) of initially circular shape released at $\xvec{x}_0$ (cyan dot). This reveals a gradual deformation into an elliptical shape during its excursion from $\xvec{x}_0$ to current position $\xvec{x}$ (cyan star) via the Lagrangian path (cyan) along the streamlines (black concentric curves). This deformation ensues from the shear flow between the streamlines and is dictated by the principal contraction ($\xvec{v}_1$; gray bar) and stretching ($\xvec{v}_2$; red bar) axes of $B_0$.
Fig.~\ref{DeformationCharacteristics1b} gives the corresponding FTLE \eqref{FTLE} and reveals an increase to a maximum about
halfway the high-shear region near the moving arc (entry and exit demarcated by the dashed lines) that is followed by a
sharp decline upon transiting into the domain interior towards a minimum at around $t\approx 3$ and a subsequent increase to a moderate level. FTLE $\sigma>0$ everywhere implies an overall net stretching yet with a partial reversal of earlier deformation in the interval between local maximum and minimum due to strong and ``unfavourable'' velocity gradients near the exit of the arc region (bottom/left dashed line).

The orientation angle $\rho$ of the principal stretching axis $\xvec{v}_2$ relative to the streamlines, defined as $\cos\rho = \xvec{v}_2^\dagger\xvec{u}/|\xvec{v}_2||\xvec{u}|$, is shown in Fig.~\ref{DeformationCharacteristics1c} and exhibits non-monotonic behaviour consistent with the FTLE: rapid alignment of $\xvec{v}_2$ with streamlines (i.e. diminishing $\rho$) within the high-shear region, followed by rapid misalignment upon exiting this region and renewed alignment while further migrating into the domain interior. Fluid elements released at other locations exhibit essentially the same behaviour upon passing through the arc region. Multiple passages of this region while circulating along closed streamlines results in progressively weaker fluctuations and eventually causes convergence on the asymptotic limit $\lim_{t\rightarrow}\rho = 0$. This exposes the stream lines as ``attractors'' for the fluid deformation in the sense that fluid parcels ultimately align with these entities. Compare this with unstable manifolds of hyperbolic points in chaotic flows and their finite-time counterparts in generic aperiodic flows, viz. attracting LCSs \cite{Speetjens2021}.

\subsubsection{Lagrangian dynamics of heat transfer}

Transformation of the energy balance \eqref{eq:ade_pde_discrete} from Eulerian ($\xvec{x}$) to Lagrangian ($\xvec{x}_0$) coordinates via \eqref{LagrangianCoord} yields the Lagrangian representation of heat transfer
\begin{equation}
    \frac{\partial \widetilde{T}}{\partial t} = -\pmb{\nabla}_0\cdot\xvec{q}_0,\quad\quad
\xvec{q}_0 = F\xvec{q} = -F\frac{1}{Pe}\xvec{\nabla}T = -F\frac{1}{Pe}F^\dagger\xvec{\nabla}_0T = -B\frac{1}{Pe}\xvec{\nabla}_0T,
\label{HeatFluxLagrangian1}
\end{equation}
with $\xvec{q}_0$ the Lagrangian representation of the diffusive heat flux $\xvec{q} = -Pe^{-1}\xvec{\nabla}T$ according to \eqref{eq:ade_pde} and $B$ the left Cauchy-Green tensor of the reversed flow
following \eqref{TensorsReversedFlow} \cite{Tang1999}. Energy balance \eqref{HeatFluxLagrangian1} describes the actual heat transfer between a moving fluid parcel at current position
$\xvec{x} = \xvec{\Phi}_t(\xvec{x}_0)$ with its neighbouring parcels. This occurs solely via diffusive flux $\xvec{q}_0$; the advective term $\xvec{v}\cdot\xvec{\nabla}T$ in the Eulerian form has vanished from \eqref{HeatFluxLagrangian1} due to
vanishing velocity in the (co-moving) Lagrangian frame. Relation \eqref{HeatFluxLagrangian1} reveals that the momentary isotropic diffusive flux $\xvec{q}=-Pe^{-1}\xvec{\nabla}T$ in a fixed position $\xvec{x}$ in
Eulerian space translates into a momentary anisotropic diffusive flux $\xvec{q}_0$, with $B$ the corresponding diffusion tensor, at the fluid parcel passing through $\xvec{x}$ and originating from $\xvec{x}_0 = \xvec{\Phi}_t^{-1}(\xvec{x})$. This anisotropy is a direct consequence of fluid deformation; $B=I$ for a non-deforming fluid and yields $\xvec{q}_0=\xvec{q}$.

The Lagrangian heat flux $\xvec{q}_0$ in \eqref{HeatFluxLagrangian1} constitutes the fundamental link between heat transfer and fluid motion and enables further investigation of the impact of the latter on
the former. Expression in terms of the principal deformation axes $\xvec{v}_{1,2}$ of tensor $B$ following \eqref{TensorsReversedFlow} yields
\begin{eqnarray}
\xvec{q}_0 = -B\frac{1}{Pe}\xvec{\nabla}_0T = -\left[\frac{\Lambda}{Pe}(\xvec{v}_1^0\cdot\xvec{\nabla}_0 T)\xvec{v}_1^0 + \frac{1}{\Lambda Pe}(\xvec{v}_2^0\cdot\xvec{\nabla}_0 T)\xvec{v}_2^0\right],
\label{HeatFluxLagrangian2}
\end{eqnarray}
and reveals that enhancement (by factor $\Lambda$) and diminution (by factor $\Lambda^{-1}$) of heat transfer occurs in the principal contraction ($\xvec{v}_1^0$) and stretching ($\xvec{v}_2^0$) directions, respectively, compared to isotropic heat diffusion $\xvec{q}_0^{\rm iso} = -Pe^{-1}\xvec{\nabla}_0T$ between non-deforming fluid parcels subject to the same flow and temperature field \cite{Tang1999}. (Symbol $\dagger$ is for brevity omitted for inner products as in \eqref{HeatFluxLagrangian2}.) However, the net effect of fluid deformation depends on the orientation of the principal axes $\xvec{v}_{1,2}^0$ relative to the temperature gradient $\xvec{\nabla}_0 T$ and this may enhance yet also diminish heat transfer between fluid parcels. This highly non-trivial process is investigated further in Sec.~\ref{HeatTransferEnhancement}.

Lagrangian energy balance \eqref{HeatFluxLagrangian1} admits a formal solution according to
\begin{eqnarray}
\widetilde{T}_D(\mathbf{x}_0,t) = \widetilde{T}(\mathbf{x}_0,0) + \frac{1}{Pe}\int_0^t \xvec{\nabla}_0\cdot\left(B|_{\xvec{\Phi}_\xi(\xvec{x}_0)}\xvec{\nabla}_0T(\xvec{x}_0,\xi)\right)d\xi \equiv \mathcal{D}_t[\xvec{x}_0]\widetilde{T}(\mathbf{x}_0,0),
\label{EvolutionOperators1}
\end{eqnarray}
with diffusion operator $\mathcal{D}_t$ as Lagrangian counterpart to the Perron-Frobenius operator $\mathcal{P}_t$ in \eqref{Spectral1}. This enables representation of simultaneous advective-diffusive heat transfer governed by \eqref{eq:ade_pde_discrete} as a composition of two successive operations: (i) anisotropic diffusion between fluid parcels at fixed positions $\xvec{x}_0$ via \eqref{EvolutionOperators1} and (ii) passive redistribution of fluid parcels to positions $\xvec{x} = \pmb{\Phi}_t(\xvec{x}_0)$, i.e.
\begin{eqnarray}
\widetilde{T}(\mathbf{x},t) = \widetilde{T}\left(\xvec{\Phi}_t(\mathbf{x}_0),t\right) = \widetilde{T}_D(\mathbf{x}_0,t).
\label{EvolutionOperators2}
\end{eqnarray}
which, upon defining an advection operator $\widetilde{T}(\mathbf{x},t) = \widetilde{T}\left(\xvec{\Phi}^{-1}_t(\xvec{x}),0\right) \equiv \mathcal{A}_t\widetilde{T}(\mathbf{x},0)$, translates into
\begin{eqnarray}
\widetilde{T}(\mathbf{x},t) = \mathcal{A}_t\mathcal{D}_t[\xvec{x}]\widetilde{T}(\mathbf{x},0) =
\mathcal{D}_t\left[\xvec{\Phi}_t^{-1}(\mathbf{x})\right]\mathcal{A}_t\widetilde{T}(\mathbf{x},0),
\label{EvolutionOperators3}
\end{eqnarray}
as two equivalent formulations in terms of Eulerian evolution operators. Thus the Perron-Frobenius operator $\mathcal{P}_t$ following \eqref{Spectral1} decomposes into an advective ($\mathcal{A}_t$) and diffusive ($\mathcal{D}_t$) factor following
\begin{eqnarray}
\mathcal{P}_t = \mathcal{A}_t\mathcal{D}_t[\xvec{x}] = \mathcal{D}_t\left[\xvec{\Phi}_t^{-1}(\mathbf{x})\right]\mathcal{A}_t,
\label{EvolutionOperators4}
\end{eqnarray}
providing a link between the Eulerian and Lagrangian representations of heat transfer.

\begin{figure}[htbp]
\centering

\begin{tabular}{cc}

\begin{tabular}{c}

\hfill\\

\includegraphics[trim=0cm 0cm 0cm 0cm,clip,width=.07\textwidth]{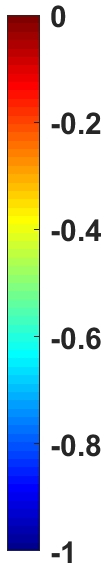}\\
\hfill
\end{tabular}

&

\begin{tabular}{ccc}
\includegraphics[trim=6cm 1cm 5cm 1cm,clip,width=.26\textwidth]{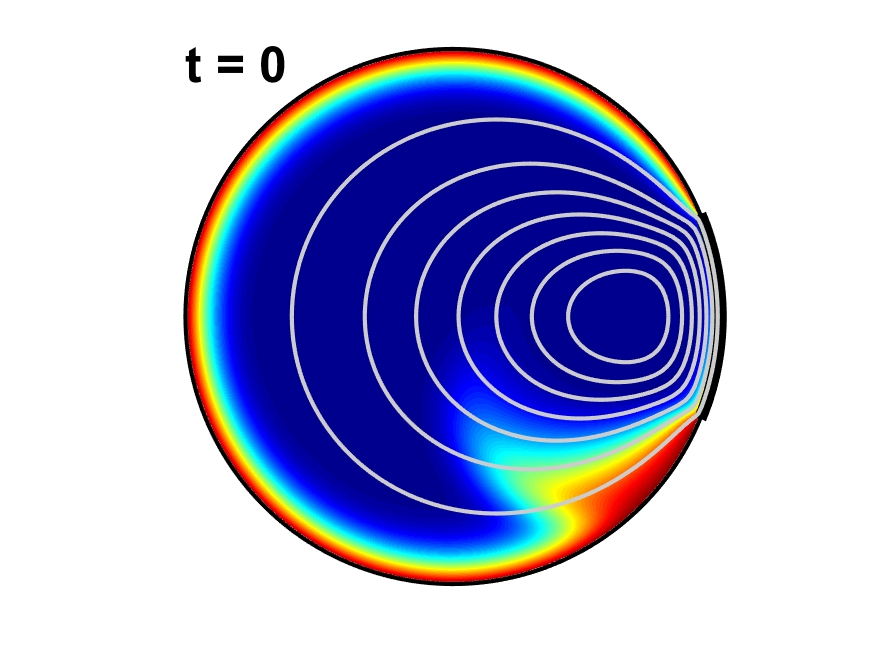}&
\includegraphics[trim=6cm 1cm 5cm 1cm,clip,width=.26\textwidth]{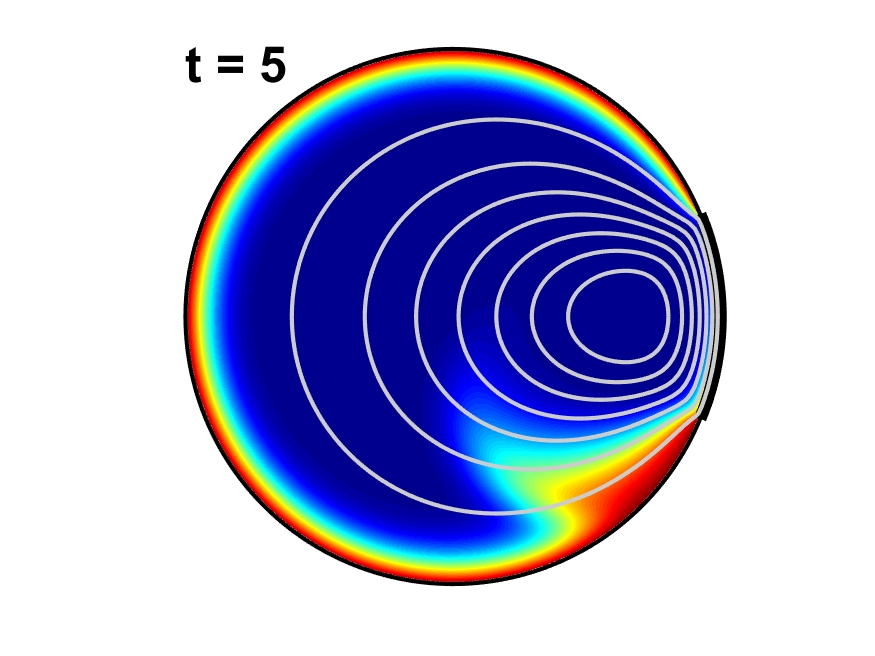}&
\includegraphics[trim=6cm 1cm 5cm 1cm,clip,width=.26\textwidth]{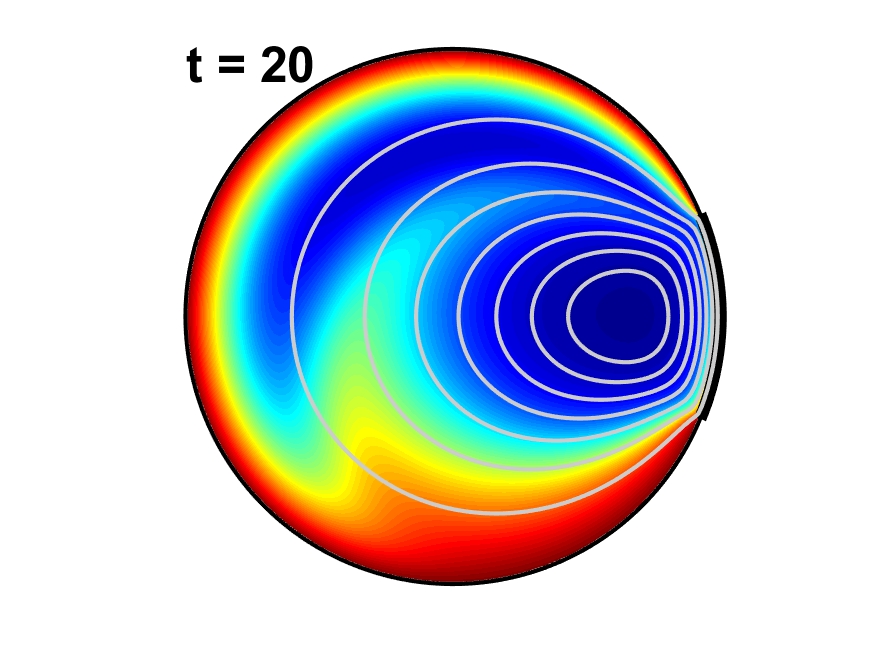}\\
\includegraphics[trim=6cm 1cm 5cm 1cm,clip,width=.26\textwidth]{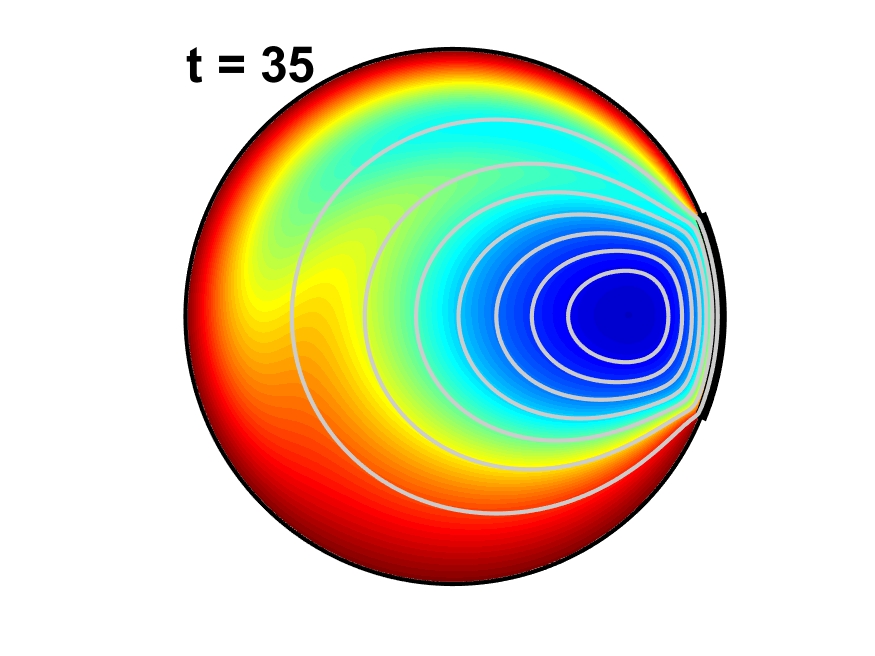}&
\includegraphics[trim=6cm 1cm 5cm 1cm,clip,width=.26\textwidth]{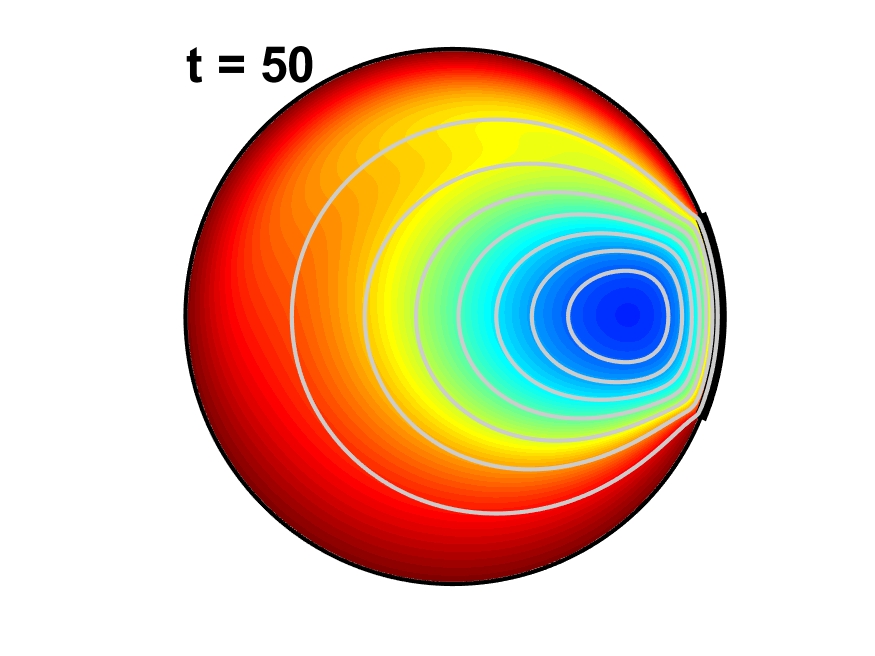}&
\includegraphics[trim=6cm 1cm 5cm 1cm,clip,width=.26\textwidth]{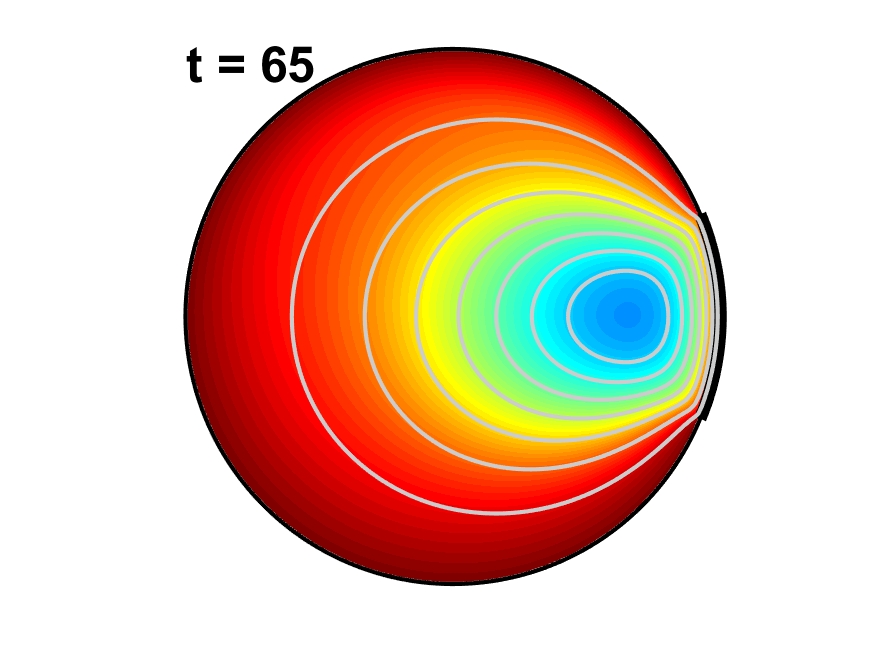}
\end{tabular}

\end{tabular}

\caption{Evolution of transient temperature $\widetilde{T}(\xvec{x},t)$ in Eulerian frame $\xvec{x}$ subject to base flow (blue: $\min(\widetilde{T}) = -1$; red: $\max(\widetilde{T}) = 0$; closed curves: streamlines).}
\label{FigEulerianLagrangian2}
\end{figure}

Fig.~\ref{FigEulerianLagrangian2} shows the temperature evolution $\widetilde{T}(\mathbf{x},t)$ in the Eulerian frame $\xvec{x}$ subject to the base flow $\xvec{v}_1$, with blue and
red indicating $\min(\widetilde{T}) = -1$ and $\max(\widetilde{T}) = 0$, respectively. (This colour coding is used throughout the remainder of this study unless noted otherwise.) Shown evolution reveals a thermal plume emanating from the bottom edge of the moving arc and propagating along the streamlines (bright closed curves) into the domain interior. The plume thus ``wraps'' itself around the center of circulation and creates a hot annular region encircling a cold ``core''. This behaviour is a direct consequence of the fluid motion: the material stretching along streamlines demonstrated in Fig.~\ref{DeformationCharacteristics1} promotes steepening and flattening of the temperature gradient transverse and parallel to the streamlines, respectively. This gradient steepening enhances transverse heat flux in the fluid heated during passage along the arc and thereby promotes its transverse thermal homogenisation while propagating and shearing along the streamlines and thus creates a thermal front (i.e. said plume). The underlying local mechanisms and corresponding impact on the global heating dynamics are investigated in Secs.~\ref{HeatTransferEnhancement}--\ref{LocalAsymptotic} and Sec.~\ref{GlobalHeatingDynamics2}, respectively.

\subsection{Local impact of fluid deformation}
\label{HeatTransferEnhancement}

Relations \eqref{HeatFluxLagrangian1} and \eqref{HeatFluxLagrangian2} reveal that fluid deformation has a twofold impact on heat transfer $\xvec{q}_0$ between fluid parcels:

{\bf 1. Change in flux density.} Consider the heat flux $\widetilde{\xvec{q}}_0 = q_1^0\xvec{v}_1^0 + q_2^0\xvec{v}_2^0$ at an undeformed rectangular $dx_1^0 \times dx_2^0$ fluid parcel in the Lagrangian frame
and the corresponding heat flux $\widetilde{\xvec{q}} = q_1\xvec{v}_1 + q_2\xvec{v}_2$ at the deformed rectangular $dx_1 \times dx_2$ fluid parcel in the Eulerian frame. (The normals of interfaces $dx_{1,2}^0$ coincide with $\xvec{v}_{2,1}^0$ and likewise for $dx_{1,2}$ versus $\xvec{v}_{2,1}$.)
The net heat exchange across the entire interface is conserved due to $|F_0|=|C_0|=1$, i.e. $dQ = dx_1^0q_2^0 + dx_2^0q_2^1 = dx_1q_2 + dx_2q_2$, yet the heat-flux
{\it density} at the {\it individual} interfaces changes as
\begin{eqnarray}
\frac{q_1^0}{q_1} = \frac{dx_2}{dx_2^0} = \sqrt{\Lambda}>1,\quad\quad \frac{q_2^0}{q_2} = \frac{dx_1}{dx_1^0} = \frac{1}{\sqrt{\Lambda}}<1,
\label{ChangeFlux}
\end{eqnarray}
due to deformation $\widetilde{\xvec{q}} = F_0\widetilde{\xvec{q}}_0$ and $d\xvec{x} = F_0d\xvec{x}_0$. Deformation thus augments (reduces) the spatial contact area $dx_2$ ($dx_1$) between fluid parcels for heat exchange in $\xvec{v}_1$ ($\xvec{v}_2$) direction, in Eulerian space; for given Eulerian heat-flux density $\widetilde{\xvec{q}}$ this increases (decreases) the heat-flux density $\widetilde{\xvec{q}}_0$ at fluid interface $dx_2^0$ ($dx_1^0$) in $\xvec{v}_1^0$ ($\xvec{v}_2^0$) direction. The net effect on the heat exchange between fluid parcels depends on the relative orientation between heat flux and principal deformation axes. The extremal cases $\widetilde{\xvec{q}}_0 = q_1\xvec{v}_1^0$ and $\widetilde{\xvec{q}}_0 = q_2\xvec{v}_2^0$ give maximum augmentation ($|\widetilde{\xvec{q}}_0|/|\widetilde{\xvec{q}}| = \sqrt{\Lambda}>1$) and reduction ($|\widetilde{\xvec{q}}_0|/|\widetilde{\xvec{q}}| = 1/\sqrt{\Lambda}<1$), respectively; generic cases give $1/\sqrt{\Lambda} \leq |\widetilde{\xvec{q}}_0|/|\widetilde{\xvec{q}}| \leq \sqrt{\Lambda}$.

{\bf 2. Change in temperature gradient.} A similar analysis for the temperature gradients $\xvec{\nabla}_0 \Ta = (\partial \Ta/\partial x_1^0)\xvec{v}_1^0 + (\partial \Ta/\partial x_2^0)\xvec{v}_2^0$ and $\xvec{\nabla} \Ta = (\partial \Ta/\partial x_1)\xvec{v}_1 + (\partial \Ta/\partial x_2)\xvec{v}_2$ across the rectangular fluid parcels introduced above yields
\begin{eqnarray}
\frac{\partial \Ta}{\partial x_1} = \frac{dx_1^0}{dx_i}\frac{\partial \Ta}{\partial x_1^0} = \sqrt{\Lambda}\frac{\partial \Ta}{\partial x_1^0},\quad\quad
\frac{\partial \Ta}{\partial x_2} = \frac{dx_2^0}{dx_i}\frac{\partial \Ta}{\partial x_2^0} = \frac{1}{\sqrt{\Lambda}}\frac{\partial \Ta}{\partial x_2^0},
\label{ChangeGradient}
\end{eqnarray}
due to deformation $\xvec{\nabla}\Ta = F^\dagger\xvec{\nabla}_0 \Ta$ and $d\xvec{x} = F_0d\xvec{x}_0$. Deformation in $\xvec{v}_1$ and $\xvec{v}_2$ directions thus steepens (i.e. $(\partial \Ta/\partial x_1)/(\partial \Ta/\partial x_1^0) = \sqrt{\Lambda}>1$) and flattens (i.e. $(\partial \Ta/\partial x_2)/(\partial \Ta/\partial x_2^0) = 1/\sqrt{\Lambda}<1$) the temperature gradient, respectively, in Eulerian space; for given temperature difference $dT$ between fluid parcels this increases (decreases) the gradient-driven heat-flux density $\widetilde{\xvec{q}} = -Pe^{-1}\xvec{\nabla}\Ta$ in Eulerian space in $\xvec{v}_1$ ($\xvec{v}_2$) direction. The net effect depends on the relative orientation between temperature gradient and principal deformation axes. The extremal cases $\xvec{\nabla}_0 \Ta = (\partial \Ta/\partial x_1^0)\xvec{v}_1^0$ and $\xvec{\nabla}_0 \Ta = (\partial T/\partial x_2^0)\xvec{v}_2^0$ give maximum augmentation ($|\xvec{\nabla}\Ta|/|\xvec{\nabla}_0\Ta| = \sqrt{\Lambda}>1$) and reduction ($|\xvec{\nabla}\Ta|/|\xvec{\nabla}_0\Ta| = 1/\sqrt{\Lambda}<1$), respectively; generic cases give $1/\sqrt{\Lambda} \leq |\xvec{\nabla}\Ta|/|\xvec{\nabla}_0\Ta| \leq \sqrt{\Lambda}$.

The anisotropic heat flux $\widetilde{\xvec{q}}_0$ between fluid parcels according to \eqref{HeatFluxLagrangian1} results from the combined effect of \eqref{ChangeFlux} and \eqref{ChangeGradient}. This yields a momentary net change in heat-flux density by a factor
\begin{eqnarray}
\beta(\xvec{x}_0) \equiv \frac{|\widetilde{\xvec{q}}_0|}{|\widetilde{\xvec{q}}_0^{\rm iso}|} =
\sqrt{\frac{(\xvec{\nabla_0}\Ta)^\dagger\cdot B^2\cdot(\xvec{\nabla_0}\Ta)}{(\xvec{\nabla_0}\Ta)^\dagger\cdot(\xvec{\nabla_0}\Ta)}}
,\quad\quad\Lambda(\xvec{x}_0)^{-1}\leq \beta(\xvec{x}_0)\leq \Lambda(\xvec{x}_0),
\label{RelativeHeatChange3}
\end{eqnarray}
compared to isotropic heat transfer $\widetilde{\xvec{q}}_0^{\rm iso} = -Pe^{-1}\xvec{\nabla}_0\Ta$ between non-deforming fluid parcels (i.e. $B=I$) for a given flow and temperature field. Relative heat-flux densities $\beta>1$ and $\beta<1$ signify relative enhancement and diminution, respectively, of momentary heat transfer between neighbouring fluid parcels at position $\xvec{x}_0$ and time $t$. The actual $\beta$, similar as before, depends essentially on the relative orientation between temperature gradient and principal deformation axes and is bounded by $\beta = \Lambda>1$ for $\xvec{\nabla}_0 \Ta = (\partial \Ta/\partial x_1^0)\xvec{v}_1^0$ and $\beta = 1/\Lambda<1$ for $\xvec{\nabla}_0 \Ta = (\partial \Ta/\partial x_2^0)\xvec{v}_2^0$.

Fluid deformation, besides the heat-flux {\it density} following \eqref{RelativeHeatChange3}, also impacts the heat-flux {\it direction}: $\Lambda>1$ namely increases and decreases the leading and trailing terms in \eqref{HeatFluxLagrangian2}, respectively, and thus promotes alignment of $\widetilde{\xvec{q}}_0$ with the principal contraction axis $\xvec{v}_1^0$. This, in turn, promotes thermal homogenisation in $\xvec{v}_1^0$-direction and, inherently, alignment of the temperature gradient $\xvec{\nabla}_0 \Ta$ with the principal stretching axis $\xvec{v}_2^0$. The net result is the emergence of a thermal front propagating in $\xvec{v}_2^0$-direction as e.g. the thermal plume in Fig.~\ref{FigEulerianLagrangian2}. These counter-acting mechanisms, viz. enhancement of the $\xvec{v}_1^0$-component of $\xvec{\nabla}_0 \Ta$ in $\widetilde{\xvec{q}}_0$ versus diminution of this same component by said thermal homogenisation, suggest a fundamental ``conflict'' between heat transfer and homogenisation in that the latter opposes the former. Angles $\varrho_q$ and $\varrho_T$ defined as
\begin{eqnarray}
\tan\,\varrho_q(\xvec{x}_0) \equiv \frac{|\xvec{v}_2^0\cdot\widetilde{\xvec{q}}_0|}{|\xvec{v}_1^0\cdot\widetilde{\xvec{q}}_0|} = \frac{|\xvec{v}_2^0\cdot\xvec{\nabla}_0\Ta|}{\Lambda^2(\xvec{x}_0)|\xvec{v}_1^0\cdot\xvec{\nabla}_0 \Ta|},\quad\quad
\tan\,\varrho_T(\xvec{x}_0) \equiv \frac{|\xvec{v}_1^0\cdot\xvec{\nabla}_0 \Ta|}{|\xvec{v}_2^0\cdot\xvec{\nabla}_0 \Ta|},
\label{ThermalAlignment}
\end{eqnarray}
express the orientation of heat flux and temperature gradient relative to principal axes $\xvec{v}_1^0$ and $\xvec{v}_2^0$, respectively, and enable examination of this process ($\varrho_{q,T}=0$ means coincidence with axes $\xvec{v}_{1,2}^0$).

\begin{figure}[htbp]
\centering

\subfigure[Trajectories]{\includegraphics[trim=2cm 2cm 2cm 1cm,clip,width=.38\textwidth]{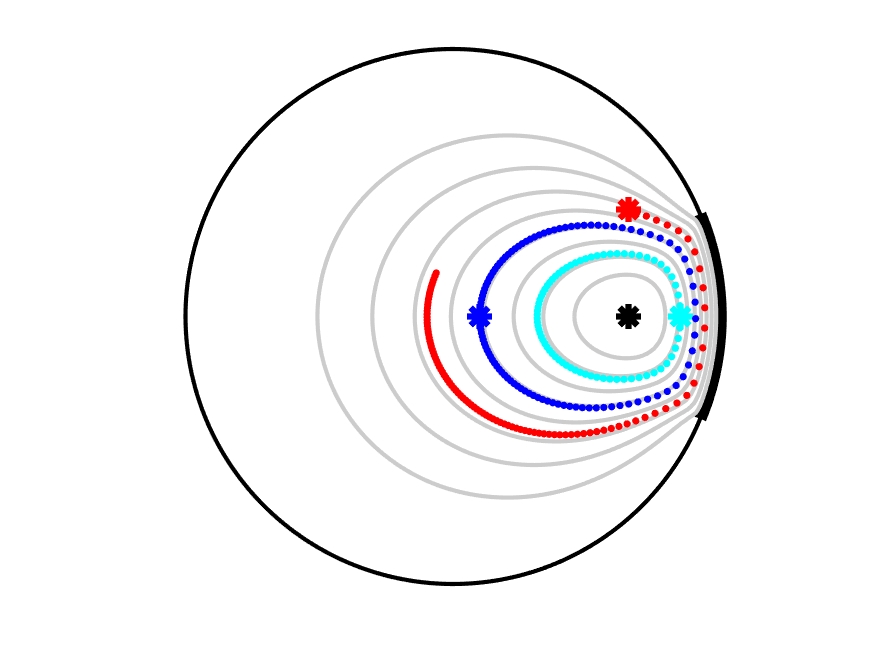}\label{HeatingDynamics0a}}
\hspace*{12pt}
\subfigure[FTLE]{\includegraphics[trim=1cm 0cm 2cm 1cm,clip,width=.32\textwidth]{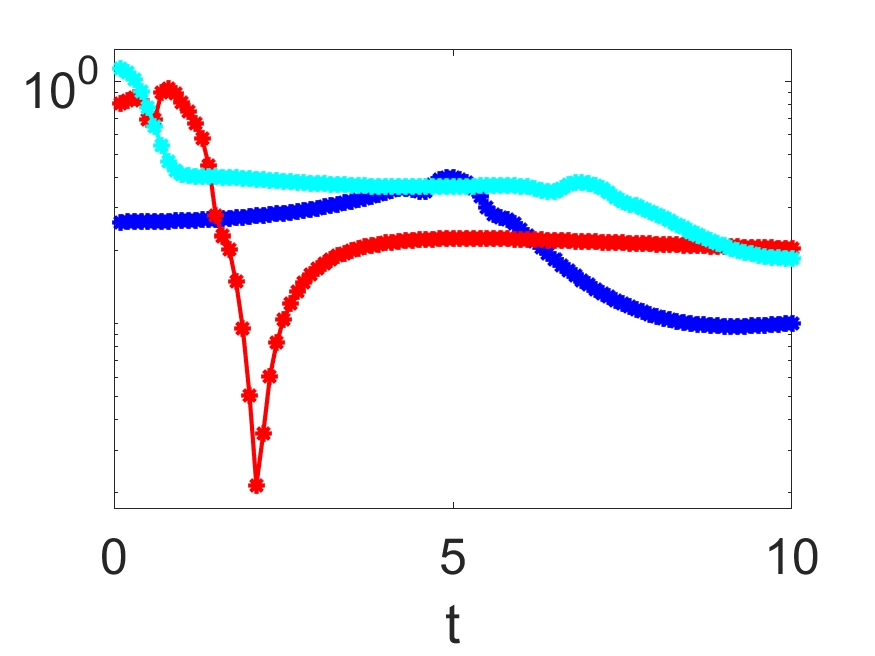}\label{HeatingDynamics0b}}

\subfigure[$\Ta(\xvec{x}_0,t)$]{\includegraphics[trim=1cm 0cm 2cm 1cm,clip,width=.32\textwidth]{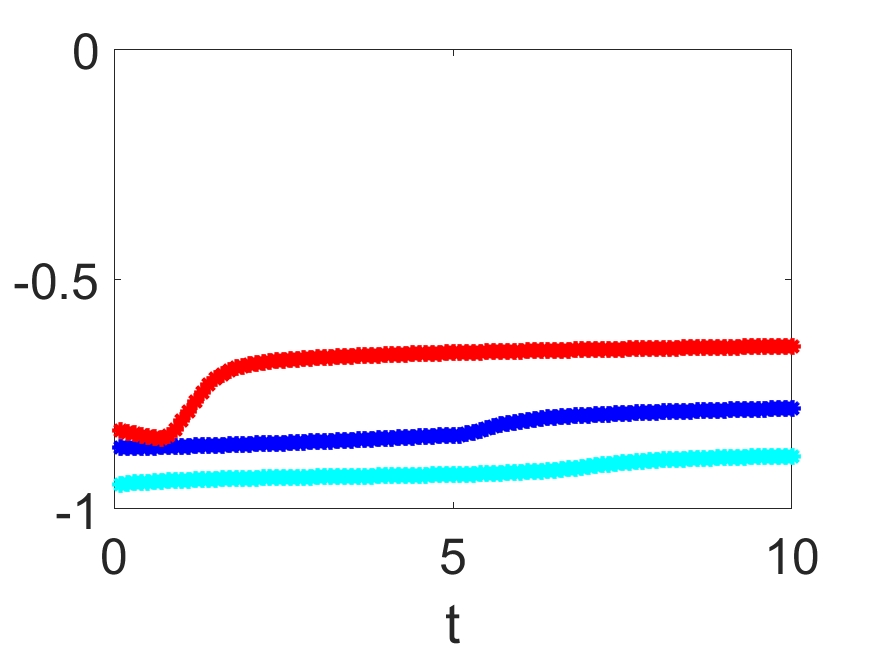}\label{HeatingDynamics0c}}
\subfigure[$\xvec{\nabla}_0\Ta$]{\includegraphics[trim=1cm 0cm 2cm 1cm,clip,width=.32\textwidth]{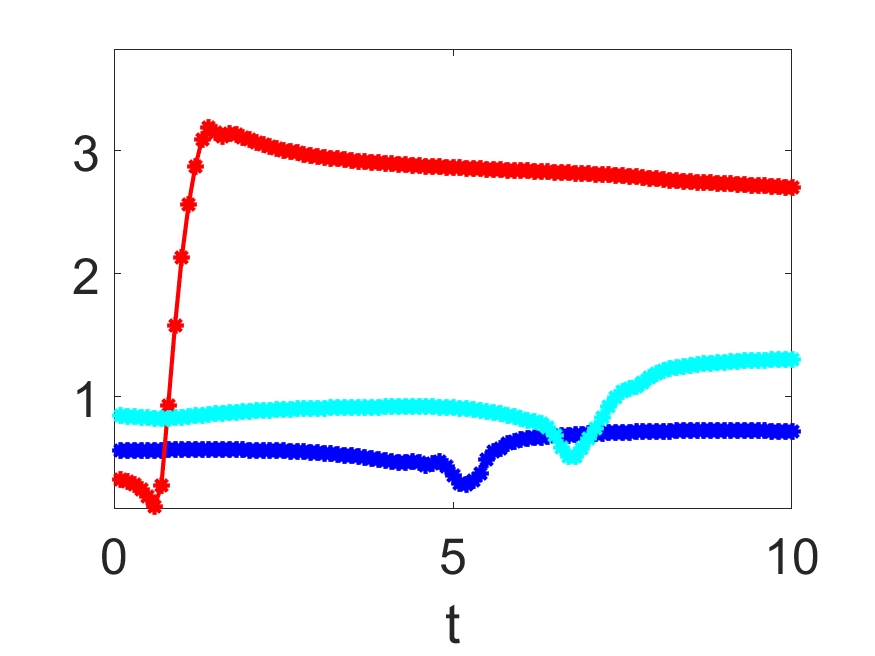}\label{HeatingDynamics1a}}
\subfigure[$\ln\beta$]{\includegraphics[trim=1cm 0cm 2cm 1cm,clip,width=.32\textwidth]{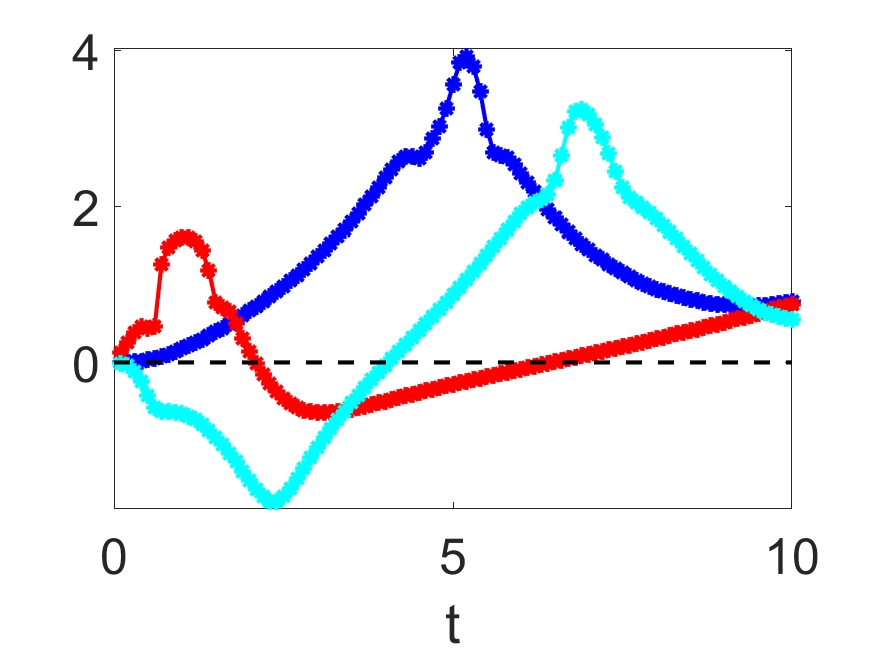}\label{HeatingDynamics1b}}

\subfigure[$\beta_\Lambda$]{\includegraphics[trim=1cm 0cm 2cm 1cm,clip,width=.32\textwidth]{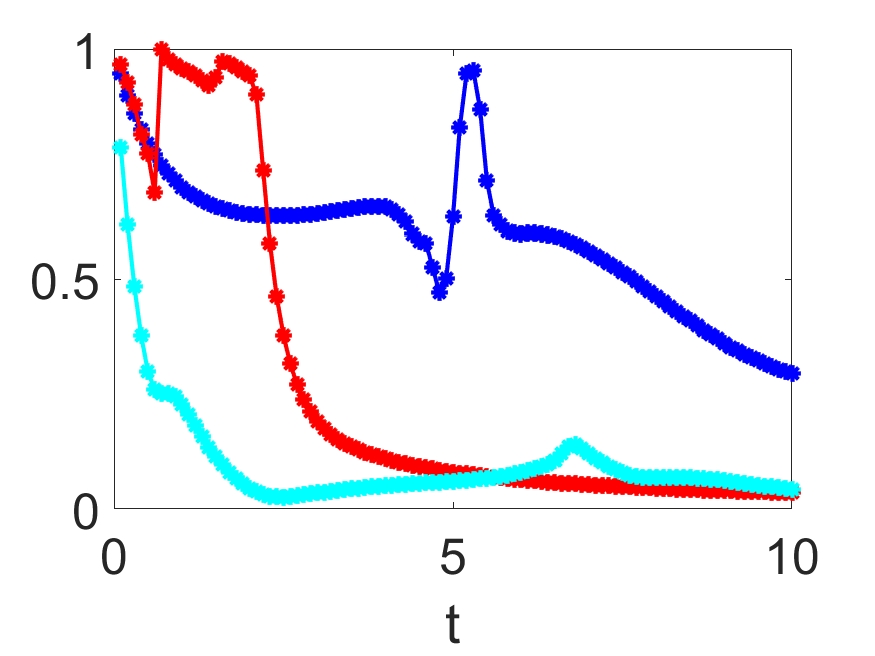}\label{HeatingDynamics1c}}
\subfigure[$\varrho_T/\pi$]{\includegraphics[trim=1cm 0cm 2cm 1cm,clip,width=.32\textwidth]{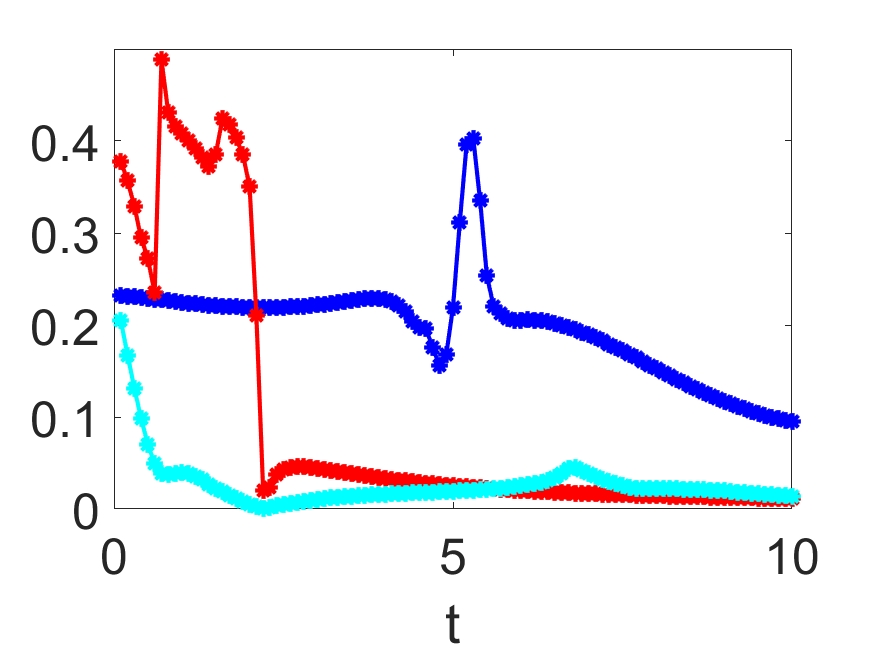}\label{HeatingDynamics1e}}
\subfigure[$\varrho_q/\pi$]{\includegraphics[trim=1cm 0cm 2cm 1cm,clip,width=.32\textwidth]{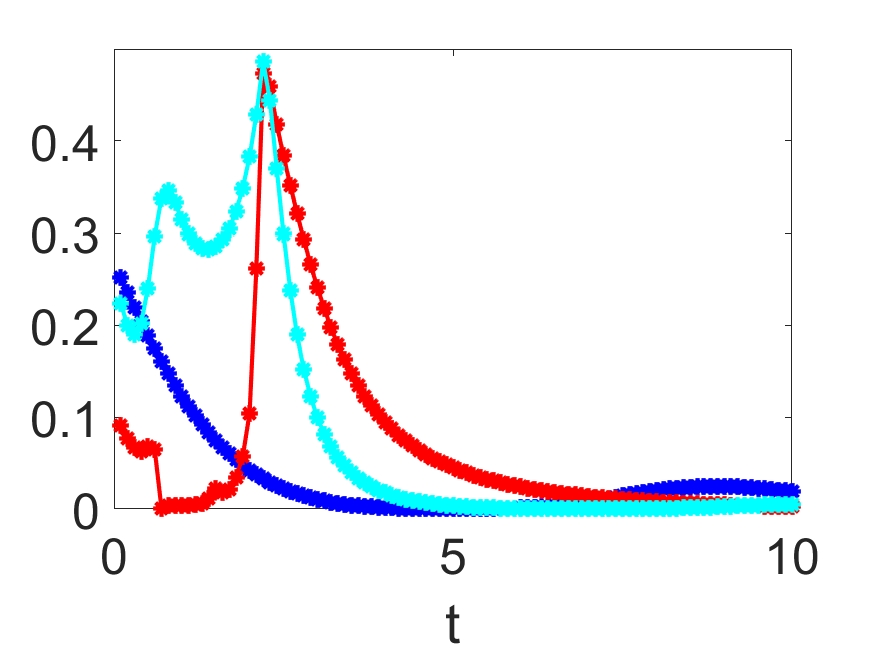}\label{HeatingDynamics1d}}

\caption{Local heating dynamics due to interplay between fluid deformation and temperature field in base flow: (a) representative fluid trajectories \RED{distinguished by red/blue/cyan (stars indicate initial positions $\xvec{x}_0$; dots indicate positions $\xvec{x}(t)$ at time levels $t = k\Delta t$ for $\Delta t = 0.1$)} versus streamlines (black); (b-d) FTLE \RED{following \eqref{FTLE}} and temperature (gradient) of fluid parcels; (e) relative heat-flux density $\beta$ \RED{following \eqref{RelativeHeatChange3}}; (f) normalised heat-flux density \RED{$\beta_\Lambda = \beta/\Lambda$}; (g) orientation of temperature gradient with principal stretching axis \RED{$\varrho_T$ following \eqref{ThermalAlignment}}; (h) orientation of heat flux with principal compression axis \RED{$\varrho_q$ following \eqref{ThermalAlignment}}.}
\label{HeatingDynamics1}
\end{figure}

The analysis below investigates the interplay between fluid deformation and thermal phenomena for three fluid parcels with initial conditions
$\xvec{x}_0$ (coloured stars) and corresponding trajectories (coloured curves) shown in Fig.~\ref{HeatingDynamics0a} relative to the streamline pattern (black curves) and central stagnation point (black star). The FTLEs in Fig.~\ref{HeatingDynamics0b} reveal a fluid deformation that is qualitatively similar to Fig.~\ref{DeformationCharacteristics1a}: peaking of the FTLE upon passage through the high-shear arc region and a subsequent decline upon entering the domain interior via the sharply deflecting streamlines at the lower arc edge. However, intensities and variations depend strongly on the initial positions. Moreover, peaks progressively weaken with multiple passages through the arc region, as demonstrated by the only minor second peak of the cyan parcel around $t\approx 7$.

The evolution of the temperature $\Ta(\xvec{x}_0,t)$ and corresponding gradient $\xvec{\nabla}_0\Ta$ along the trajectories using the temperature field at $t=20$ in Fig.~\ref{FigEulerianLagrangian2} as initial condition are shown in Fig.~\ref{HeatingDynamics0c} and Fig.~\ref{HeatingDynamics1a}, respectively, and reveal a significant thermal heterogeneity. The impact of the fluid deformation on heat transfer is shown in Fig.~\ref{HeatingDynamics1b} in terms of the momentary change in heat-flux density $\beta$ following
\eqref{RelativeHeatChange3} (expressed as $\ln\beta$ for greater legibility). This exposes peaks $\beta>1$ ($\ln\beta>0$) that coincide with the FTLE-peaks in Fig.~\ref{HeatingDynamics0b} and thus signify momentary heat-transfer enhancement due to passage through the arc region. (The $\beta$-peak for the cyan parcel corresponds with the beforementioned second passage of the arc region around $t\approx 7$.) However, the quantitative correlations are non-trivial in that the weak FTLE-peaks of the blue/cyan parcels yield pronounced $\beta$-peaks of comparable magnitude while the high FTLE-peak of the red parcel yields a substantially smaller $\beta$-peak. Moreover, both cyan and red parcels exhibit significant heat-transfer diminution ($\beta<1$ or $\ln\beta<0$) during the transition from arc region to flow interior, signifying an unfavourable orientation of fluid deformation versus temperature gradient $\xvec{\nabla}_0\Ta$.

Normalised heat-flux density $\beta_\Lambda\equiv \beta/\Lambda$ quantifies to what extent heat-transfer enhancement reaches its (theoretical) ceiling $\max \beta = \Lambda$ following \eqref{RelativeHeatChange3} and the corresponding evolutions
of this measure are shown in Fig.~\ref{HeatingDynamics1c}. This reveals short episodes of $\beta_\Lambda$ close to unity -- signifying maximum beneficial impact of fluid deformation -- that closely correlate with the (first) FTLE-peaks in Fig.~\ref{HeatingDynamics0b} followed by a dramatic breakdown. The breakdown in $\beta_\Lambda$ reflects a sudden reduction of the impact of fluid deformation on heat transfer and ensues from the diminution of heat flux $\xvec{q}_0$ by thermal homogenisation due to the abovementioned conflict between heat transfer and homogenisation. Such local thermal homogenisation, once reached, strongly diminishes the impact of future fluid deformation and thus explains why the second FTLE peak of the cyan parcel around $t\approx 7$, in contrast with the pronounced $\beta$-peak in Fig.~\ref{HeatingDynamics1b}, induces an only weak second peak in $\beta_\Lambda$. This has the fundamental implication that significant impact of fluid deformation on heat transfer and thermal homogenisation is for a given fluid parcel primarily restricted to the first passage of the arc region. These phenomena are further examined in Sec.~\ref{LocalAsymptotic}.

The dynamics of angles $\varrho_q$ and $\varrho_T$ of the heat flux $\widetilde{\xvec{q}}_0$ and temperature gradient $\xvec{\nabla}_0\Ta$, respectively, with the principal axes $\xvec{v}_{1,2}^0$ following \eqref{ThermalAlignment} substantiate the above findings. Fig.~\ref{HeatingDynamics1e} gives $\varrho_T$ and exposes highly dynamic and non-monotonic behaviour during episodes/peaks
of $\beta_\Lambda$ close to unity followed by a rapid decline -- signifying progressive alignment of $\xvec{\nabla}_0\Ta$ with principal stretching axis $\xvec{v}_2^0$ and, inherently, thermal homogenisation in $\xvec{v}_1$-direction -- during the breakdown of $\beta_\Lambda$. The relative orientation angle $\varrho_q$ of heat flux $\widetilde{\xvec{q}}_0$ in Fig.~\ref{HeatingDynamics1d} overall exhibits the same correlation with $\beta_\Lambda$ in that also here progressive alignment sets in upon breakdown of $\beta_\Lambda$.

\subsection{Local breakdown of heat-transfer enhancement}
\label{LocalAsymptotic}

\begin{figure}[htbp]
\centering

\includegraphics[trim=0cm 0cm 0cm 0cm,clip,width=.45\textwidth,height=.25\textwidth]{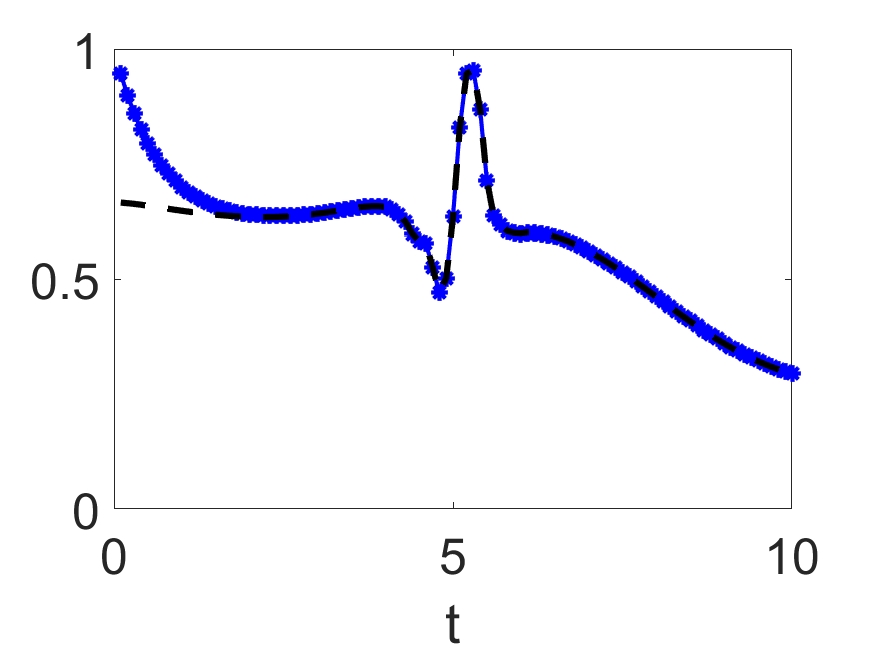}
\hspace*{12pt}
\includegraphics[trim=0cm 0cm 0cm 0cm,clip,width=.45\textwidth,height=.25\textwidth]{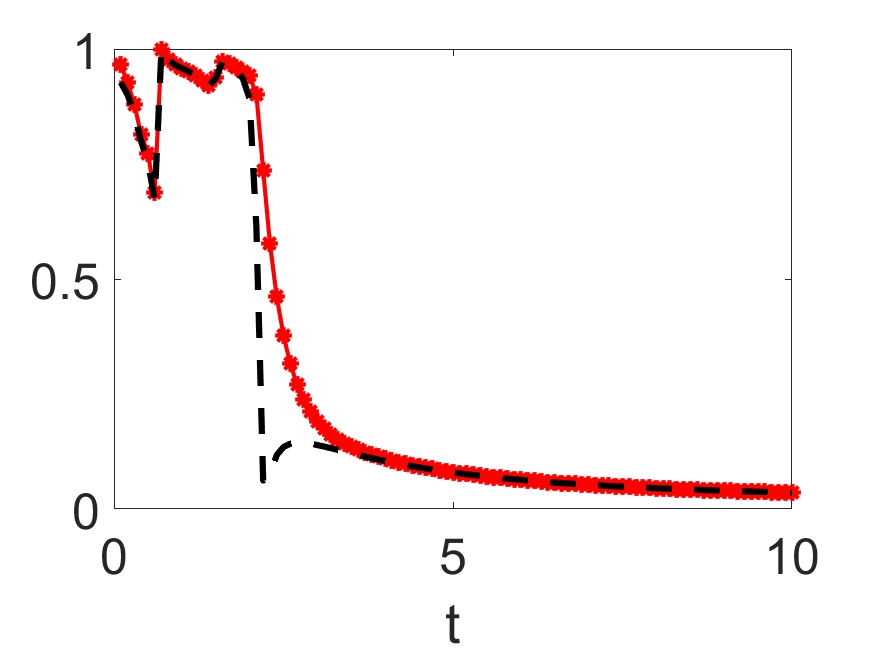}
\includegraphics[trim=0cm 0cm 0cm 0cm,clip,width=.45\textwidth,height=.25\textwidth]{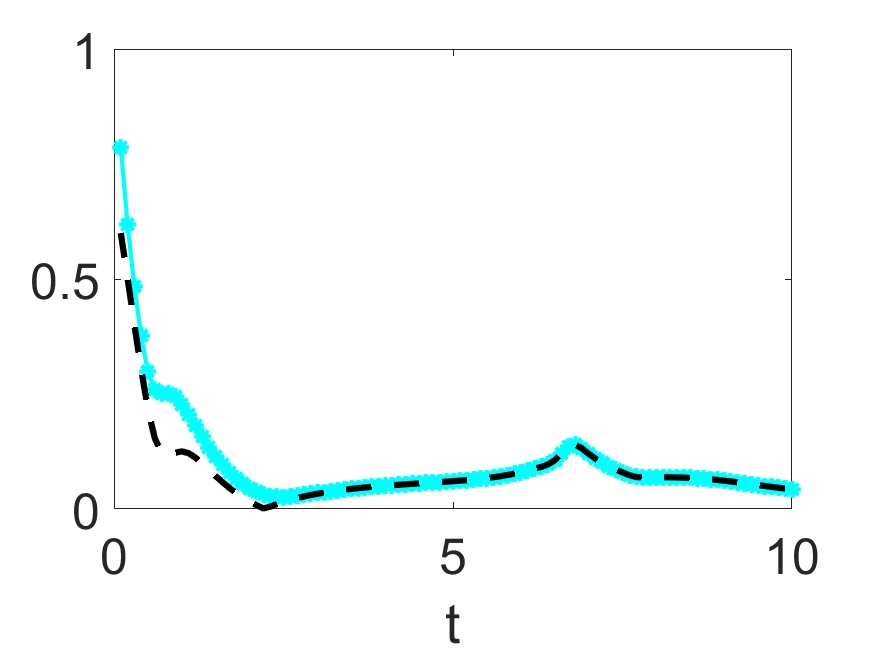}

\caption{Local breakdown of heat-transfer enhancement during transition from $\beta_\Lambda\sim\mathcal{O}(1)$ to $\beta_\Lambda\ll 1$ of normalised heat-flux density $\beta_\Lambda = \beta/\Lambda$ (solid) versus intermediate state $\beta_\Lambda^\ast$ (dashed) for fluid trajectories in Fig.~\ref{HeatingDynamics0a}.}
\label{HeatingDynamics2}
\end{figure}

The breakdown of maximum heat-transfer enhancement after the first FTLE peak demonstrated with the normalised heat-flux density $\beta_\Lambda$  in Fig.~\ref{HeatingDynamics1c} ensues from the beforementioned conflict between heat transfer and homogenisation and signifies a highly non-trivial role of fluid deformation in the thermal transport. Consider for a generic analysis the simplified case of a constant FTLE $\sigma$ in \eqref{FTLE}, resulting in $\Lambda(t) = \exp(2\sigma t)$, and temperature gradient $\xvec{\nabla}_0\Ta = c_1\exp(-\sigma_T t)\xvec{v}_1^0 + c_2\xvec{v}_2^0$, with $(c_1,c_2)$ the initial temperature gradient and $\sigma_T>0$ the thermal-homogenisation rate. Substitution in \eqref{HeatFluxLagrangian2} gives
\begin{eqnarray}
\widetilde{\xvec{q}}_0(\xvec{x}_0,t) = -\frac{1}{Pe}e^{(2\sigma - \sigma_T)t}\left[c_1\xvec{v}_1^0 + c_2e^{(\sigma_T-4\sigma)t}\xvec{v}_2^0\right],
\label{HeatFluxLagrangianAsymptotic1}
\end{eqnarray}
and restricts the thermal-homogenisation rate to $2\sigma < \sigma_T < 4\sigma$ to maintain bounded $\xvec{q}_0$. Through
\begin{eqnarray}
\tan\,\varrho_q = \frac{c_2}{c_1}e^{(\sigma_T-4\sigma)t},\quad
\tan\,\varrho_T = \frac{c_1}{c_2}e^{-\sigma_T t},\quad
\beta_\Lambda \equiv
\frac{\beta}{\Lambda} = \sqrt{\frac{c_1^2\exp(-2\sigma_T t) + c_2^2\exp(-8\sigma t)}{c_1^2\exp(-2\sigma_T t) + c_2^2}},
\label{HeatFluxLagrangianAsymptotic2}
\end{eqnarray}
this indeed yields the conjectured alignment of heat flux and temperature gradient with principal contraction ($\xvec{v}_1^0$) and stretching ($\xvec{v}_2^0$) axes, respectively, i.e. $\lim_{t\rightarrow\infty}\varrho_{q,T}=0$ and, in consequence,
\begin{eqnarray}
\lim_{t\rightarrow\infty}\widetilde{\xvec{q}}_0(\xvec{x}_0,t) = -\frac{\Lambda}{Pe}(\xvec{v}_1^0\cdot\xvec{\nabla}_0 \Ta)\xvec{v}_1^0,\quad\quad
\lim_{t\rightarrow\infty}\xvec{\nabla}_0 \Ta = (\xvec{v}_2^0\cdot\xvec{\nabla}_0 \Ta)\xvec{v}_2^0,
\label{HeatFluxLagrangianAsymptotic3}
\end{eqnarray}
thus demonstrating the generic mechanisms behind said conflict. Relations $\xvec{\nabla}\Ta = F\xvec{\nabla}_0\Ta = \exp(-\sigma t)\left[c_1\exp((2\sigma-\sigma_T)t)\xvec{v}_1 + c_2\xvec{v}_2\right]$ and $2\sigma-\sigma_T<0$ imply that the $\xvec{v}_1^0$-wise thermal homogenisation in the Lagrangian frame causes corresponding $\xvec{v}_1$-wise thermal homogenisation in the Eulerian frame and, despite the non-zero limit \eqref{HeatFluxLagrangianAsymptotic3} for $\xvec{\nabla}_0\Ta$, subsequent full thermal homogenisation $\lim_{t\rightarrow\infty}\xvec{\nabla}\Ta = \xvec{0}$. This results in rapid convergence of $\beta_\Lambda$ on the intermediate state
\begin{eqnarray}
\beta_\Lambda \approx \beta_\Lambda^\ast \equiv \sqrt{\frac{c_1^2\exp(-2\sigma_T t)}{c_1^2\exp(-2\sigma_T t) + c_2^2}} = \frac{|\xvec{v}_1^0\cdot\xvec{\nabla}_0 \Ta|}{|\xvec{\nabla}_0 \Ta|},
\label{HeatFluxLagrangianAsymptotic4}
\end{eqnarray}
for $t\gtrsim\mathcal{O}(\tau_\beta)$ and subsequent progression to limit $\lim_{t\rightarrow\infty}\beta_\Lambda = 0$. Here
\begin{eqnarray}
\tau_\beta = 1/2(4\sigma-\sigma_T),
\label{RelaxationTime}
\end{eqnarray}
is the characteristic decay time of ratio $\exp(-t/\tau_\beta) = \exp(-8\sigma t)/\exp(-2\sigma_T t)$ between the ``fast'' and ``slow'' unsteady terms in $\beta_\Lambda$. Decay rate $d\beta^\ast_\Lambda/dt = \sigma_T\beta^\ast_\Lambda({\beta^\ast_\Lambda}^2-1)\ll 0$ due to $\sigma_T\gg 1$ around the first FTLE peaks and $\beta^\ast_\Lambda<1$ causes a rapid diminution of $\beta_\Lambda^\ast$ and thus explains the dramatic breakdown from $\beta_\Lambda\sim\mathcal{O}(1)$ to $\beta_\Lambda\ll 1$ in Fig.~\ref{HeatingDynamics1c} upon reaching the intermediate state $\beta_\Lambda^\ast$ at $t\sim\mathcal{O}(\tau_\beta)$. Fig.~\ref{HeatingDynamics2} overlays shown $\beta_\Lambda$ with the generic form of $\beta_\Lambda^\ast$ following \eqref{HeatFluxLagrangianAsymptotic4} (dashed black curve) and the close agreement beyond the first FTLE peaks -- as well as during the build-up towards these peaks for the red and blue parcels -- indeed substantiates these findings.

The above exposes the local breakdown of heat-transfer enhancement observed in Sec.~\ref{HeatTransferEnhancement} as generic (and consistent with the ``diffusive relaxation'' during chaotic advection observed in \cite{Tang1999}). This renders the common belief of efficient heat transfer being automatic with fluid mixing and chaotic advection fallacious. Fluid deformation namely plays a dual role by, on the one hand, enhancing heat transfer and thermal homogenisation between/of fluid parcels (provided favourable orientations with the temperature gradient) yet, on the other hand, restricting its beneficial impact to short-lived episodes in the transient. Their specific occurrence, duration and intensity for a given fluid parcel may vary greatly, though (Fig.~\ref{HeatingDynamics2}). However, the local thermal equilibration of fluid parcels progressively lagging behind with faster fluid deformation strongly suggests that $\sigma_T$ and, consequently, $\tau_\beta$ tend towards their lower bounds
$\min\sigma_T = 2\sigma$ and $\min \tau_\beta = 1/4\sigma$, respectively, with increasing $\sigma$. This advances $\tau_\beta\simeq\mathcal{O}(0.3-3)$ as estimated duration of said episodes for fluid parcels passing through the high-shear arc region using $\sigma\simeq\mathcal{O}(0.1-1)$ from Fig.~\ref{DeformationCharacteristics1b}. The arc region is to emerge as dominant for the global behaviour in Sec.~\ref{GlobalHeatingDynamics2}.

\subsection{Global metrics for heating and their general dynamics}
\label{GlobalHeatingDynamics1}

Essential for effective control is an adequate description of the global dynamics associated with the above local behaviour. Key to this is that the control target, viz. fast heating of the cold fluid towards the hot equilibrium $\widetilde{T}_\infty=0$, entails concurrent accomplishment of two goals: (i) fast increase of the total energy content (denoted ``energising'' hereafter) and (ii) fast homogenisation of the internal temperature distribution. Decomposition of the transient temperature as
\begin{eqnarray}
\widetilde{T}(\xvec{x},t) = \bar{T}(t) + T'(\xvec{x},t),\quad\quad \bar{T}(t) = \frac{1}{A}\int_{\mathcal{D}}\widetilde{T}(\xvec{x},t)d^2\xvec{x},\quad\quad
T'(\xvec{x},t) \equiv \widetilde{T}(\xvec{x},t) - \bar{T}(t),
\label{DecompositionHeatingHomogenisation}
\end{eqnarray}
with here $A = \pi$, isolates the temperature contributions corresponding with these goals. Average temperature $\bar{T}(t) = \widetilde{E}(t)/\pi$ represents the total energy content $\widetilde{E}$ of field $\widetilde{T}$ (here $\widetilde{E}\leq 0$); ``heterogeneity'' $T'(\xvec{x},t)$ represents the departure of $\widetilde{T}$ from the momentary homogeneous state $\bar{T}$.

Decomposition \eqref{DecompositionHeatingHomogenisation} exposes the far greater dynamic complexity of the current heating problem compared to homogenisation problems commonly considered in literature (Sec.~\ref{Intro}). The latter generically concern adiabatic domains, implying $\bar{T}(t)=0$, and thus depend solely on the single field $\widetilde{T}=T'$. Fluid heating, on the other hand, encompasses two processes, viz. energising and homogenisation, and relevant to its dynamics are therefore the total transient field $\widetilde{T}$ as well as its individual components $\bar{T}(t)$ and $T'$ according to \eqref{DecompositionHeatingHomogenisation}.

The above advances 3 measures for the global dynamic behaviour of the heating process, viz.
\begin{eqnarray}
\Ja(t) \equiv \frac{1}{\pi}\int_{\mathcal{D}}\widetilde{T}^2(\xvec{x},t)d^2\xvec{x},\quad
\Jb(t) \equiv \frac{1}{\pi}\int_{\mathcal{D}}\widetilde{T}(\xvec{x},t)d^2\xvec{x},\quad
\Jc(t) \equiv \frac{1}{\pi}\int_{\mathcal{D}}{T'}^2(\xvec{x},t)d^2\xvec{x},
\label{HeatingNorms}
\end{eqnarray}
with $\Ja$ the global departure from equilibrium, $\Jb = \bar{T}$ the normalised energy content of the transient state and $\Jc$ the global heterogeneity. These measures via \eqref{DecompositionHeatingHomogenisation} and $\int_{\mathcal{D}}T'(\xvec{x},t)d^2\xvec{x}=0$ relate as
\begin{eqnarray}
\Ja(t) = \Jb^2(t) + \Jc(t),
\label{RelationMetrics}
\end{eqnarray}
and thus effectively constitute two independent degrees of freedom.

The evolution of $\Jb$ is governed by the integral energy balance corresponding with \eqref{eq:ade_pde_discrete}. Property $\mathbf{v}\cdot\pmb{\nabla}\Ta = \pmb{\nabla}\cdot(\xvec{v}\Ta)$ for $\pmb{\nabla}\cdot\xvec{v}=0$
\RED{upon elimination of time derivative $\partial \Ta/\partial t$ with \eqref{eq:ade_pde_discrete} and application of Gauss'} divergence theorem \cite{Kreyszig2011} yields
\begin{eqnarray}
\frac{d\Jb}{dt} = \frac{1}{\pi}\int_{\mathcal{D}}\frac{\partial \Ta}{\partial t} d^2\xvec{x} = -\frac{1}{\pi}\int_{\mathcal{D}}\pmb{\nabla}\cdot(\xvec{v}\Ta + \widetilde{\xvec{q}})d^2\xvec{x}
= -\frac{1}{\pi}\int_{\Gamma}(\xvec{v}\Ta + \widetilde{\xvec{q}})\cdot\xvec{n}ds,
\label{HeatingFluid1}
\end{eqnarray}
with $\widetilde{\xvec{q}}=-Pe^{-1}\xvec{\nabla}\Ta$. Here the convective flux $\xvec{v}\cdot\xvec{n}\,\Ta$ vanishes by virtue of both the impenetrable boundary ($\xvec{v}\cdot\xvec{n} = 0$) and the homogeneous Dirichlet condition $\Ta|_{\Gamma}=0$. This reveals that energising effectively occurs only by the diffusive flux $\widetilde{\xvec{q}}$ normal to the boundary, i.e.
\begin{eqnarray}
\frac{d\Jb}{dt} = -\frac{1}{\pi}\int_{\Gamma}\widetilde{\xvec{q}}\cdot\xvec{n}ds = \frac{1}{\pi Pe}\int_{\Gamma}\xvec{n}\cdot\pmb{\nabla}\widetilde{T}ds,
\label{HeatingFluid2}
\end{eqnarray}
and flow $\xvec{v}$ only indirectly affects the heat-flux density and temperature gradient via motion and deformation of fluid parcels by the mechanisms from Sec.~\ref{HeatTransferEnhancement}. The evolution of $\Ja$ is governed by
\begin{eqnarray}
\frac{d\Ja}{dt} = \frac{2}{\pi}\int_{\mathcal{D}}\widetilde{\xvec{q}}\cdot\pmb{\nabla}\widetilde{T}d^2\xvec{x}
 = -\frac{2}{\pi Pe}\int_{\mathcal{D}}|\pmb{\nabla}\widetilde{T}|^2d^2\xvec{x},
\label{HeatingFluid3}
\end{eqnarray}
which follows in a similar way as \eqref{HeatingFluid2} from the energy equation (\ref{EvolutionL2norm}). The evolution of $\Jc$ via relation \eqref{RelationMetrics} depends on \eqref{HeatingFluid2} and \eqref{HeatingFluid3} according to $d\Jc/dt = d\Ja/dt - 2\Jb(d\Jb/dt)$.

Relation \eqref{HeatingFluid3} via inequality $|\pmb{\nabla}\Ta|> 0$ in at least one non-zero subset $\xvec{x}\in\mathcal{D}$ for any non-uniform $\Ta$ implies $d\Ja/dt < 0$ and thus a monotonic decay
of measure $\Ja$ from $\Ja(0)>0$ to $\lim_{t\rightarrow\infty}\Ja(t) = 0$. This is a consequence of the Second Law of Thermodynamics, which dictates that diffusive heat flux always acts against the temperature gradient, i.e. $\widetilde{\xvec{q}} = -Pe^{-1}\pmb{\nabla}\Ta$, and thereby yields the particular RHS of \eqref{HeatingFluid3}. This implies that {\it any} non-zero transient temperature $\widetilde{T}\left(\mathbf{x},t\right)$, irrespective of the flow, always evolves towards the final equilibrium $\widetilde{T}_\infty=0$ and the system is intrinsically stable. The decay {\it rate} of $\Ja$ depends essentially on the flow, however. The principal goal of the control strategy in Sec.~\ref{ControlStrategy} is finding the {\it fastest} route towards this equilibrium.

Measure $\Jc$, on the other hand, is inherently {\it non-monotonic} due to identical initial and asymptotic conditions $\Jc(0)=\lim_{t\rightarrow\infty}\Jc(t)=0$ and this constitutes a further essential departure from the beforementioned homogenisation problems (where $\Jc=\Ja$ implies monotonic behaviour). Said conditions namely dictate an initial growth of heterogeneity
($d\Jc/dt|_{t=0}>0$) that through an inevitable $\Jc(t)>0$ at intermediate times $t$ settles on an eventual decline ($d\Jc/dt < 0$) towards the final state. Through relation \eqref{RelationMetrics} this in principle admits non-monotonic evolutions from $\Jb(0)<0$ to $\lim_{t\rightarrow\infty}\Jb(t)=0$ for measure $\Jb$ as well. However, here the uniform boundary condition $\Ta|_\Gamma=0$ together with the uniform initial condition $\Ta(\xvec{x},0)=-1$ implies $\Ta(\xvec{x},t)\leq 0$ for all $\xvec{x}$ and $t$ and
thus $\partial\Ta/\partial r|_\Gamma>0$ and, in consequence, monotonic $d\Jb/dt>0$.

\subsection{Global impact of fluid deformation}
\label{GlobalHeatingDynamics2}

The evolution of metrics $\Jb$ and $\Ja$ -- and, via relation \eqref{RelationMetrics}, indirectly also of $\Jc$ -- is determined by \eqref{HeatingFluid2} and \eqref{HeatingFluid3}, respectively. Lagrangian coordinates admit, similar as in Sec.~\ref{HeatTransferEnhancement}, investigation of the impact of fluid flow on these evolutions. Consider to this end said metrics for a material region with current Eulerian position $\mathcal{D}(t)$ and its boundary $\Gamma(t)$, i.e.
\begin{eqnarray}
\Jbx(t) = \frac{1}{\pi}\int_{\mathcal{D}_0}\widetilde{T}(\xvec{x}_0,t)d^2\xvec{x}_0,\quad\quad
\Jax(t) = \frac{1}{\pi}\int_{\mathcal{D}_0}\widetilde{T}^2(\xvec{x}_0,t)d^2\xvec{x}_0,
\label{HeatingFluid4}
\end{eqnarray}
with $\mathcal{D}_0$ and $\Gamma_0$ the corresponding initial (fixed) position in the Eulerian (Lagrangian) frame. The evolutions of \eqref{HeatingFluid4} are in Lagrangian coordinates described by
\begin{eqnarray}
\frac{d\Jbx}{dt} = -\frac{1}{\pi}\int_{\Gamma_0}\widetilde{\xvec{q}}_0\cdot\xvec{n}_0ds_0,\quad\quad
\frac{d\Jax}{dt} = \frac{2}{\pi}\int_{\mathcal{D}_0}\widetilde{\xvec{q}}_0\cdot\pmb{\nabla}_0\widetilde{T}d^2\xvec{x}_0 - \frac{2}{\pi}\int_{\Gamma_0}\widetilde{T}\widetilde{\xvec{q}}_0\cdot\xvec{n}_0ds_0,
\label{HeatingFluid5}
\end{eqnarray}
where the boundary integral in $d\Jax/dt$ emerges for generic material interfaces $\Gamma(t)$ evolving with the flow. This integral can be omitted in the present analysis for reasons explained below.

The impact of the flow on the global dynamics is investigated via the rates-of-change of the metric evolutions \eqref{HeatingFluid5}, which, upon omitting said boundary integral, are governed by
\begin{eqnarray}
\frac{d^2\Jbx}{dt^2} &=& \frac{1}{\pi Pe}\int_{\Gamma_0}\left(\frac{\partial B}{\partial t}\xvec{\nabla}_0 \widetilde{T}\right)\cdot\xvec{n}_0ds_0
+ \frac{1}{\pi Pe}\int_{\Gamma_0}\left(B\xvec{\nabla}_0\left(\frac{\partial\widetilde{T}}{\partial t}\right)\right)\cdot\xvec{n}_0ds_0\nonumber\\
\frac{d^2\Jax}{dt^2} &=& -\frac{2}{\pi Pe}\int_{\mathcal{D}_0}\left(\frac{\partial B}{\partial t}\xvec{\nabla}_0 \widetilde{T}\right)\cdot\xvec{\nabla}_0 \widetilde{T}d^2\xvec{x}_0
-\frac{4}{\pi Pe}\int_{\mathcal{D}_0}(B\xvec{\nabla}_0 \widetilde{T})\cdot\xvec{\nabla}_0\left(\frac{\partial\widetilde{T}}{\partial t}\right)d^2\xvec{x}_0,
\label{HeatingFluid6}
\end{eqnarray}
using $\widetilde{\xvec{q}}_0 = -Pe^{-1}B\xvec{\nabla}_0 \widetilde{T}$. The link with the corresponding behaviours of global metrics $\Jb$ and $\Ja$ in \eqref{HeatingNorms} in the Eulerian frame is established via leading-order approximations of \eqref{HeatingFluid6}. This embarks on considering (i) an initial material boundary $\Gamma_0$ of circular shape with radius $r_0 = 1 - \epsilon_r$ very close to the circular boundary $\Gamma$ of radius $r=1$ (i.e. $\epsilon_r \ll 1$) and (ii) incremental Lagrangian motion from $\xvec{x}(t)$ to $\xvec{x}(t+dt) = \xvec{x}(t) + d\xvec{x} = \xvec{x}(t) + \xvec{u}dt$. Thus $\mathcal{D}(t) \approx \mathcal{D}_0 \approx \mathcal{D}$ and $\Gamma(t) \approx \Gamma_0 \approx \Gamma$, which via $\widetilde{T}|_{\Gamma(t)} \approx \widetilde{T}|_{\Gamma} = 0$ indeed to good approximation implies vanishing of the boundary integral in the evolution of $\Jax$ in \eqref{HeatingFluid5}. The leading-order approximations of the relevant quantities and variables, i.e. $\Jbx = \Jb$ and $\Jax = \Ja$, $F = F_0 =I$, $B = I$, $d\xvec{x}=d\xvec{x}_0$, $\xvec{n}=\xvec{n}_0$ and $ds = ds_0$, combined with the rate-of-change $\partial B/\partial t \equiv \dot{B} = -2D$ of the left Cauchy-Green tensor as per \cite{Chadwick1999} subsequently yield
\begin{eqnarray}
\frac{d^2\Jb}{dt^2} &=& \frac{1}{\pi Pe}\int_{\Gamma}(\dot{B}\xvec{\nabla} \widetilde{T})\cdot\xvec{n}ds + \frac{1}{\pi}\int_{\Gamma}\xvec{n}\cdot\xvec{\nabla}\left(\frac{\partial\widetilde{T}}{\partial t}\right)ds\nonumber\\
\frac{d^2\Ja}{dt^2} &=& -\frac{2}{\pi Pe}\int_{\mathcal{D}}(\dot{B}\xvec{\nabla}\widetilde{T})\cdot\xvec{\nabla} \widetilde{T}d^2\xvec{x} -\frac{4}{\pi Pe}\int_{\mathcal{D}}
\xvec{\nabla}\widetilde{T}\cdot\xvec{\nabla}\left(\frac{\partial\widetilde{T}}{\partial t}\right)d^2\xvec{x},
\label{HeatingFluid7}
\end{eqnarray}
with $D = 1/2[\xvec{\nabla}\xvec{u}+ (\xvec{\nabla}\xvec{u})^\dagger]$ the strain-rate tensor.

Relations \eqref{HeatingFluid7} describe the rates-of-change of the metric evolutions \eqref{HeatingFluid2}
and \eqref{HeatingFluid3} and via the leading integrals incorporate the momentary impact of fluid deformation on the heating dynamics for given temperature $T(\xvec{x},t)$.
The corresponding integrands describe the rate-of-change of integrands $\bar{f}\equiv\xvec{n}\cdot\xvec{\nabla}\Ta|_\Gamma = \partial\Ta/\partial r|_\Gamma$ and $\widetilde{f}\equiv|\xvec{\nabla}\Ta|^2$ in \eqref{HeatingFluid2} and \eqref{HeatingFluid3}, respectively, via
\begin{eqnarray}
\bar{g}(\xvec{x},t) &\equiv& \left.\frac{d\bar{f}}{dt}\right|_{\Ta} = (\dot{B}\xvec{\nabla}\widetilde{T})\cdot\xvec{n}
= 2\mu\left[(\xvec{w}_1\cdot\xvec{\nabla}\widetilde{T})\xvec{w}_1\cdot\xvec{n} - (\xvec{w}_2\cdot\xvec{\nabla}\widetilde{T})\xvec{w}_2\cdot\xvec{n}\right] \nonumber\\
\widetilde{g}(\xvec{x},t) &\equiv& \left.\frac{d\widetilde{f}}{dt}\right|_{\Ta} = (\dot{B}\xvec{\nabla} \widetilde{T})\cdot\xvec{\nabla} \widetilde{T}
= 2\mu\left[(\xvec{w}_1\cdot\xvec{\nabla}\widetilde{T})^2 - (\xvec{w}_2\cdot\xvec{\nabla}\widetilde{T})^2\right],
\label{HeatingFluid8a}
\end{eqnarray}
with $\mu > 0$ ($-\mu <0$) the stretching (contraction) rate along principal axis $\xvec{w}_2$ ($\xvec{w}_1$) of tensor $D = \mu\xvec{w}_2\xvec{w}_2 - \mu\xvec{w}_1\xvec{w}_1$. Here the impact of fluid deformation on heat transfer, analogous to \eqref{HeatFluxLagrangian2}, again depends on the relative orientation between principal deformation axes and temperature gradient. The qualitative global impact is determined by the sign of the leading integrals in \eqref{HeatingFluid7}:
\begin{itemize}

\item {\boldmath$\bar{G}(t) \equiv \int_\Gamma\bar{g}(\xvec{x},t)ds > 0$} momentarily accelerates the growth from $\Jb<0$ to $\lim_{t\rightarrow\infty}\Jb(t)=0$ compared to a non-deforming
fluid (i.e. $d^2\Jb/dt^2|_{\mu>0} > d^2\Jb/dt^2|_{\mu=0}$) and thus accomplishes faster energising. Conversely, $\bar{G}(t)<0$ delays global energising.

\item {\boldmath$\widetilde{G}(t) \equiv\int_\mathcal{D}\widetilde{g}(\xvec{x},t)d^2\xvec{x} > 0$} momentarily accelerates the decay from $\Ja>0$ to $\lim_{t\rightarrow\infty}\Ja(t)=0$ compared to a non-deforming fluid (i.e. $d^2\Ja/dt^2|_{\mu>0} < d^2\Ja/dt^2|_{\mu=0}$) and thus accomplishes faster equilibration. Conversely, $\widetilde{G}(t)<0$ delays global equilibration.

\end{itemize}
Integrals $\bar{G}(t)$ and $\widetilde{G}(t)$ incorporate the net impact on global energising and equilibration, respectively, by local enhancement (i.e. $\bar{g}(\xvec{x},t)>0$ and $\widetilde{g}(\xvec{x},t)>0$) or diminution (i.e. $\bar{g}(\xvec{x},t)<0$ and $\widetilde{g}(\xvec{x},t)<0$). However, the local contributions to the global behaviour may, consistent with Fig.~\ref{HeatingDynamics1}, vary significantly. This admits investigation via the corresponding relative change rates
\begin{eqnarray}
\bar{\beta} \equiv \frac{\bar{g}}{\bar{f}}  = \xvec{e}_r^\dagger\cdot \dot{B}\cdot \xvec{e}_r = -2\xvec{e}_r^\dagger\cdot D\cdot \xvec{e}_r = -2\frac{\partial v_r}{\partial r},\quad\quad
\widetilde{\beta} \equiv \frac{\widetilde{g}}{\widetilde{f}} = \xvec{e}_T^\dagger\cdot\dot{B}\cdot \xvec{e}_T = -2\xvec{e}_T^\dagger\cdot D\cdot \xvec{e}_T,
\label{HeatingFluid8b}
\end{eqnarray}
with $\xvec{e}_T = \xvec{\nabla}T/|\xvec{\nabla}T|$, and constitute Eulerian counterparts to $\beta$ following \eqref{RelativeHeatChange3}. Here $\bar{\beta}$ follows from uniformity of the wall
temperature, which implies $\xvec{\nabla T} = (\partial T/\partial r)\xvec{e}_r$ at boundary $\Gamma$ and to leading-order approximation on the beforementioned circles $\Gamma_0$. This furthermore implies $\xvec{e}_T = \xvec{e}_r$ and thus $\widetilde{\beta} = \bar{\beta}$ at $\Gamma$. Moreover, both measures are bounded as $-2\mu\leq\bar{\beta}\leq2\mu$ and $-2\mu\leq\widetilde{\beta}\leq2\mu$.

\begin{figure}[H]

\centering

\subfigure[$\bar{\beta}$ at $r_0 = 0.99$.]{
\includegraphics[trim=1cm 0cm 2cm 1cm,clip,width=.45\textwidth,height=.25\textwidth]{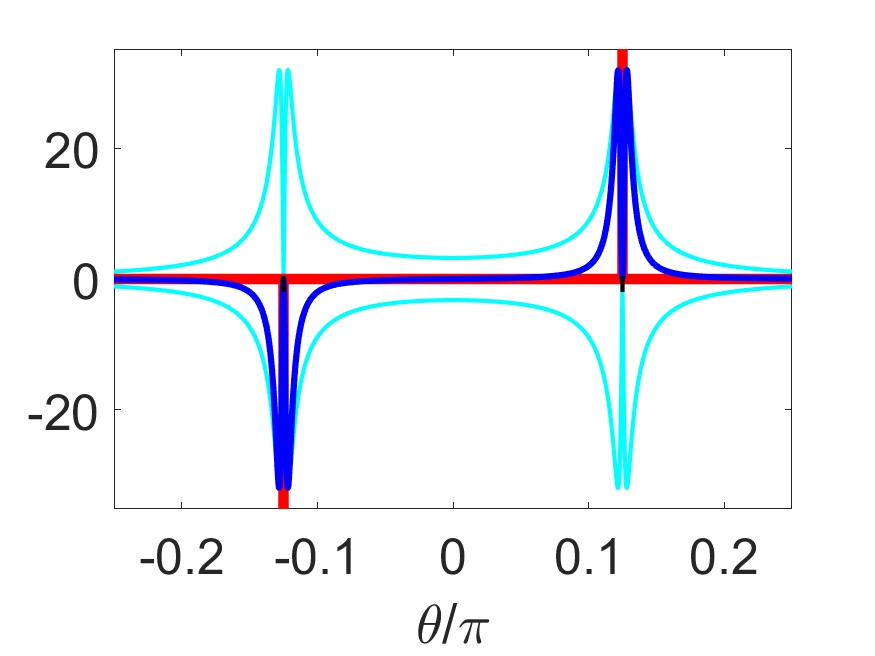}
\label{HeatingDynamics3a}
}
\hspace*{12pt}
\subfigure[$\bar{f}$ at $r=1$.]{
\includegraphics[trim=1cm 0cm 2cm 1cm,clip,width=.45\textwidth,height=.25\textwidth]{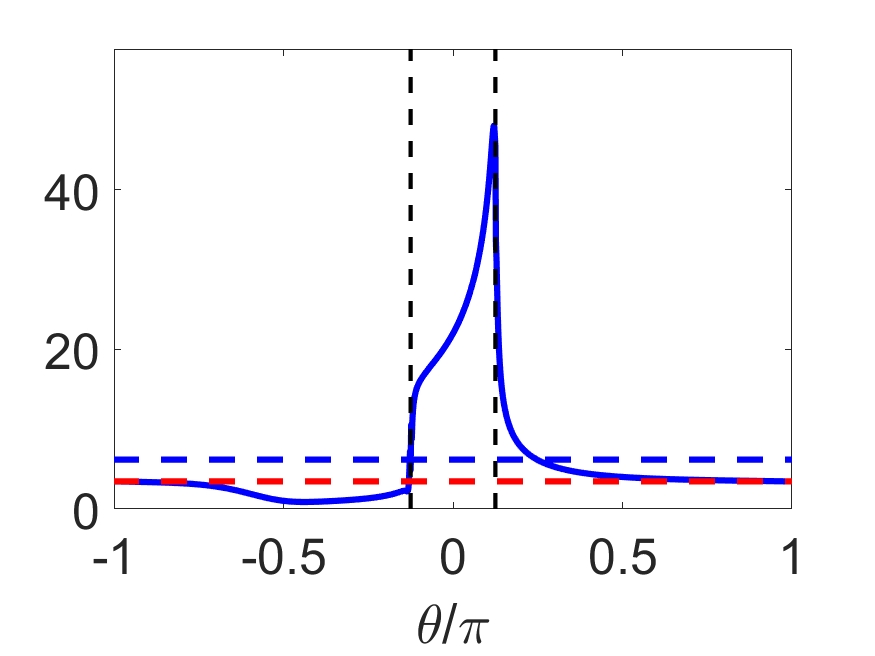}
\label{HeatingDynamics3b}
}
\subfigure[$\bar{g}/\max|\bar{g}|$ at $r_0 = 0.99$.]{
\includegraphics[trim=1cm 0cm 2cm 1cm,clip,width=.45\textwidth,height=.25\textwidth]{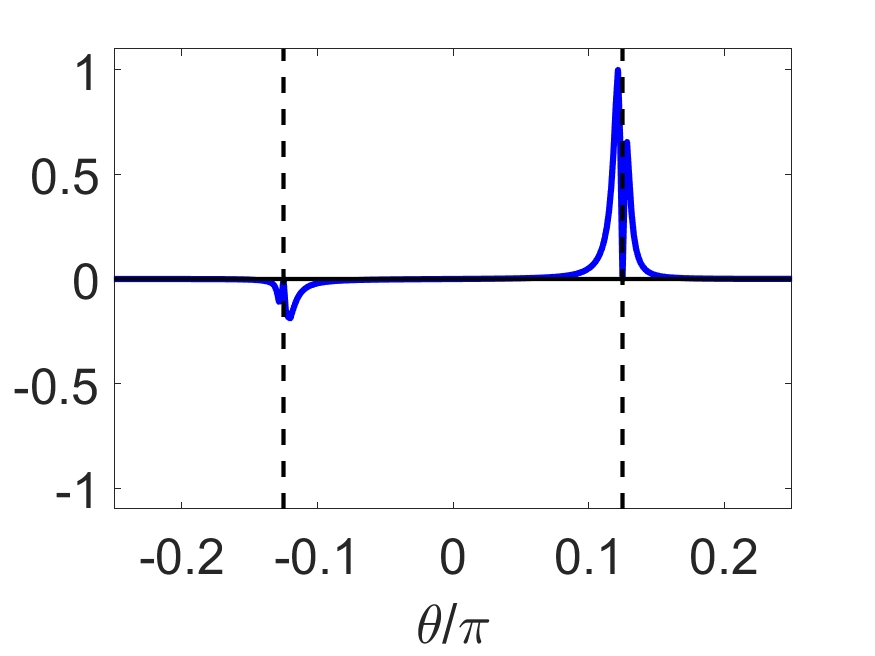}
\label{HeatingDynamics3c}
}

\caption{Impact of fluid deformation on energising for temperature at $t=20$ in Fig.~\ref{FigEulerianLagrangian2}: (a) relative change rate $\bar{\beta} = \bar{g}/\bar{f}$ on circle $r_0 = 0.99$ (blue) including limit on $r=1$ (red) and bounds (cyan); (b) wall temperature gradient
$\bar{f} = \partial \Ta/\partial r|_\Gamma$ (blue) including average (blue dashed) and diffusive limit (red dashed); (c) change rate $\bar{g}$ at $r_0 = 0.99$ (blue). Vertical dashed lines indicate leading ($\theta = \Delta/2$) and trailing ($\theta = -\Delta/2$) arc edges.}
\label{HeatingDynamics3}
\end{figure}

Incompressibility $\xvec{\nabla}\cdot\xvec{v}=0$ yields $\partial v_r/\partial r = -r^{-1}\partial v_\theta/\partial \theta$ and via \eqref{HeatingFluid8b} links the energising dynamics to the boundary velocity $\xvec{v}_1|_\Gamma = v_\theta\xvec{e}_\theta$. Through $v_\theta = H(\theta-\Delta/2)-H(\theta+\Delta/2)$ this gives $\bar{\beta} = 2\delta(\theta-\Delta/2) - 2\delta(\theta+\Delta/2)$ as singular limit for $\bar{\beta}$ on $\Gamma$, with $H$ and $\delta$ the well-known Heaviside and Dirac functions, respectively. This, in turn, yields
\begin{eqnarray}
\bar{G}(t) = r_0\int_0^{2\pi} \bar{\beta}(\theta;r_0) \frac{\partial \Ta}{\partial r} d\theta \stackrel{r_0=1}{=} 2\left\{\left.\frac{\partial \Ta}{\partial r}\right|_{\theta = \Delta/2} - \left.\frac{\partial \Ta}{\partial r}\right|_{\theta = -\Delta/2}\right\},
\label{HeatingFluid9}
\end{eqnarray}
as leading-order approximation for integral $\bar{G}$ on said $\Gamma_0$ and its corresponding limit on $\Gamma$. Fig.~\ref{HeatingDynamics3a} shows $\bar{\beta}$ for $r_0=0.99$ (blue) including its limit (red) and bounds (cyan) and reveals that significant contributions to $\bar{G}$ are indeed restricted to the direct proximity of the arc edges $\theta = \pm\Delta/2$ (dashed lines). The rapid material stretching along streamlines and subsequent relaxation near arc edges $\theta = \Delta/2$ and $\theta = -\Delta/2$, respectively, demonstrated in Fig.~\ref{DeformationCharacteristics1} causes a relative steepening (flattening) of wall temperature gradient $\bar{f} = \partial \Ta/\partial r|_\Gamma$ -- or equivalently, a local rate of change $\bar{\beta}>0$ ($\bar{\beta}<0$) -- near former (latter) arc edge by the mechanism according to Sec.~\ref{HeatTransferEnhancement}. This implies
\begin{eqnarray}
0<\left.\frac{\partial \Ta}{\partial r}\right|_{\theta=-\Delta/2,t>0}\quad<\quad\left.\frac{\partial \Ta}{\partial r}\right|_{\Gamma,t=0}\quad<\quad\left.\frac{\partial \Ta}{\partial r}\right|_{\theta=\Delta/2,t>0},
\label{HeatingFluid10}
\end{eqnarray}
due to the uniform initial gradient $\partial \Ta/\partial r|_{\Gamma,t=0} > 0$ for the initial/boundary conditions in \eqref{eq:ade_pde_discrete} and via \eqref{HeatingFluid9} thus always accelerates energising ($\bar{G}>0$). This behaviour is demonstrated in Fig.~\ref{HeatingDynamics3b} by $\partial \Ta/\partial r|_\Gamma$ (blue curve) and corresponding integrand $\bar{g}$ in Fig.~\ref{HeatingDynamics3c} for the
temperature field at $t=20$ in Fig.~\ref{FigEulerianLagrangian2} and, in accordance with $\bar{G}>0$, yields an average temperature gradient (blue dashed line in Fig.~\ref{HeatingDynamics3b}) and associated boundary heat flux that is significantly larger than the diffusive limit (red dashed line). However, consistent with the local behaviour exposed in Sec.~\ref{HeatTransferEnhancement}, unfavourable orientations of fluid deformation versus temperature gradient may via \eqref{HeatingFluid9} also cause $\bar{G}<0$ and thus (temporary) diminution of global energising.

\begin{figure}[htbp]
\centering

\subfigure[$\mu_{\rm log}$]{\includegraphics[trim=8cm 1.8cm 7cm 1cm,clip,width=.29\textwidth]{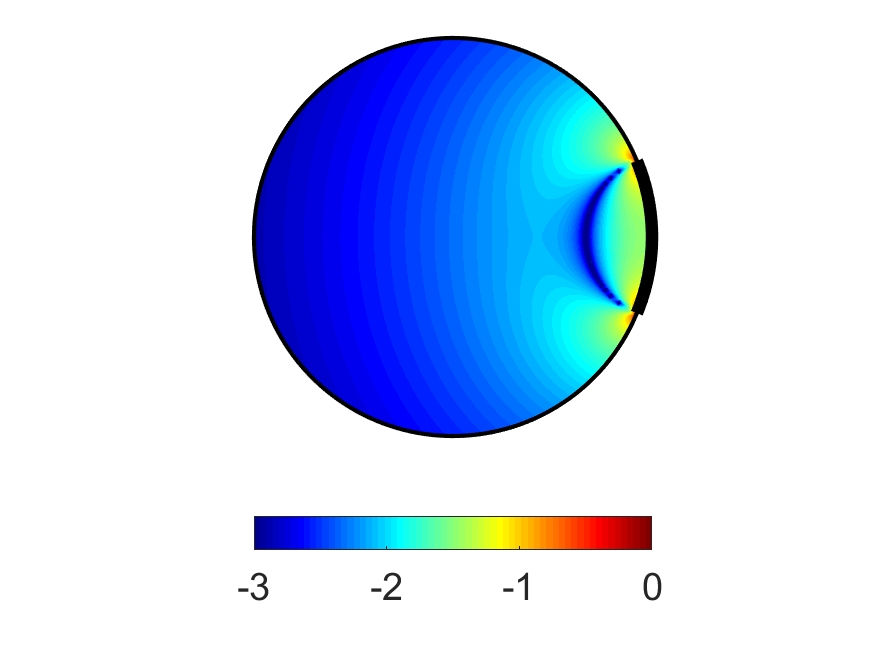}\label{HeatingDynamics4a}}
\hspace*{16pt}
\subfigure[$\widetilde{\beta}_{\rm log}$]{\includegraphics[trim=0cm 0cm 0cm 0cm,clip,width=.6\textwidth]{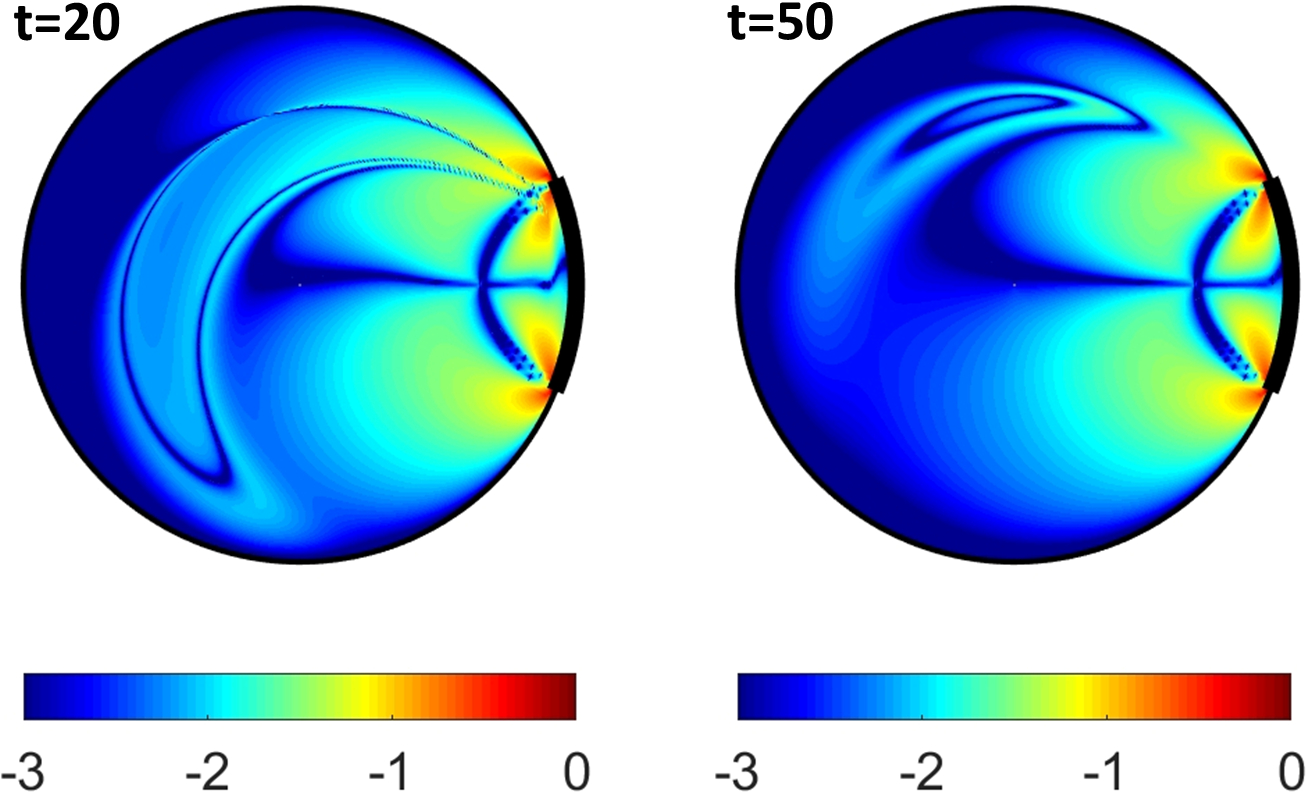}\label{HeatingDynamics4b}}

\subfigure[Partition of $\widetilde{\beta}$ into impact levels.]{
\includegraphics[trim=0cm 0cm 0cm 0cm,clip,width=.53\textwidth]{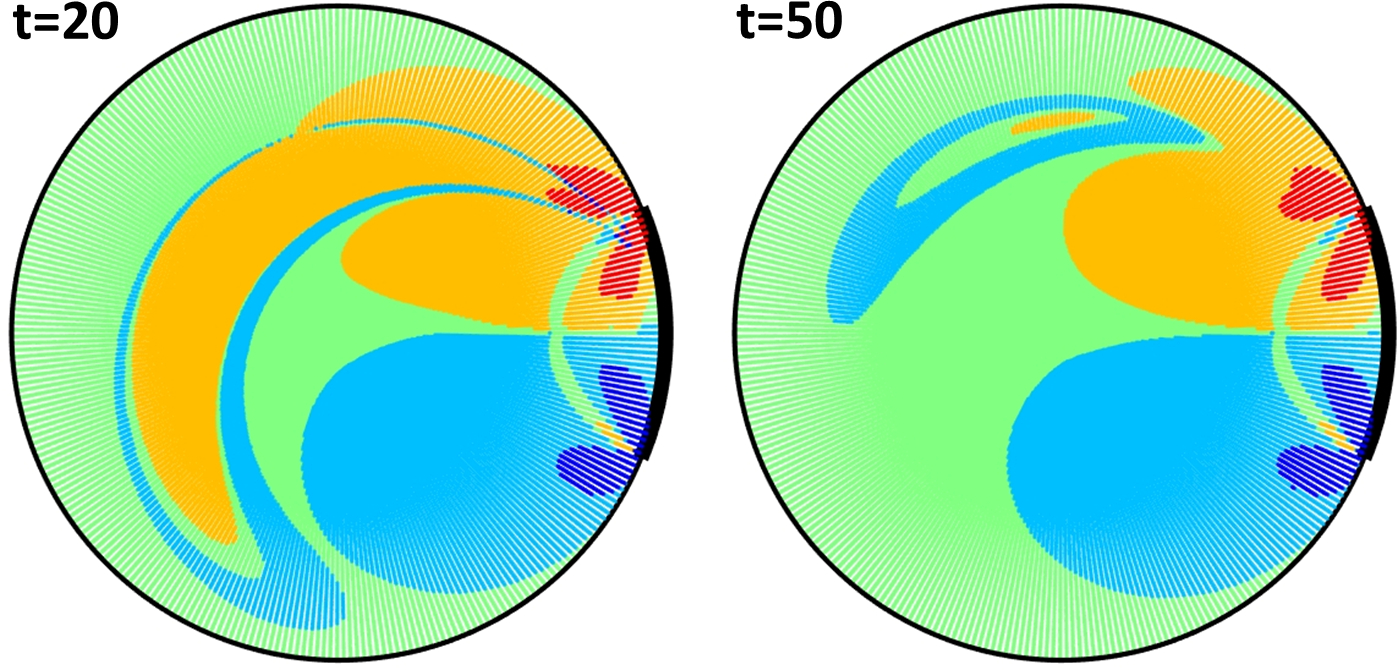}
\hspace*{6pt}
\includegraphics[trim=0cm 0cm 0cm 0cm,clip,width=.1\textwidth]{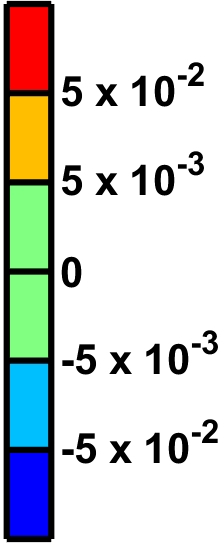}
\label{HeatingDynamics4c}}
\hspace*{16pt}
\subfigure[Reoriented $\Ta$.]{\includegraphics[trim=6cm 2cm 5cm 1cm,clip,width=.27\textwidth]{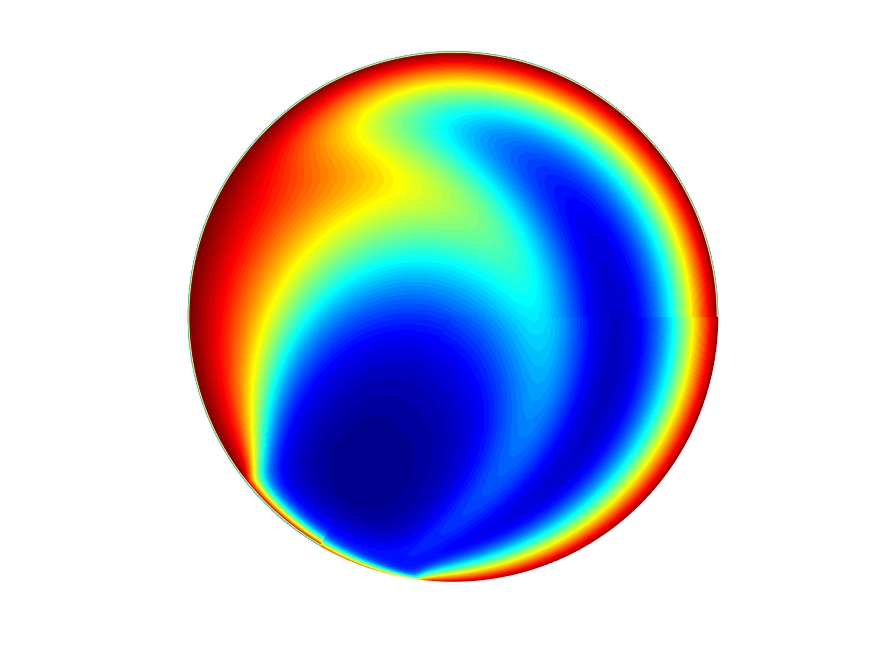}\label{HeatingDynamics4e}}

\subfigure[$\widetilde{\beta}_{\rm log}$ for reoriented $\Ta$.]{\includegraphics[trim=0cm 0cm 0cm 0cm,clip,width=.47\textwidth]{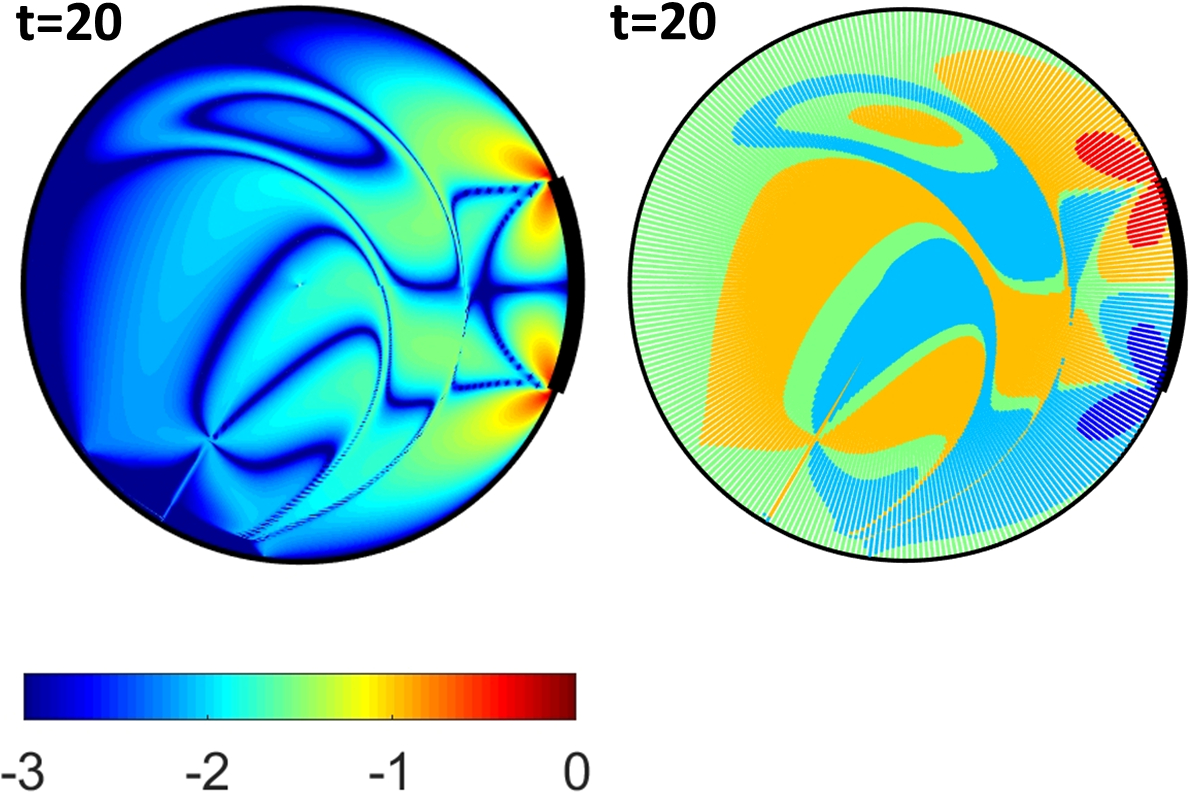}\label{HeatingDynamics4f}}
\hspace*{12pt}
\subfigure[$\widetilde{\beta}_{\rm log}^\ast$]{\includegraphics[trim=0cm 0cm 0cm 0cm,clip,width=.47\textwidth]{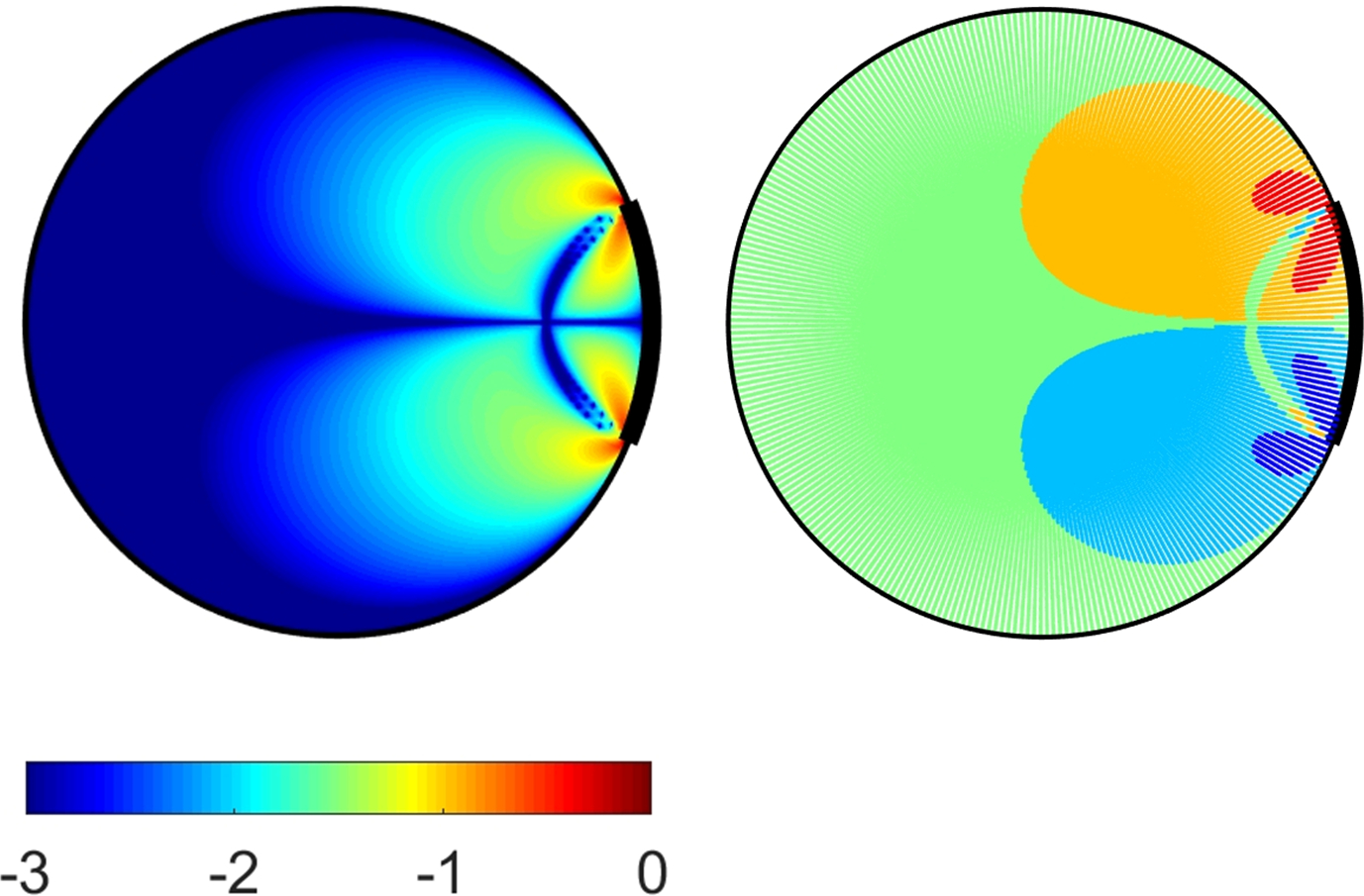}\label{HeatingDynamics4d}}

\caption{Impact of fluid deformation on equilibration: (a) stretching rate $\mu$ of fluid parcels expressed as $\mu_{\rm log} =\log_{10}(\mu/\max\mu)$; (b) relative change rate $\widetilde{\beta} = \widetilde{g}/\widetilde{f}$ expressed as $\widetilde{\beta}_{\rm log} =\log_{10}(|\widetilde{\beta}|/\max|\widetilde{\beta}|)$; (c) partition of $\widetilde{\beta}$ via $\widetilde{\beta}_{\rm rel}\equiv \widetilde{\beta}/\max|\widetilde{\beta}|$ into regions with strong enhancement ($\widetilde{\beta}_{\rm rel}\geq\epsilon_2$; red) versus strong diminution ($\widetilde{\beta}_{\rm rel}\leq-\epsilon_2$; blue), moderate enhancement ($\epsilon_1\leq \widetilde{\beta}_{\rm rel}<\epsilon_2$; yellow) versus moderate diminution ($-\epsilon_2< \widetilde{\beta}_{\rm rel}\leq -\epsilon_1$; cyan) and insignificant impact ($|\widetilde{\beta}_{\rm rel}|<\epsilon_1$; green) using thresholds $(\epsilon_1,\epsilon_2)=(5\times 10^{-3},5\times 10^{-2})$; (d) reoriented $\Ta$ at $t=20$; (e) $\widetilde{\beta}_{\rm log}$ for reoriented $\Ta$; (f) approximation $\widetilde{\beta}^\ast$ expressed as $\widetilde{\beta}_{\rm log}^\ast = \log_{10}(|\widetilde{\beta}^\ast|/\max|\widetilde{\beta}^\ast|)$.}

\label{HeatingDynamics4}
\end{figure}

Global equilibration shows similar characteristics due to the fact that significant fluid deformation within the interior also concentrates primarily around the arc edges. Fig.~\ref{HeatingDynamics4a} demonstrates this by the stretching rate $\mu$ (shown as $\mu_{\rm log} = \log_{10}(\mu/\max\mu)$ to enhance contrast) and reveals a rapid decay from its peaks near said edges by several orders of magnitude within a confined boundary region. Typical corresponding magnitudes of $\widetilde{\beta}$ are shown in Fig.~\ref{HeatingDynamics4b} for $t=20$ and $t=50$ and Fig.~\ref{HeatingDynamics4c} distinguishes subregions of different impact levels using $\widetilde{\beta}_{\rm rel}\equiv \widetilde{\beta}/\max|\widetilde{\beta}|$ and thresholds $(\epsilon_1,\epsilon_2)=(5\times 10^{-3},5\times 10^{-2})$: strong enhancement ($\widetilde{\beta}_{\rm rel}\geq\epsilon_2$; red) versus strong diminution ($\widetilde{\beta}_{\rm rel}\leq-\epsilon_2$; blue); moderate enhancement ($\epsilon_1\leq \widetilde{\beta}_{\rm rel}<\epsilon_2$; yellow) versus moderate diminution ($-\epsilon_2< \widetilde{\beta}_{\rm rel}\leq -\epsilon_1$; cyan); insignificant ($|\widetilde{\beta}_{\rm rel}|<\epsilon_1$; green). (Regions $\widetilde{\beta}>0$ and $\widetilde{\beta}<0$ correspond with local enhancement and diminution, respectively, of equilibration due to $\widetilde{f}>0$ and thus fully capture the equilibration dynamics.) This reveals that the confinement of significant fluid deformation to the arc region carries over to its impact on the equilibration dynamics. The red/blue regions in Fig.~\ref{HeatingDynamics4c} with the strongest enhancement/diminution of equilibration remain largely stationary and closely correlate with $\mu$. Significant temporal changes in $\widetilde{\beta}$ occur mainly in the yellow/cyan regions and for the present temperature field manifest themselves in the diminishing banana-shaped region emerging from arc edge $\theta = \Delta/2$. However, their intensity is 1-2 orders of magnitude smaller compared to the arc region and thus of secondary importance only.

The relative rate of change $\widetilde{\beta}$ for a reorientation of the temperature field at $t=20$ following Fig.~\ref{HeatingDynamics4e} is shown in Fig.~\ref{HeatingDynamics4f} and exposes high-impact regions (red/blue) at the arc edges that are basically reorientations of their base-flow counterparts in Fig.~\ref{HeatingDynamics4b} -- and thus (nearly) independent of the momentary $\Ta$ -- coexisting with interior regions (yellow/cyan) that depend significantly on $\Ta$ yet have an only moderate impact. This behaviour is typical of any reorientation and diminishes the (potential) impact of internal fluid mixing and chaotic advection, thereby further eroding the beforementioned common belief of an invariably beneficial role of these conditions.

The rapid alignment of isothermals with streamlines in the arc region (Fig.~\ref{FigEulerianLagrangian2}) suggests that $\widetilde{\beta}$ here approximately behaves as $\widetilde{\beta}^\ast = -2\xvec{e}_v^\dagger\cdot D\cdot\xvec{e}_v$, with $\xvec{e}_v = (v_y,-v_x)/|\xvec{v}_1|$ the normal to the streamlines. (Property $\xvec{e}_v|_\Gamma = \xvec{e}_r|_\Gamma$ ensures consistency with the previous relation $\widetilde{\beta}|_\Gamma=\bar{\beta}|_\Gamma$.) Fig.~\ref{HeatingDynamics4d} gives the magnitude of $\widetilde{\beta}^\ast$ (left) and corresponding partition by impact levels (right). This indeed reveals a close agreement of the approximated strong-impact regions (red/blue) with their actual counterparts in Figs.~\ref{HeatingDynamics4c} and \ref{HeatingDynamics4e}; deviations are restricted to the moderate-impact regions (yellow/cyan). Thus $\widetilde{G}$ to good approximation collapses on a form akin to \eqref{HeatingFluid9}, i.e.
\begin{eqnarray}
\widetilde{G}(t) \approx \int_{(r,\theta)\in\mathcal{D}^+} \widetilde{\beta}^\ast(r,\theta)\left[|\xvec{\nabla}\Ta|^2_{(r,\theta)} - |\xvec{\nabla}\Ta|^2_{(r,-\theta)}\right] rdrd\theta
\label{HeatingFluid11}
\end{eqnarray}
with $\mathcal{D}^+$ the strong-enhancement region (red) in Fig.~\ref{HeatingDynamics4d} (right) and using the anti-symmetry $\widetilde{\beta}^\ast(r,\theta) = -\widetilde{\beta}^\ast(r,-\theta)$ about the centerline $\theta=0$ of the arc. Similar reasoning as that underlying \eqref{HeatingFluid10} gives $0<|\xvec{\nabla}\Ta|^2_{(r,-\theta),t>0}<|\xvec{\nabla}\Ta|^2_{(r,\pm\theta),t=0}<|\xvec{\nabla}\Ta|^2_{(r,\theta),t>0}$
and in conjunction with $\widetilde{\beta}^\ast|_{\mathcal{D}^+}>0$ implies $\widetilde{G}>0$. So fluid deformation for (at least) base flow $\xvec{v}_1$ and given initial/boundary conditions in \eqref{eq:ade_pde_discrete} always accelerates, besides energising, also equilibration. However, unfavourable reorientations may, as before, also cause (temporary) diminution of global equilibration ($\widetilde{G}<0$).

\section{Adaptive flow reorientation}
\label{ControlStrategy}

\subsection{Control strategy revisited}
\label{ControlStrategy2}

Sec.~\ref{DynamicsOfHeating} reveals that the impact of flow on thermal transport is highly non-trivial and depends essentially on the interplay between fluid deformation and temperature field. This may benefit as well as deteriorate said transport and thus conclusively demonstrates that efficient fluid mixing by chaotic advection does not imply enhanced thermal performance. Decomposition \eqref{EvolutionOperators4} of the Perron-Frobenius operator $\mathcal{P}_t$ clearly illustrates this: the temperature evolution depends on factors $\mathcal{A}_t$ (advection) and
$\mathcal{D}_t$ (diffusion) yet mixing/advection-based optimisation effectively omits $\mathcal{D}_t$. Moreover, significant impact of fluid deformation is restricted to the active arc region (Figs.~\ref{HeatingDynamics3} and \ref{HeatingDynamics4}) and the underlying processes become exponentially complex and intractable with consecutive flow reorientations. These issues render (i) optimisation of heat transfer based solely on mixing/advection characteristics inefficient and potentially ineffective and (ii) control strategies based directly on fundamental thermal mechanisms as exposed in Secs.~\ref{LagrangianHeatTransfer} and \ref{HeatTransferEnhancement} impractical.

\begin{figure}[!hbt]
    \centering
    \includegraphics[width=\textwidth,height=0.27\textwidth]{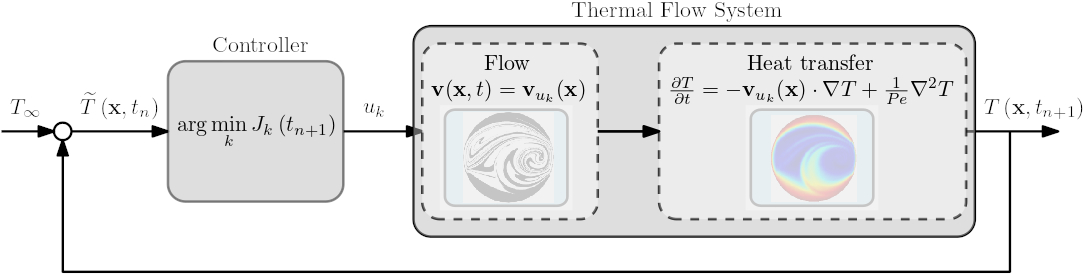}

    \caption{Schematic of control loop for adaptive flow reorientation in the RAM.}
    \label{fig:ControlStrategySchematic}
\end{figure}

The global metrics for heating defined in Sec.~\ref{GlobalHeatingDynamics1} incorporate the global impact of the flow (Sec.~\ref{GlobalHeatingDynamics2}) and thus offer a workable alternative to determine the ``best'' reorientation scheme $\mathcal{U}$ in the general control strategy in Sec.~\ref{System1}. Such metrics namely admit its formulation as the minimisation of a cost function $J(t)$ from $J(0) > 0$ to $J(t_\epsilon) \leq \epsilon$ in the shortest possible time $t_\epsilon>0$, with $\epsilon>0$ a preset tolerance. The control action consists of step-wise activating flow $\mathbf{v}_{u_k}$ in \eqref{eq:reorientated_flow} at $t_n$ that minimises $J$ for the finite horizon $t_{n+1} = t_n+\tau$, i.e.
\begin{equation}
\arg \min_k J_k\left(t_{n+1}\right),\quad J_k(t) \equiv J\left(\widetilde{T}_k(\xvec{x},t)\right),\quad
\widetilde{T}_k(\xvec{x},t) \equiv\widetilde{T}(\xvec{x},t;\mathbf{v}_{u_k}(\xvec{x}),\widetilde{T}(\xvec{x},t_n)),
	\label{eq:discrete_cost}
\end{equation}
from predicted future states $\widetilde{T}_k(\xvec{x},t_{n+1})$ of the momentary temperature $\widetilde{T}(\xvec{x},t_n)$. The schematic in Fig.~\ref{fig:ControlStrategySchematic} shows this control loop, which, by
relying on (real-time) feedback from the actual state at $t_n$ and its evolution, enables
%
control that is robustness to disturbances accumulated up to $t_n$.

Critical for an effective controller is an adequate cost function $J$ in \eqref{eq:discrete_cost}. The dynamics of global metrics \eqref{HeatingNorms} (Sec.~\ref{GlobalHeatingDynamics1}) advance metric $\Ja$ as the most suitable candidate, i.e. $J\equiv\Ja$, for the following reasons. First, $\Ja$ incorporates both energising (represented by $\Jb$) and homogenisation (represented by $\Jc$) by virtue of relation \eqref{RelationMetrics} and thus accounts for the two fundamental processes in the heating process. Second, the fundamental conflict between heat transfer and homogenisation (Sec.~\ref{HeatTransferEnhancement}) precludes an overall effective controller based on either $\Jb$ (energising) or $\Jc$ (homogenisation). Third, the invariably monotonic decay of $\Ja$ dictated by \eqref{HeatingFluid3} ensures convergence and regularity of the minimisation procedure \eqref{eq:discrete_cost}. The inherently (potentially) non-monotonic behaviour of $\Jc$ ($\Jb$), on the other hand, may compromise this procedure.

The above effectively yields the step-wise optimal control adopted in \cite{lensvelt2020} including metric $\Ja$ as cost function. However, the study in \cite{lensvelt2020} thus followed standard practice in control theory by using the well-known L$^2$-norm of state $\widetilde{T}$ (identifying with $\Ja$) \cite{khalil2014}. The present study provides the hitherto missing scientific support for adopting $\Ja$ as cost function in the current thermal context. The practical control action involves repetition of the following steps at time levels $t_n$:
\begin{enumerate}
	\item[] {\bf Step 1} Predict future temperature $\widetilde{T}_k(\xvec{x},t_{n+1})$ from $\widetilde{T}\left(\mathbf{x},t_{n}\right)$ for each flow $\mathbf{v}_{u_k}$.
	\item[] {\bf Step 2} Select flow $\mathbf{v}_{u_k}(\xvec{x})$ that minimises cost function $J=\Ja$ according to \eqref{eq:discrete_cost}.
	\item[] {\bf Step 3} Activate the selected aperture $k$ for time interval $t_n\leq t\leq t_{n+1}$.
\end{enumerate}
This procedure is terminated at time level $t_{n+1}$ for which $J_{n+1} = J(t_{n+1}) < \epsilon$. The corresponding transient time $t_\epsilon$,
defined as $J(t_\epsilon)=\epsilon$, is interpolated via $t_\epsilon = t_n + \tau (J_n-\epsilon)/(J_{n}-J_{n+1})$. Prediction using the actual intermediate state $\widetilde{T}\left(\mathbf{x},t_{n}\right)$ provides robustness to disturbances accumulated up to $t_{n}$ and this positions the control strategy within the realm of Model Predictive Control (MPC) \cite{Camacho2013}.
A difference compared to common MPC is that here the prediction horizon coincides with the duration of the control action (prediction horizons usually are far ahead into the future). A strong point is that the current strategy involves the full spatio-temporal state described by the conservation laws (MPC of thermo-fluidic systems usually concerns global behaviour of bulk quantities described by integral models \cite{Corriou2018}).

\RED{Moreover, cost functions $J$ in \eqref{eq:discrete_cost}, instead of solely including the state $\Ta$ in the current $J = \Ja$, generically also account for the effort required to execute the control action \cite{Corriou2018}. In the RAM this consists of the energy consumption $E(t)$ for driving the arcs. However, given reorientation schemes U here always concern step-wise activation of a single arc at constant speed, said energy consumption for any $\mathcal{U}$ always increases linearly in time following $E(t) = Wt$, with $W$ the constant power consumption by the flow forcing. Hence $E(t)$ is irrelevant for the minimisation procedure \eqref{eq:discrete_cost} and reaching the control target (i.e. the minimum transient time $t_\epsilon$) automatically yields the minimum energy consumption in the present approach. Advanced control strategies involving e.g.
simultaneous activations of multiple arcs at variable speeds, on the other hand, result in energy consumption that depends explicitly on the activation scheme and must therefore generically be
taken into account through cost function $J$ for an overall optimal performance.}

\subsection{Compact model for fast predictions}
\label{Predictor}

Essential for a useful controller is sufficiently fast prediction of the $2N+1$ future states $\widetilde{T}_k(\xvec{x},t_{n+1})$ in
\eqref{eq:discrete_cost} at each $t_n$ with an acceptable computational effort. Backbone for this fast predictor is the property that spectral decomposition \eqref{eq:spectral1} for the temperature evolution in the reoriented flow $\xvec{v}_u$ is a linear transformation from spectral decomposition \eqref{Spectral1} of the base flow following \eqref{eq:spectral1b}. This enables a fact predictor via the numerical approximation of Perron-Frobenius operator $\mathcal{P}_t$ in \eqref{Spectral1} and embarks on spatial discretisation of ADE \eqref{eq:ade_pde_discrete} for $\xvec{v}=\xvec{v}_1$, i.e.
\begin{equation}
\frac{d\Tb\left(t\right)}{d t} = \mathbf{A}\Tb\left(t\right),
\label{StateSpace1}
\end{equation}
with $\Tb\left(t\right) = [\Ta\left(\mathbf{x}_0,t\right),\dots,\Ta\left(\mathbf{x}_M,t\right)]^\dagger$ the temperatures in the nodes of the computational grid $\mathbf{X} = [\mathbf{x}_0,\dots,\mathbf{x}_M]^\dagger$ and system matrix $\mathbf{A}$ the discrete approximation of the advection-diffusion operator. Steady base flow $\mathbf{v}_1$ implies a constant $\mathbf{A}$ and thus admits
\begin{equation}
\Tb\left(t\right) = \mathbf{P}_t\Tb_0 = \sum_{m=0}^M \alpha_m\pmb{\phi}_me^{\lambda_m t},\quad
\pmb{\alpha} = \mathbf{V}^{-1}\Tb_0,\quad
\mathbf{P}_t = \mathbf{V} e^{\pmb{\Lambda}_1 t}\mathbf{V}^{-1},
\label{StateSpace2}
\end{equation}
as semi-analytical solution for \eqref{StateSpace1} using the spectral decomposition $\mathbf{A} = \mathbf{V}\pmb{\Lambda}\mathbf{V}^{-1}$, with $\mathbf{V} = [\pmb{\phi}_0,\dots,\pmb{\phi}_M]$ and $\pmb{\Lambda} = \mbox{diag}(\lambda_0,\dots,\lambda_M)$ the eigenvector and eigenvalue matrices, respectively.

Relation \eqref{StateSpace2} constitutes the discrete approximation of spectral decomposition \eqref{Spectral1} of the Perron-Frobenius operator $\mathcal{P}_t$. The reorientation property of the continuous system following Sec.~\ref{System2} carries over to its discrete approximation and yields
\begin{equation}
\Tb\left(t\right) = \mathbf{P}_t^{(u)}\Tb_0 = \sum_{m=0}^M \alpha_m^{(u)}\pmb{\psi}_m^{(u)}e^{\mu_m t},\quad
\pmb{\alpha}^{(u)} = \mathbf{V}_u^{-1}\Tb_0,\quad
\mathbf{P}_t^{(u)} = \mathbf{V}_u e^{\pmb{\Omega}_u t}\mathbf{V}_u^{-1},
\label{StateSpace3}
\end{equation}
as discrete approximation of \eqref{eq:spectral1}. Eigenvalues $\pmb{\Omega}_u = \mbox{diag}(\mu_0,\dots,\mu_M)$ and eigenvectors $\mathbf{V}_u = [\pmb{\psi}_0^{(u)},\dots,\pmb{\psi}_M^{(u)}]$ are following \eqref{eq:spectral1b} given by $(\pmb{\Omega}_u,\xvec{V}_u) = (\pmb{\Lambda},\xvec{G}_u\xvec{V})$ and $(\pmb{\Omega}_u,\xvec{V}_u) = (\pmb{\Lambda}^0,\xvec{V}^0)$ for $u\neq 0$ and $u=0$, respectively. The reorientation operator is following \eqref{eq:spectral1c} composed of the discrete rotation ($\xvec{R}_u$) and reflection ($\xvec{S}$) operators: $\xvec{G}_u = \xvec{R}_u$ for $u>0$ and $\xvec{G}_u = \xvec{S}\xvec{R}_u$ for $u<0$.

The above discrete approximation translates the evolution and reorientation into standard matrix-vector multiplications using pre-constructed system matrices and thus tremendously reduces the computational effort compared to a conventional time-marching scheme for \eqref{StateSpace1} and its counterpart for reoriented flows. Further substantial reduction relies on the exponential decay with characteristic time scale $\tau_m = -1/Re(\mu_m)$ of the individual modes $m$ in \eqref{StateSpace3}. This renders contributions by modes $m$ for which $\tau_m/\tau\ll 1$ insignificant and admits close approximation of the temperature evolution by a truncation of expansion \eqref{StateSpace3} at $Q\ll M$, i.e.
\begin{equation}
\Tb(t)\approx\Tc(t) \equiv \widehat{\mathbf{P}}_t^{(u)}\Tb_0 = \sum_{m=0}^Q \widehat{\alpha}_m^{(u)}\pmb{\psi}_m^{(u)}\exp(\mu_m t),\quad
\widehat{\pmb{\alpha}}^{(u)} = \widehat{\mathbf{V}}_u^\ast\Tb_0,\quad \widehat{\mathbf{P}}_t^{(u)} = \widehat{\mathbf{V}}_u e^{\widehat{\pmb{\Omega}}_u t}\widehat{\mathbf{V}}_u^\ast,
\label{StateSpace4}
\end{equation}
with $\widehat{\mathbf{V}}_u^\ast$ the Moore-Penrose inverse of the reduced eigenvector basis $\widehat{\mathbf{V}}_u = [\pmb{\psi}_0^{(u)},\dots,\pmb{\psi}_Q^{(u)}]$.

The total reduction in computational effort afforded by compact model \eqref{StateSpace4} amounts to 3-4 orders of magnitude compared to conventional time-marching of \eqref{StateSpace1} \cite{lensvelt2020}. Moreover, system matrix $\xvec{A}$ can be constructed from any spatial discretisation method including e.g. FEM and FVM (and even directly exported from some commercial CFD packages) and
decomposed by linear-algebra tools as e.g. {\tt MATLAB}. Thus the compact model enables the fast temperature predictions necessary to make the control strategy in Sec.~\ref{ControlStrategy2} viable for practical applications.

\section{Computational performance analysis}
\label{sec:example_and_numerical_results}

\subsection{Performance of minimisation procedure and predictor}
\label{sec:numerical_method}

The general performance and robustness of the minimisation procedure \eqref{eq:discrete_cost} and the compact model \eqref{StateSpace3} are investigated for a typical RAM consisting of $N=3$ apertures at $Pe=500$ and $\tau = 5$ and involves the following steps. First, simulation of both the actual temperature evolution $\widetilde{T}(\xvec{x},t)$ and the step-wise predictions $\widetilde{T}_k(\xvec{x},t)$ by full resolution of ADE \eqref{eq:ade_pde_discrete} using the dedicated spectral scheme for the RAM of \cite{lester2008a}. This serves as reference for the performance analysis. Second, simulation of $\widetilde{T}(\xvec{x},t)$ by full resolution of ADE \eqref{eq:ade_pde_discrete} using a conventional FVM scheme \cite{thomas1995} and predictions $\widetilde{T}_k(\xvec{x},t)$ via approximation \eqref{StateSpace3} based on truncation of the spectral decomposition of the FVM system matrix $\xvec{A}$ in \eqref{StateSpace1}. This emulates a predictor-based minimisation in a practical implementation. The FVM grid consists of a cylindrical mesh with equidistant inter-nodal spacings $(\Delta r,\Delta\theta)$ and here includes $M=51,121$ elements (convergence verified by mesh-refinement tests); truncation is at $Q=500$ according to typical reductions $Q/M\sim\mathcal{O}(10^{-1})$ and thus accomplishes the total reduction in computational effort as estimated in Sec.~\ref{Predictor} \cite{lensvelt2020}.

\begin{figure}[!hbt]
    \centering
\subfigure[Reference temperature $\widetilde{T}(\xvec{x},t)$]{\includegraphics[width=\textwidth]{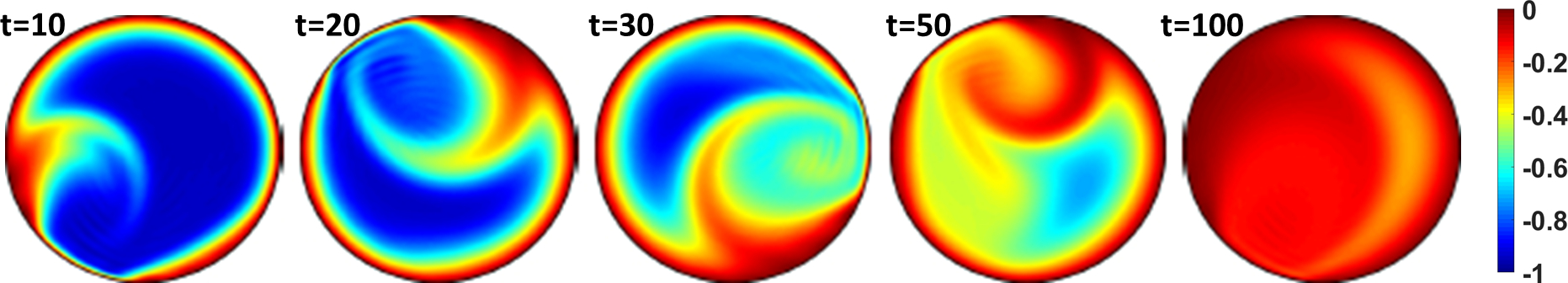}\label{fig:FVM_SPECa}}

\subfigure[Adaptive reorientation scheme $\mathcal{U}$.]{\includegraphics[width=.47\textwidth,height=.27\textwidth]{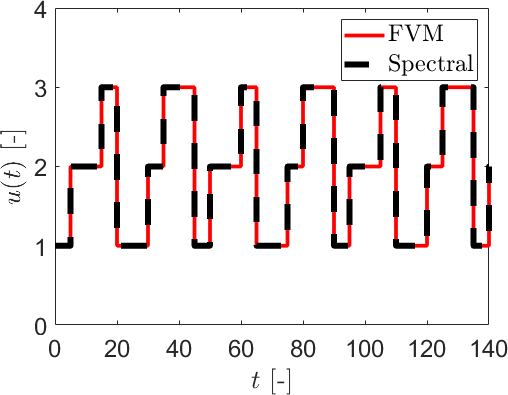}\label{fig:FVM_SPECb}}
\hspace*{12pt}
\subfigure[$\log_{10} J$]{\includegraphics[width=.47\textwidth,height=.27\textwidth]{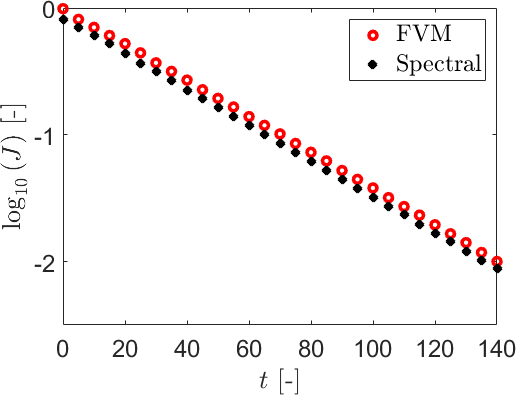}\label{fig:FVM_SPECc}}

\caption{Performance of minimisation procedure and FVM-based predictor for RAM with three apertures ($N=3$) at $(Pe,\tau)=(500,5)$: (a) reference temperature $\widetilde{T}(\xvec{x},t)$ simulated by spectral method; (b) adaptive reorientation scheme \RED{$\mathcal{U}=\{u_0,u_1,\dots\}$ accomplishing switching between reorientations $u_n$ of the base flow following \eqref{eq:reorientated_flow} at time
levels $t_n = n\tau$} for spectral versus FVM-based method; (c) cost function $J$ for spectral versus FVM-based method.}
\label{fig:FVM_SPEC}
\end{figure}

The temperature evolution $\widetilde{T}(\xvec{x},t)$ and adaptive reorientation scheme $\mathcal{U}$ (here restricted to $u\geq 0$) determined via \eqref{eq:ade_pde_discrete} by simulation with the spectral method are shown in Fig.~\ref{fig:FVM_SPECa} and Fig.~\ref{fig:FVM_SPECb}, respectively. This results in sequential activation of the apertures reminiscent of periodic reorientation, i.e. $\mathcal{U} = \{1,2,3,1,2,3,\dots\}$, yet with variable duration, i.e. $t_{\rm active}=\tau$ or $t_{\rm active}=2\tau$, signifying an essentially aperiodic reorientation scheme. The reorientation is clearly visible in the temperature field via the emergence of multiple thermal plumes. Fig.~\ref{fig:FVM_SPECc} gives the evolution of the cost function $J=\Ja$ as $\log_{10} J$ and reveals, consistent with \eqref{HeatingFluid3}, a monotonic and exponential decay that reaches the termination criterion $J(t_\epsilon)\leq\epsilon$ at transient time $t_\epsilon\approx 140$ for $\epsilon=10^{-2}$. Simulations for different $(Pe,\tau)$ yield reorientation schemes that comprise of switching between apertures akin to Fig.~\ref{fig:FVM_SPECb} and cost functions that exponentially converge similar to Fig.~\ref{fig:FVM_SPECc} and terminate the control action at comparable transient times $t_\epsilon$. This demonstrates an overall functioning control strategy that converges and identifies an effective reorientation scheme.

The reorientation scheme $\mathcal{U}$ obtained with the FVM-based method is included in Fig.~\ref{fig:FVM_SPECb} (red solid) and coincides with the reference scheme (black dashed) found via the spectral method. The corresponding evolution of the cost function in Fig.~\ref{fig:FVM_SPECc} (red circle) slightly deviates from its reference (black cross), though, yet this must be attributed to different characteristics of the numerical schemes. (FVMs are e.g. known to suffer from numerical diffusion \cite{thomas1995}.) However, attainment of identical $\mathcal{U}$ implies that the controller is insensitive to (at least) disturbances of this magnitude, caused by numerical effects or otherwise. Simulations for different cases consolidate these findings. This demonstrates that the FVM-based predictor admits reliable and robust performance of minimisation \eqref{eq:discrete_cost} and the control strategy indeed is viable for practical applications.

\subsection{Performance of adaptive flow reorientation for process enhancement}
\label{PerformanceScheme1}

The effectiveness of adaptive flow reorientation is determined by comparing the transient time $t_\epsilon$ required to reach equilibrium $\widetilde{T}_\infty$ with that for conventional periodic reorientation consisting of repetition of the sequence $\mathcal{U}=\{1,2,\dots,N\}$ at given $(Pe,\tau)$. The performance indicator
\begin{equation}
		\chi(Pe,\tau) = \frac{t_{\varepsilon,p}(Pe,\tau)}{t_{\varepsilon,a}(Pe,\tau)},
\label{eq:homogen_ratio}
\end{equation}
where subscripts ``p'' and ``a'' denote periodic and adaptive schemes, respectively, quantifies this effectiveness as follows: $\chi>1$ indicates a shorter transient and, inherently, a faster equilibration rate -- and thus superior performance -- of the adaptive scheme compared to the periodic scheme; $\chi<1$ indicates a relatively inferior performance of the adaptive scheme.

The performance analysis is carried out for aperture configurations $N = 2,3,4$ and involves evaluation of $\chi$ in $(Pe,\tau)$-space for $5\times 10^2\leq Pe \leq 5\times 10^3$ and $3\leq\tau\leq 30$ using tolerance $\varepsilon=10^{-2}$ introduced above. The range for $Pe$ encompasses the values $Pe\sim\mathcal{O}(10^3)$ typical of practical systems within the present scope (Sec.~\ref{Intro}); the range for $\tau$ captures the transition from regular towards chaotic advection with conventional periodic reorientation schemes in the RAM for given $N$ \cite{baskan2016}. Performance indicator $\chi$ is computed on an equidistant grid of $50\times 50 = 2500$ points inside the considered parameter range using the spectral method of Sec.~\ref{sec:numerical_method}.

\begin{figure}[!hbt]
\subfigure[$N=2$]{\includegraphics[width=.36\textwidth,height=.3\textwidth]{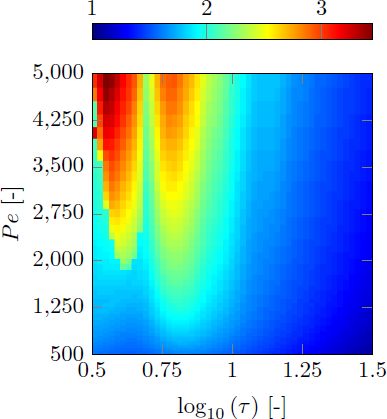}\label{fig:L_homogenisationsa}}
\hspace*{3pt}
\subfigure[$N=3$]{\includegraphics[width=.30\textwidth,height=.3\textwidth]{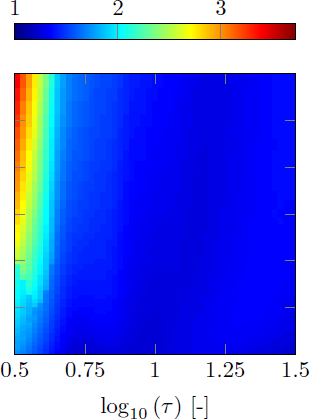}\label{fig:L_homogenisationsb}}
\hspace*{3pt}
\subfigure[$N=4$]{\includegraphics[width=.30\textwidth,height=.3\textwidth]{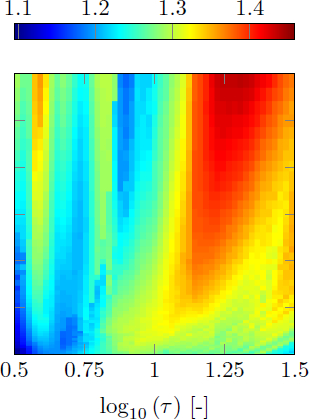}\label{fig:L_homogenisationsc}}		
\caption{Performance indicator $\chi$ versus $(Pe,\tau)$ for RAM with number of apertures $N$ as indicated.}
\label{fig:L_homogenisations}
\end{figure}

Fig.~\ref{fig:L_homogenisations} gives $\chi$ versus $(Pe,\tau)$ for $N = 2$ (left), $N = 3$ (center) and $N = 4$ (right) using a logarithmic
scale for $\tau$ to enhance contrast. This yields $\chi > 1$ everywhere and thus exposes -- at least in the considered parameter range -- the adaptive scheme as superior to the conventional periodic scheme. The difference in performance exhibits substantial variation, however. Peak performance is for cases $N=2$ and $N=3$ confined to relatively narrow bands near the lower bound $\tau = 3$ ($\log_{10}\tau = 0.5$) and within this regime reaches $\max\chi = 3.5$ for $N=2$ and $\max\chi = 3.9$ for $N=3$ at the upper bound $Pe=5\times 10^3$; performance rapidly declines to $\chi\simeq\mathcal{O}(1)$ upon increasing
$\tau$. Case $N=4$ is devoid of pronounced high-performance bands and exhibits more erratic behaviour with a considerably lower $\max\chi = 1.5$. Sec.~\ref{PerformanceScheme3} reconciles
shown performance with the advection characteristics and fluid deformation.

\begin{figure}[!hbt]

\includegraphics[trim=0cm 0cm 0cm 0cm,clip,width=\textwidth]{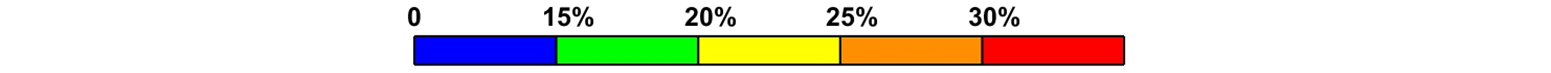}\vspace*{6pt}

\subfigure[$N=2$]{\includegraphics[width=.32\textwidth,height=.28\textwidth]{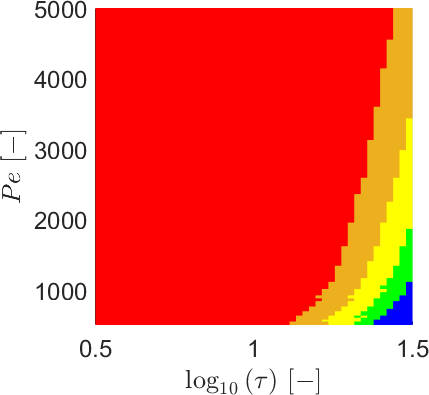}\label{fig:L_homogenisations2a}}
\hspace*{2pt}
\subfigure[$N=3$]{\includegraphics[width=.32\textwidth,height=.28\textwidth]{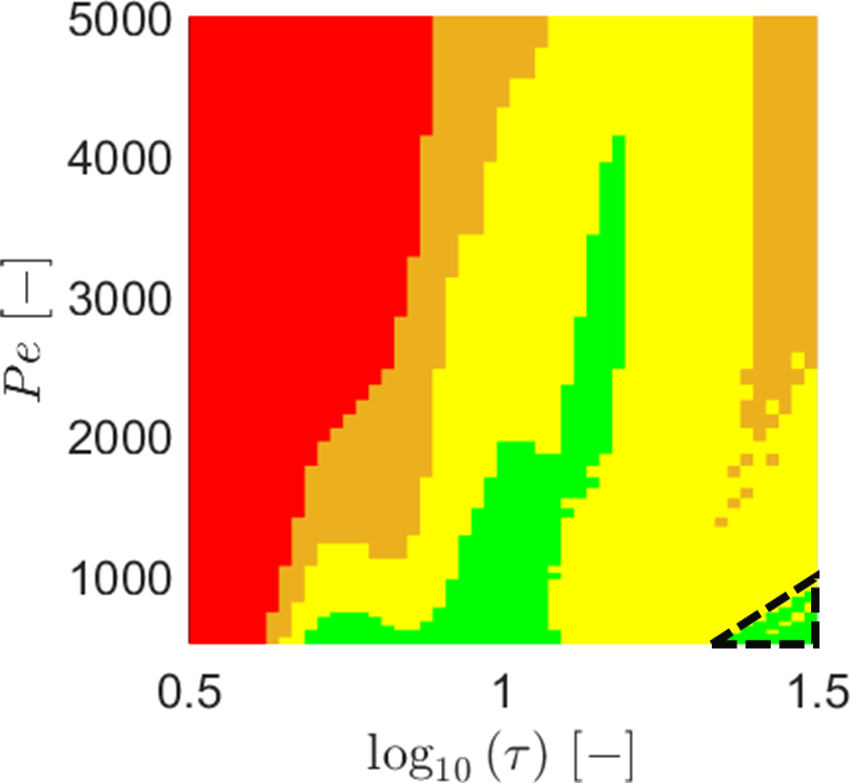}\label{fig:L_homogenisations2b}}
\hspace*{2pt}
\subfigure[$N=4$]{\includegraphics[width=.32\textwidth,height=.28\textwidth]{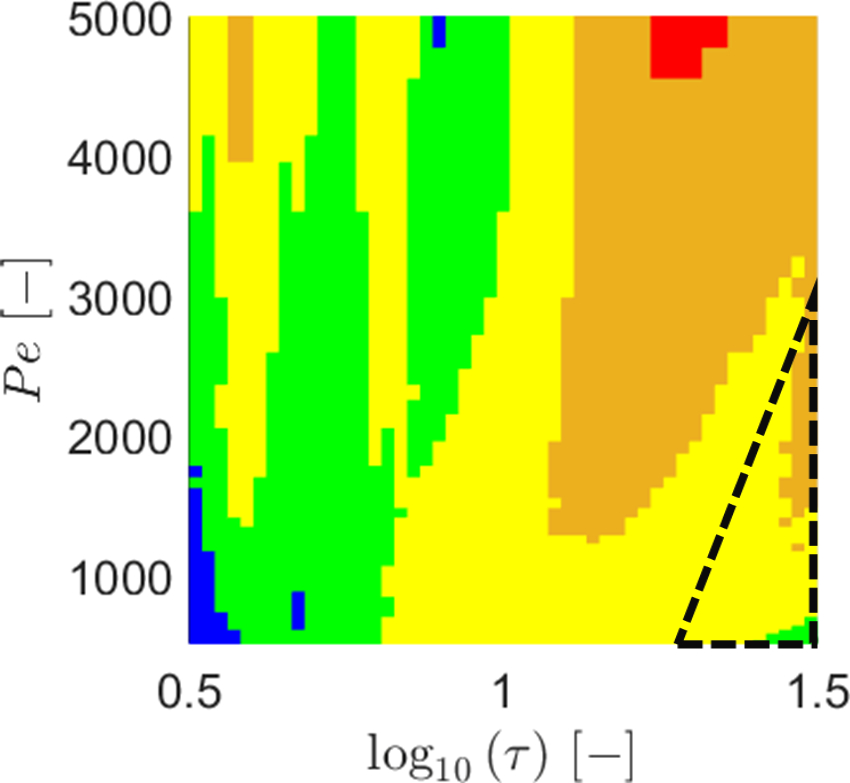}\label{fig:L_homogenisations2c}}

\caption{Relative acceleration of fluid heating $\chi_{\rm rel} = (\chi-1)\times 100\%$ by adaptive flow reorientation for
RAM with number of apertures $N$ as indicated: $0<\chi_{\rm rel}\leq15\%$ (blue), $15\%<\chi_{\rm rel}\leq 20\%$ (green), $20\%<\chi_{\rm rel}\leq 25\%$ (yellow), $25\%<\chi_{\rm rel}\leq 30\%$ (orange), $\chi_{\rm rel}> 30\%$ (red). \RED{Dashed black triangles in panels (b) and (c) approximately outline regions with adaptive schemes consisting of periodic repetitions
of $\mathcal{U}=\{1,3,2\}$ and $\mathcal{U}=\{1,4,3,2\}$, respectively.}}

\label{fig:L_homogenisations2}
\end{figure}

Important for practical purposes is that the adaptive scheme also in regions $\chi\simeq\mathcal{O}(1)$ outside the high-performance bands significantly outperforms the conventional periodic scheme. Fig.~\ref{fig:L_homogenisations2} highlights this by partition of parameter space into regions with a relative acceleration of the fluid heating $\chi_{\rm rel} = (\chi-1)\times 100\%$ of $0<\chi_{\rm rel}\leq15\%$ (blue), $15\%<\chi_{\rm rel}\leq 20\%$ (green), $20\%<\chi_{\rm rel}\leq 25\%$ (yellow), $25\%<\chi_{\rm rel}\leq 30\%$ (orange) and $\chi_{\rm rel}> 30\%$ (red). This reveals accelerations by at least $15\%$ in nearly
the entire parameter range (save some localised regions for $N=2$ and $N=4$) and $25\%$ or more in large areas. This has the major (practical) implication that adaptive flow reorientation enables the same process with the same device yet at considerably lower effort compared to conventional operation (energy consumption for flow forcing is proportional to $t_\epsilon$). Process enhancement of this magnitude namely constitutes a dramatic reduction in energy and (potentially also) resource consumption in industries as those highlighted in Sec.~\ref{Intro}.

\subsection{Optimal reorientation schemes and heating dynamics}
\label{PerformanceScheme2}

\begin{figure}[!hbt]
\subfigure[Essentially aperiodic schemes]{\includegraphics[width=.49\textwidth,height=.28\textwidth]{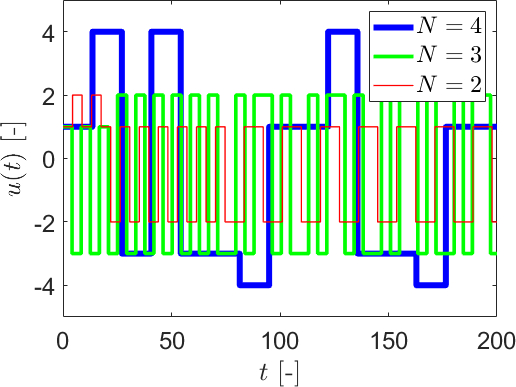}\label{fig:aperiodic_sequences}}
\hspace*{6pt}
\subfigure[Schemes with prolonged periodic episodes]{\includegraphics[width=.49\textwidth,height=.28\textwidth]{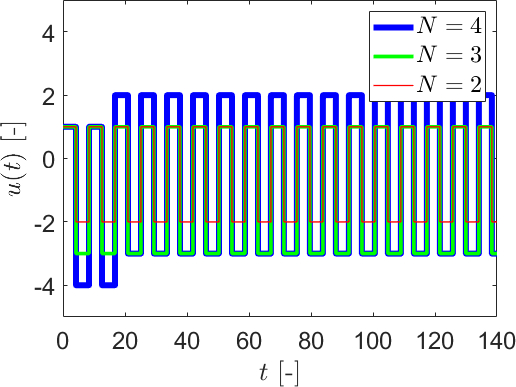}\label{fig:plume_sequences}}
\caption{Typical adaptive reorientation schemes \RED{$\mathcal{U}=\{u_0,u_1,\dots\}$ accomplishing switching between reorientations $u_n$ of the base flow following \eqref{eq:reorientated_flow} at time
levels $t_n = n\tau$}: (a) essentially aperiodic schemes for $(Pe,\tau)_{N=2}=(2471,4.4)$, $(Pe,\tau)_{N=3}=(3125,4.2)$ and $(Pe,\tau)_{N=4}=(1012,13.6)$; (b) schemes with prolonged periodic episodes for $(Pe,\tau)=(3275,4.2)$ and $N$ as indicated.}
	\label{fig:adaptive_sequences}
\end{figure}

Control actions in the considered parameter range invariably involve aperture activation and frequent switching ($u\neq0$); intermediate diffusion-only steps ($u=0$) are non-existent in the adaptive reorientation schemes. This implies that flow reorientation, despite inevitable local diminution (Sec.~\ref{HeatTransferEnhancement}), on a global scale always enhances the heating process. Striking, though, is that the controller never selects the conventional periodic reorientation scheme introduced in Sec.~\ref{PerformanceScheme1} or\RED{, generically,} repetitions of other systematic progressions along all apertures such as e.g. $\mathcal{U}=\{-1,-2,\dots,-N\}$ or $\mathcal{U}=\{N,N-1,\dots,1\}$. \RED{The sole exceptions are periodic repetitions of $\mathcal{U}=\{1,3,2\}$ for $N=3$ and $\mathcal{U}=\{1,4,3,2\}$ for $N=4$ found in the localised parameter regimes approximately
bounded by the dashed black triangles in Fig.~\ref{fig:L_homogenisations2b} and Fig.~\ref{fig:L_homogenisations2c}, respectively.} Fig.~\ref{fig:FVM_SPECb} demonstrates that the adaptive scheme, even upon restriction to $u>0$, exhibits aperiodic deviation from said conventional scheme. Adaptive reorientation schemes generically are essentially aperiodic and Fig.~\ref{fig:adaptive_sequences} shows typical cases for $N=2$ (red), $N=3$ (black) and $N=4$ (blue) yielding $\chi = 2.49$, $\chi = 2.15$ and $\chi = 1.25$, respectively. However, adaptive reorientation schemes in a significant portion of the parameter space encompass prolonged periodic episodes (including some cases that are entirely periodic) as illustrated in Fig.~\ref{fig:plume_sequences} for $(Pe,\tau)=(3275,4.2)$. Here the reorientation schemes settle on the periodic sequence
\begin{eqnarray}
\mathcal{U}_{N=2} = \{1,-2,1,-2,\dots\},\quad\mathcal{U}_{N=3} = \{1,-3,1,-3,\dots\},\quad\mathcal{U}_{N=4} = \{2,-3,2,-3,\dots\},
\label{PeriodicSchemes}
\end{eqnarray}
either directly from the start ($N=2$) or after a short transient ($N=3,4$) and for given $(Pe,\tau)$ yield $\chi_{N=2}=2.81$, $\chi_{N=3}=2.15$ and $\chi_{N=4}=1.31$.

\begin{figure}[!htbp]
	\centering
\subfigure[Single thermal plume for $N=2$ (top) and $N=3$ (bottom).]{\includegraphics[width=\textwidth]{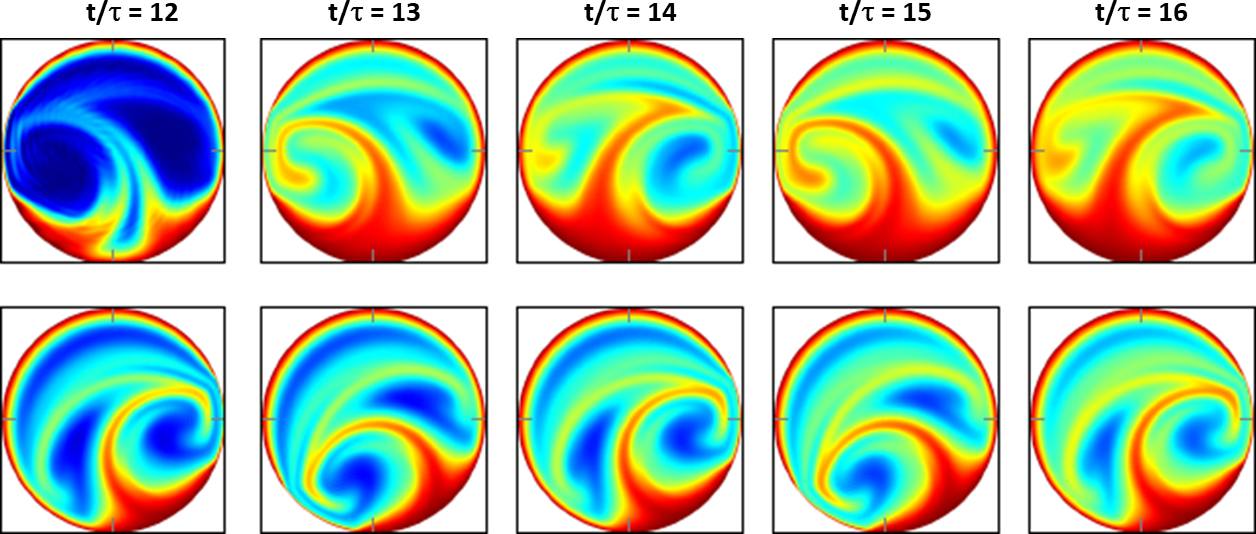}\label{fig:plume_apertures_a}}
\subfigure[Pair of thermal plumes for $N=4$.]{\includegraphics[width=\textwidth]{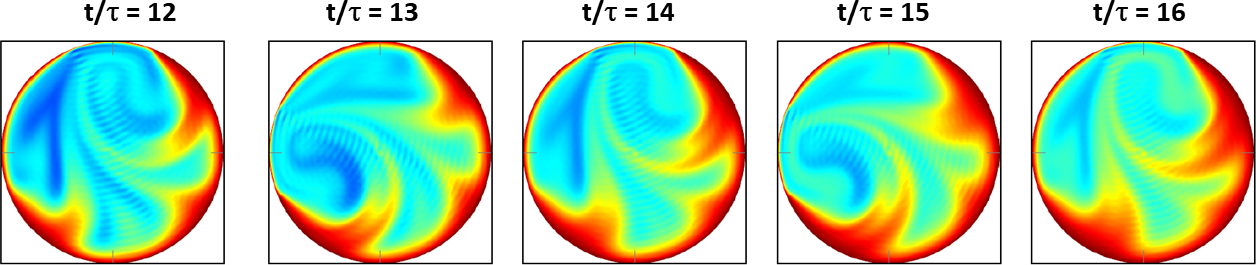}\label{fig:plume_apertures_b}}
	
\caption{Formation of thermal plumes due to periodic (episodes of) reorientation schemes selected by adaptive flow reorientation demonstrated for case $(Pe,\tau)=(3275,4.2)$ in Fig.~\ref{fig:plume_sequences}: (a) single thermal plume driven by two adjacent arcs alternately moving towards each other for $N=2$ (top) and $N=3$ (bottom); (b) pair of thermal plumes driven by two adjacent arcs alternately moving away from each other for $N=4$.}
	\label{fig:plume_apertures}
\end{figure}

Key difference between said periodic sequences and the conventional periodic schemes is that the former always consist of switching between two adjacent arcs that move in opposite directions. This sets up two alternating and counter-rotating circulations and, depending on the relative arc movement, results in the following thermal behaviour. Arcs moving {\it towards} each other create a single thermal plume between the arcs that sways back and forth with the alternating circulations as demonstrated by the temperature evolutions in Fig.~\ref{fig:plume_apertures_a} corresponding with case $N=2$ (top) and $N=3$ (bottom) in \eqref{PeriodicSchemes}. Arcs moving {\it away from} each other create two thermal plumes, emanating from the leading edges of the arcs, that are periodically reinvigorated by the alternating circulations as demonstrated in Fig.~\ref{fig:plume_apertures_b} for case $N=4$ in \eqref{PeriodicSchemes}.

Aperiodic and (largely) periodic schemes as illustrated in Fig.~\ref{fig:adaptive_sequences} may emerge throughout parameter space yet the controller generally tends towards aperiodic (periodic) schemes for ``lower'' (``higher'') $\tau$. Such aperiodic schemes often exhibit intermittent behaviour by comprising of plume-forming periodic episodes interspersed with aperiodic intervals; this e.g. occurs typically in the high-performance regions for $N=2$ and $N=3$ in Fig.~\ref{fig:L_homogenisations}. Moreover, persistent single plumes as shown in Fig.~\ref{fig:plume_apertures_a} mainly emerge for $N=2$ and $N=3$; case $N=4$ overall tends more towards aperiodic behaviour. However, a direct link between type of adaptive reorientation and parameter regimes appears absent yet generic correlations with the advection characteristics exist (Sec.~\ref{PerformanceScheme3}).

\begin{figure}[!hbt]

\subfigure[$N=2$]{\includegraphics[width=.5\textwidth,height=.27\textwidth]{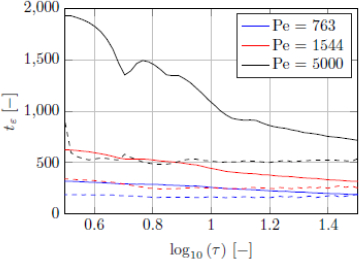}\label{fig:L_slicesa}}
\hspace*{12pt}
\subfigure[$N=3$]{\includegraphics[width=.45\textwidth,height=.26\textwidth]{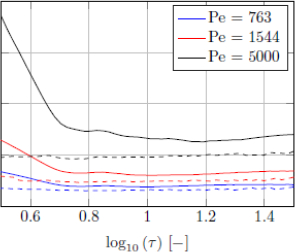}\label{fig:L_slicesb}}

\caption{Transient time $t_{\varepsilon,a}$ (dashed) and $t_{\varepsilon,p}$ (solid) of adaptive and conventional periodic schemes, respectively, versus $\tau$ for selected $Pe$ and RAM with number of apertures $N$ as indicated.}
    \label{fig:L_slices}
\end{figure}

The dependence of performance indicator $\chi$ on parameters $(Pe,\tau)$ shown in Fig.~\ref{fig:L_homogenisations} is inextricably linked to that of the transient times for the adaptive ($t_{\varepsilon,a}$) and conventional periodic ($t_{\varepsilon,p}$) schemes. Fig.~\ref{fig:L_slices} gives $t_{\varepsilon,a}$ (dashed) and $t_{\varepsilon,p}$ (solid) for $N=2$ and $N=3$ at the indicated $Pe$ and reveals a significant decline of $t_{\varepsilon,p}$ with increasing $\tau$ that grows more pronounced with larger $Pe$ versus a nearly uniform $t_{\varepsilon,a}$ in $\tau$-direction that increases moderately with $Pe$. Case $N=4$ (not shown) exhibits similar behaviour yet with a mildly-fluctuating (instead of significantly declining) $t_{\varepsilon,p}$. These observations imply that variation of $\chi$ in Fig.~\ref{fig:L_homogenisations} stems primarily from variable $t_{\epsilon,p}$ and, inherently, variable performance of the conventional periodic scheme.

Uniform $t_{\epsilon,a}$ dependent only on $Pe$ in Fig.~\ref{fig:L_slices}, on the other hand, signifies a consistent performance
of the adaptive scheme that (save the localised spike at the lower bound for $\tau$ at $Pe=5000$ in Fig.~\ref{fig:L_slicesa}) is virtually independent of both $\tau$ and $N$. This implies that the controller systematically identifies the (on average) optimal orientation between deformation and temperature gradient at the fluid parcels and heat transfer is limited primarily by the thermo-physical conditions at the molecular level for given $Pe$. Heat transfer between fluid parcels namely occurs at time scales proportional to the diffusive time scale $t_{\rm diff} = R^2/\nu$, which in the non-dimensional formulation of Sec.~\ref{System1} (i.e. relative to the advective time scale $t_{\rm adv} = R/U$) depends linearly on $Pe$ via $t_{\rm diff}' = t_{\rm diff}/t_{\rm adv}= Pe$. Linear correlation $t_{\epsilon,a}\propto Pe$, for (at least) $N=2$ and $N=3$ in Fig.~\ref{fig:L_homogenisations} implies that the controller (within the constraint of the step-wise optimisation of Sec.~\ref{ControlStrategy2}) consistently identifies the ``best'' reorientation scheme for molecular transfer rates at given $Pe$.

\subsection{Role of fluid deformation and chaotic advection}
\label{PerformanceScheme3}

Fluid deformation impacts the global heat transfer by changing the rates of global energising (quantified by metric $\Jb$) and equilibration (quantified by metric $\Ja$) via the leading integrals in \eqref{HeatingFluid7} represented by functions $\bar{G}$ and $\widetilde{G}$ according to \eqref{HeatingFluid9} and \eqref{HeatingFluid11}, respectively. The spatial distribution of the corresponding integrands $\bar{g}$ ($\widetilde{g}$), shown in Fig.~\ref{HeatingDynamics3} (Fig.~\ref{HeatingDynamics4}) in terms of $\bar{\beta}$ ($\widetilde{\beta}$) following \eqref{HeatingFluid8b}, reveals that this impact is primarily restricted to the direct vicinity of the arc edges and (at least) for temperature evolutions in the base flow $\xvec{v}_1$ from the uniform initial condition $T_0$ accelerates energising and equilibration. The generality of this behaviour is investigated below.

\begin{figure}[!hbt]

\centering

\subfigure[$\tau=2$]{
\includegraphics[trim=1cm 0cm 1cm 1cm,clip,width=.48\textwidth,height=.24\textwidth]{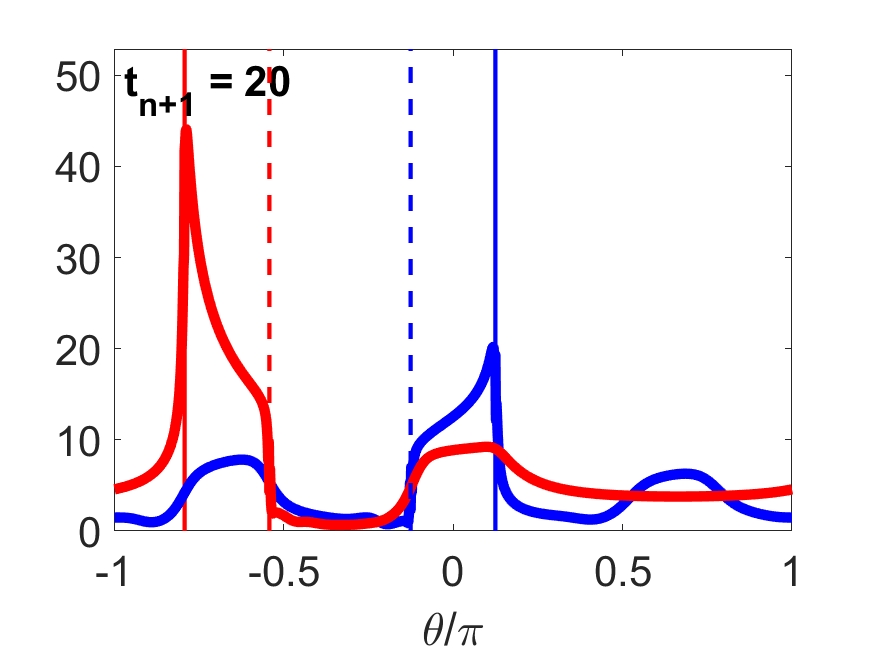}
\hspace*{12pt}
\includegraphics[trim=1cm 0cm 1cm 1cm,clip,width=.48\textwidth,height=.24\textwidth]{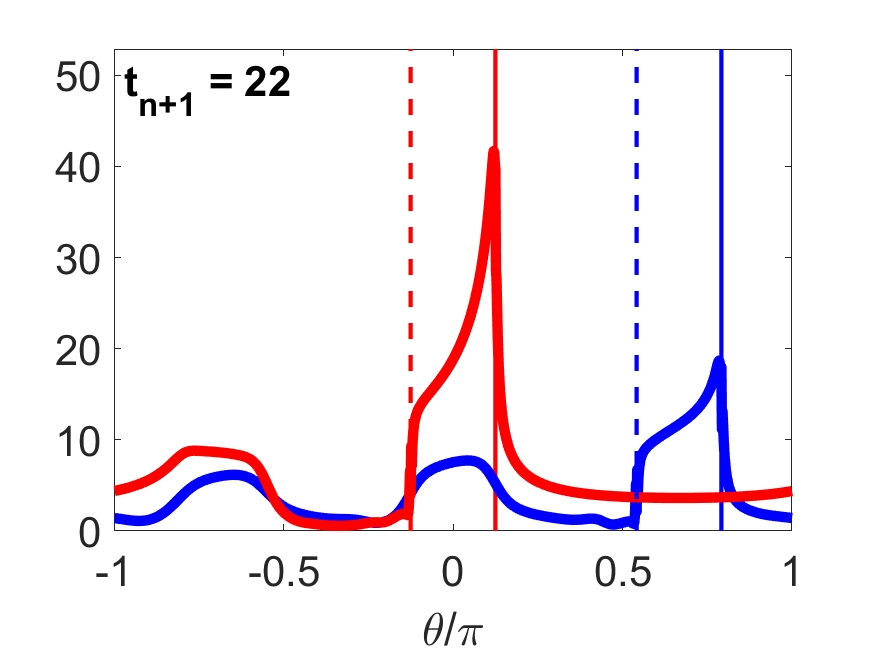}
}

\subfigure[$\tau=5$]{
\includegraphics[trim=1cm 0cm 1cm 1cm,clip,width=.48\textwidth,height=.24\textwidth]{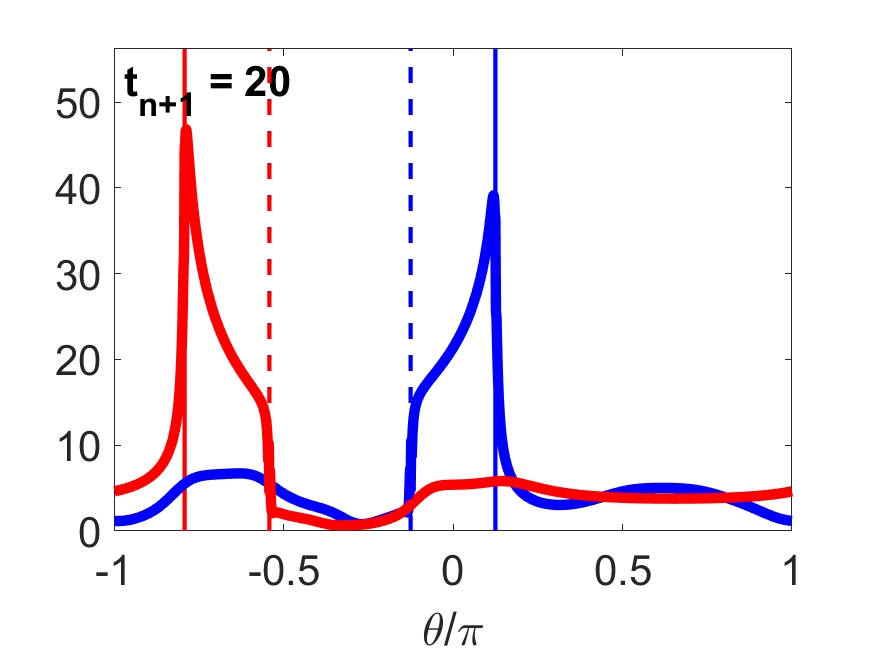}
\hspace*{12pt}
\includegraphics[trim=1cm 0cm 1cm 1cm,clip,width=.48\textwidth,height=.24\textwidth]{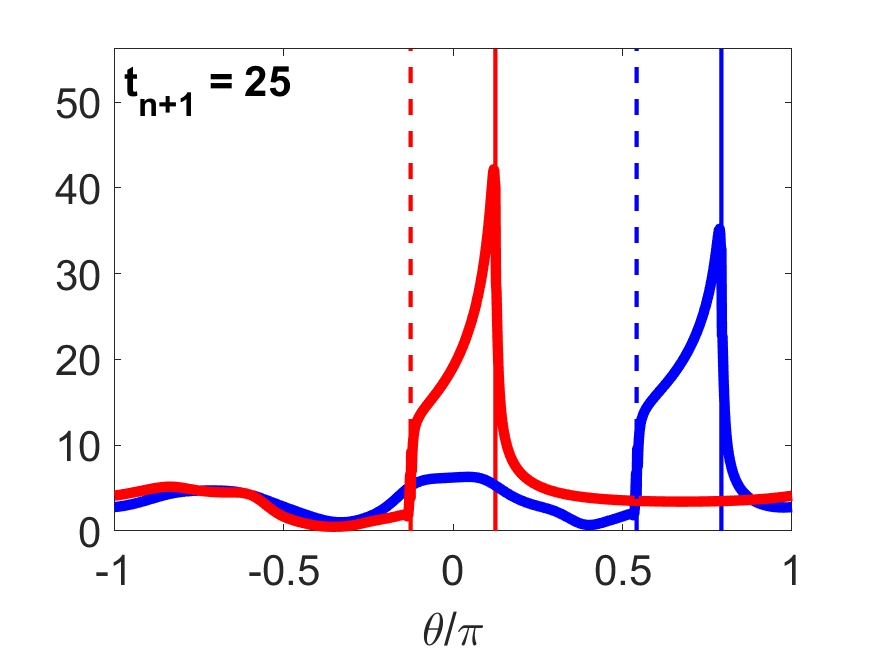}
}

\subfigure[$\tau=10$]{
\includegraphics[trim=1cm 0cm 1cm 1cm,clip,width=.48\textwidth,height=.24\textwidth]{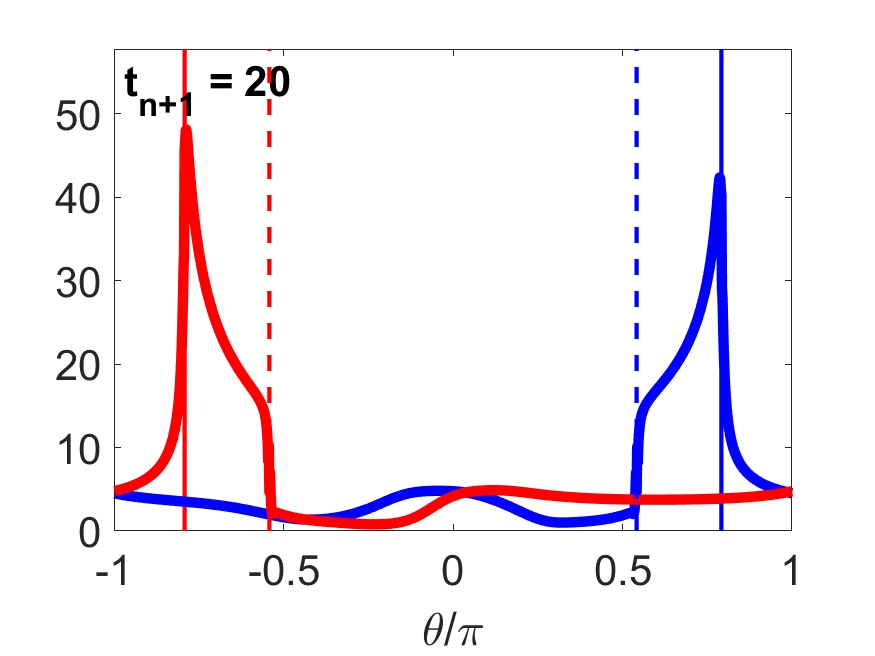}
\hspace*{12pt}
\includegraphics[trim=1cm 0cm 1cm 1cm,clip,width=.48\textwidth,height=.24\textwidth]{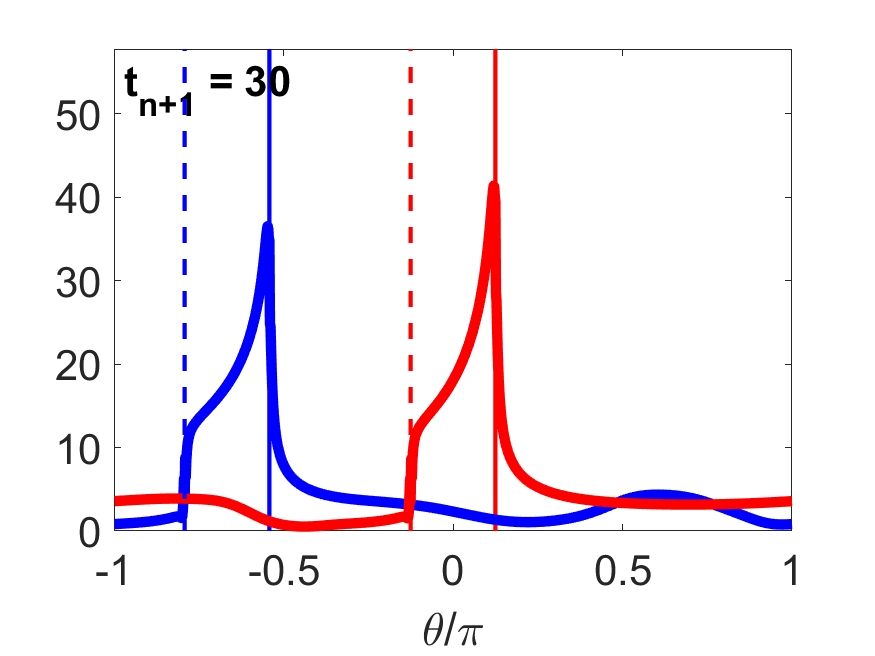}
}

\caption{Impact of fluid deformation on energising demonstrated by wall temperature gradient $\bar{f} = \partial \Ta/\partial r|_\Gamma$ for adaptive (red) versus conventional periodic (blue) scheme at time
levels $t_{n+1}$ for $N=3$ and $Pe=1000$ at given $\tau$. Dashed (solid) vertical lines indicate leading (trailing) edges of active arc.}
\label{TemperatureGradBoundary}
\end{figure}

Key to energising is the wall temperature gradient $\bar{f} = \partial \Ta/\partial r|_\Gamma$. Fig.~\ref{TemperatureGradBoundary} shows $\bar{f}$ for the adaptive (red) versus conventional periodic (blue) scheme at time levels $t_{n+1}$ during an intermediate episode of the transient for the typical case $N=3$ and $Pe=1000$ at given $\tau$. (The adaptive scheme corresponds with $\mathcal{U}_{N=3} = \{1,-3,1,-3,\dots\}$ following \eqref{PeriodicSchemes}.) Gradient $\bar{f}$ significantly steepens at the active arc during step $t_n\leq t\leq t_{n+1}$, with corresponding leading (trailing) edges indicated by dashed (solid) vertical lines, and peaks at the trailing egde. This reveals that fluid deformation, despite considerable changes in initial conditions at $t_n$, step-wise creates essentially the same situation as for the base flow in Fig.~\ref{HeatingDynamics3b} and -- upon identifying $\theta = \Delta/2$ ($\theta = -\Delta/2$) in \eqref{HeatingFluid9} with the trailing (leading) edge of the active arc -- implies $\bar{G}>0$ and thus accelerated energising compared to a non-deforming fluid. Both schemes accomplish this yet the consistently higher peaks at the trailing edges for the adaptive scheme render this effect more pronounced and thereby signify a systematically superior performance over the conventional scheme.

\begin{figure}[!hbt]

\includegraphics[trim=0cm 0cm 0cm 0cm,clip,width=\textwidth]{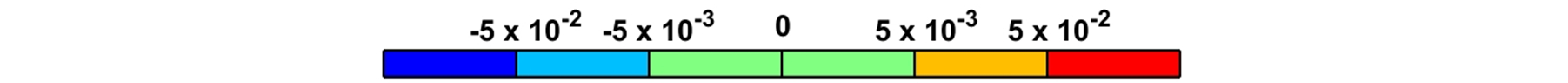}\vspace*{6pt}

\subfigure[Adaptive scheme $\tau=2$]{
\includegraphics[trim=5cm 1cm 5cm 1cm,clip,width=.25\textwidth]{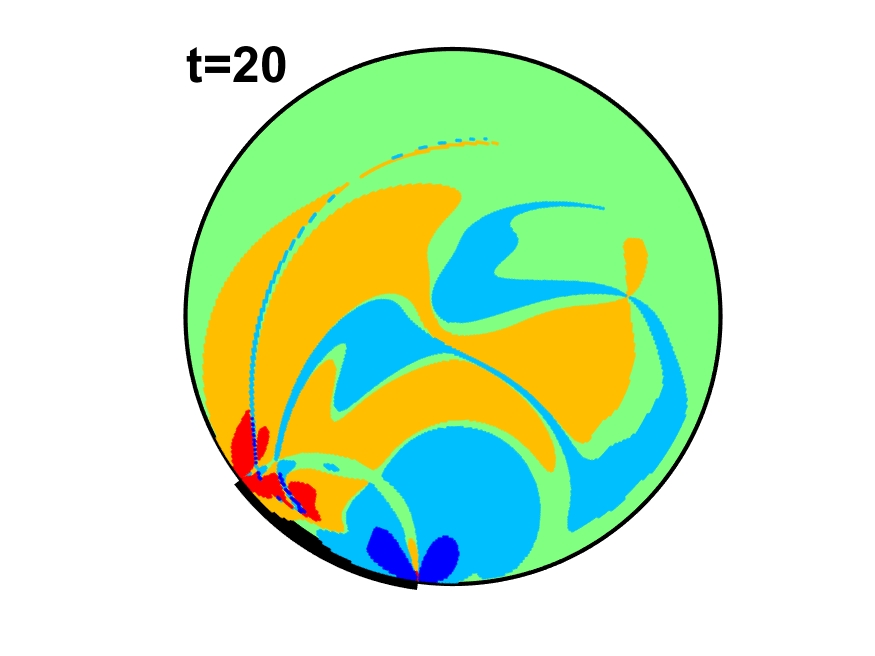}
\includegraphics[trim=5cm 1cm 5cm 1cm,clip,width=.25\textwidth]{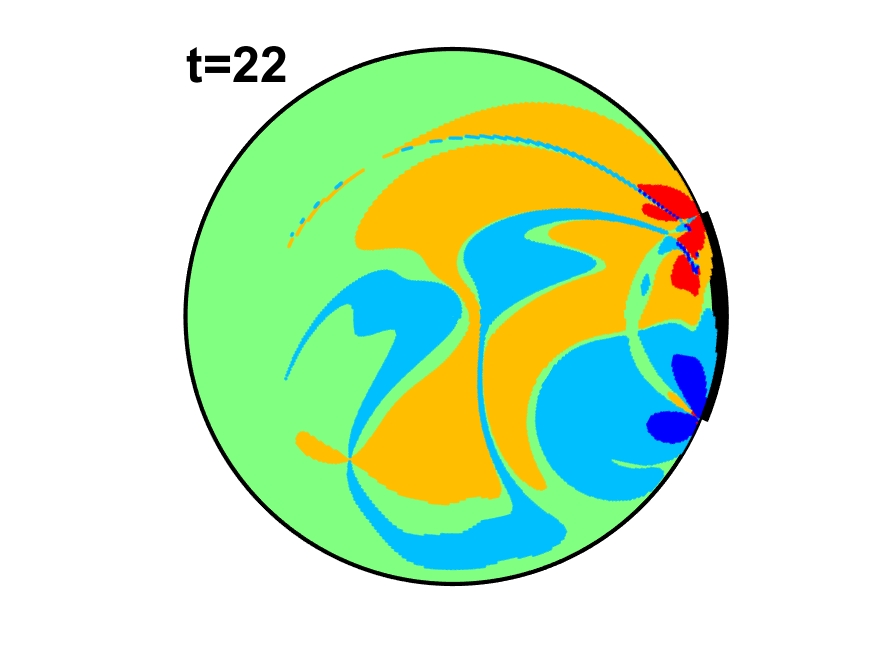}
}
\subfigure[Conventional scheme $\tau=2$]{
\includegraphics[trim=5cm 1cm 5cm 1cm,clip,width=.25\textwidth]{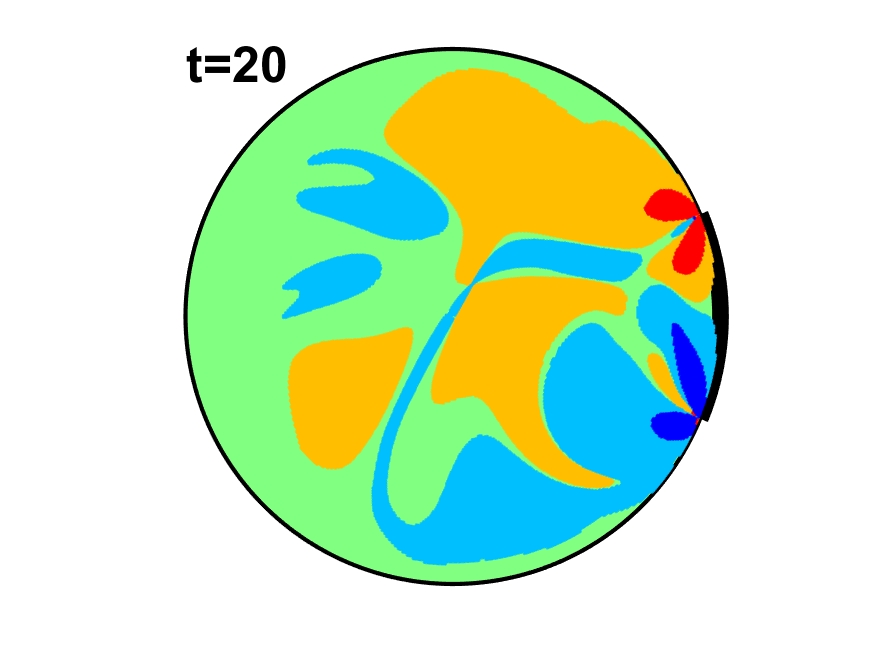}
\includegraphics[trim=5cm 1cm 5cm 1cm,clip,width=.25\textwidth]{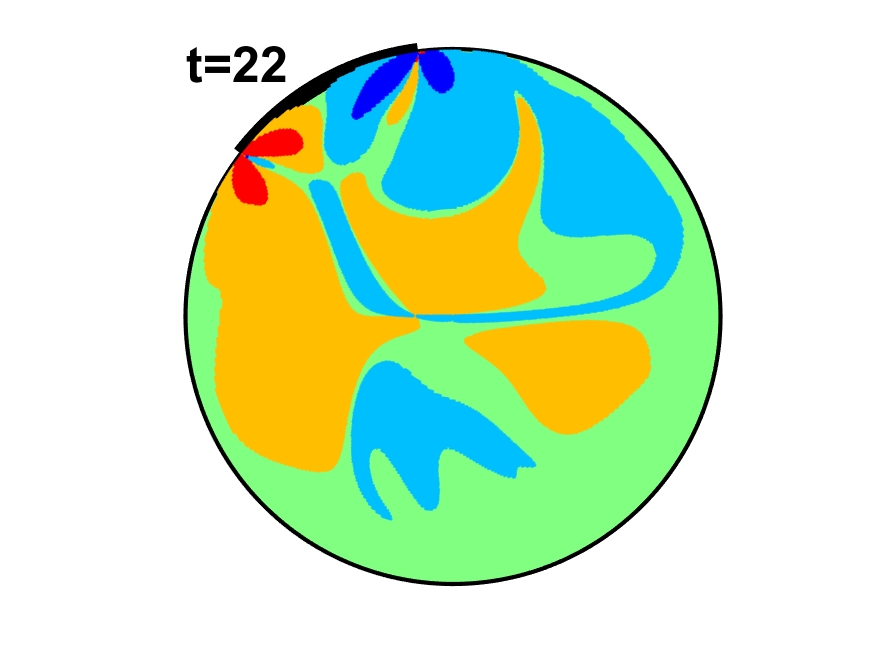}
}

\subfigure[Adaptive scheme $\tau=5$]{
\includegraphics[trim=5cm 1cm 5cm 1cm,clip,width=.25\textwidth]{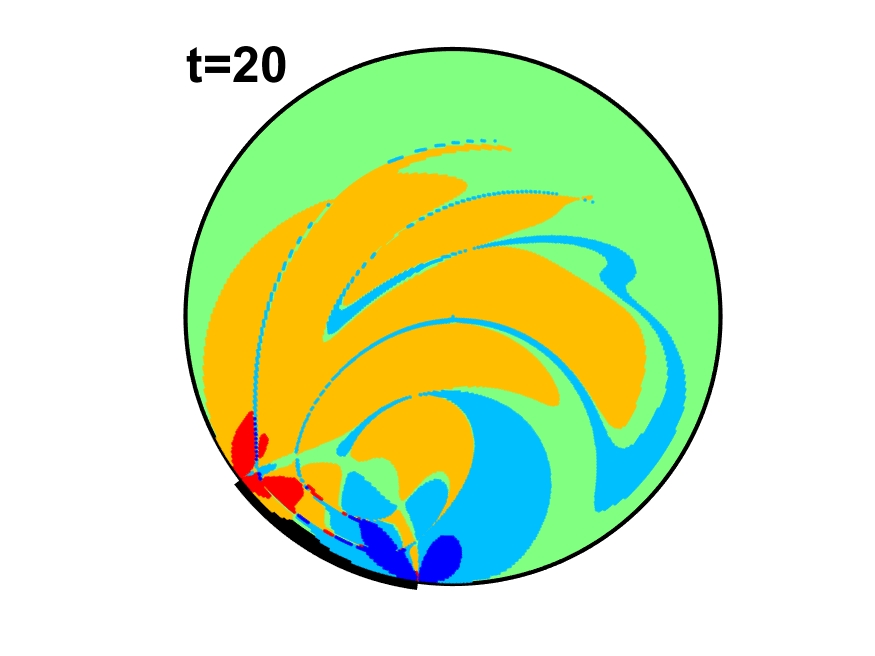}
\includegraphics[trim=5cm 1cm 5cm 1cm,clip,width=.25\textwidth]{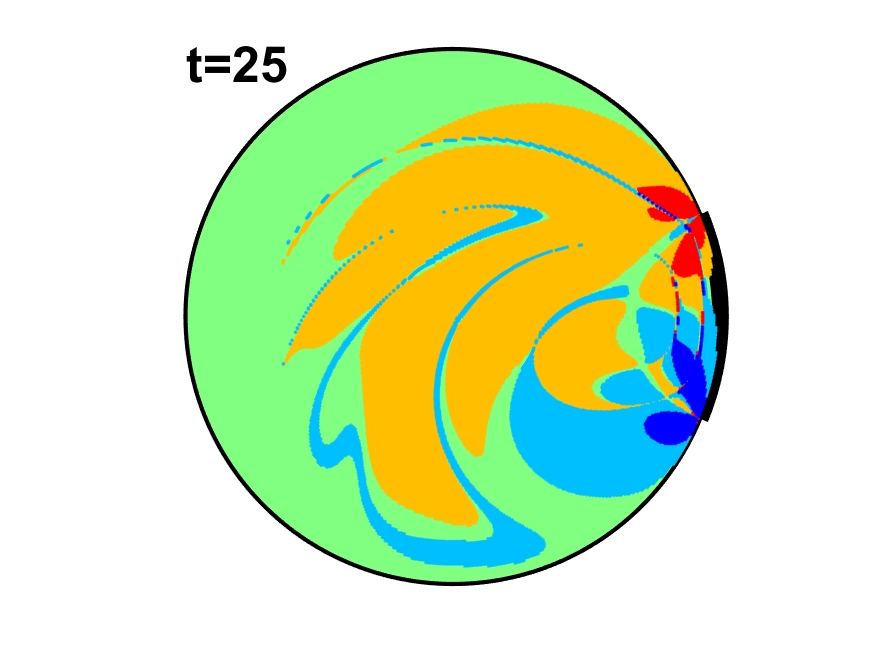}
}
\subfigure[Conventional scheme $\tau=5$]{
\includegraphics[trim=5cm 1cm 5cm 1cm,clip,width=.25\textwidth]{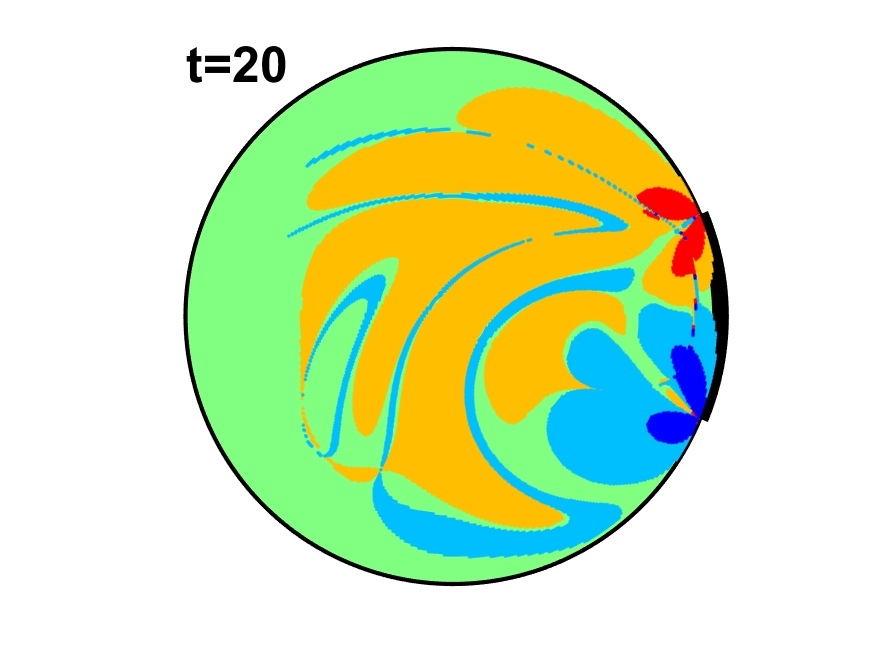}
\includegraphics[trim=5cm 1cm 5cm 1cm,clip,width=.25\textwidth]{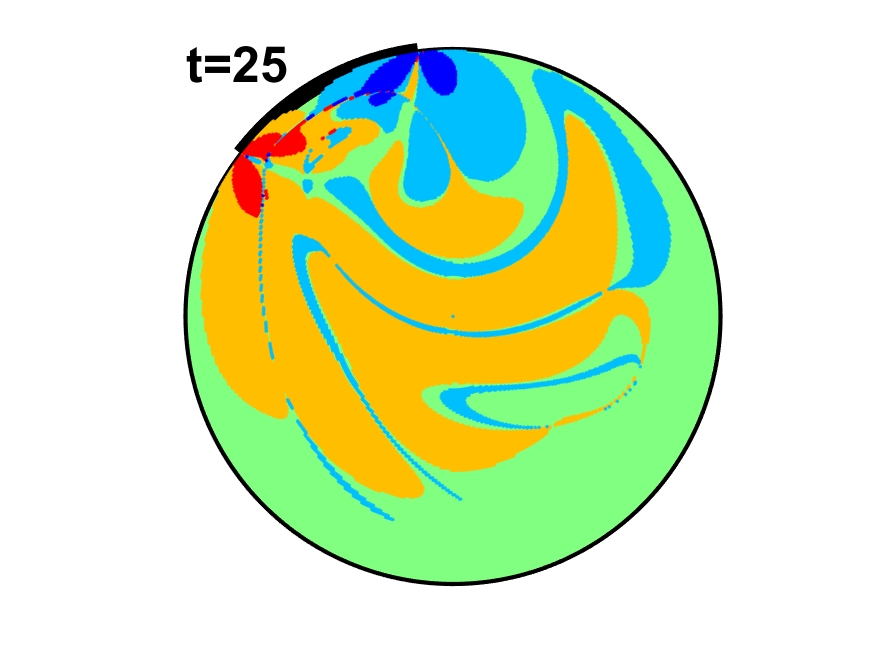}
}

\subfigure[Adaptive scheme $\tau=10$]{
\includegraphics[trim=5cm 1cm 5cm 1cm,clip,width=.25\textwidth]{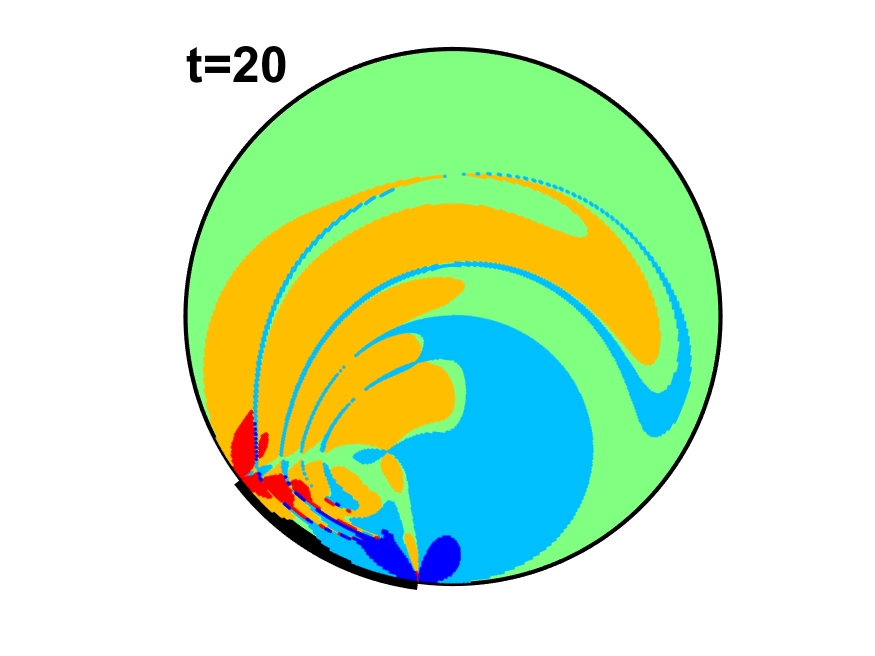}
\includegraphics[trim=5cm 1cm 5cm 1cm,clip,width=.25\textwidth]{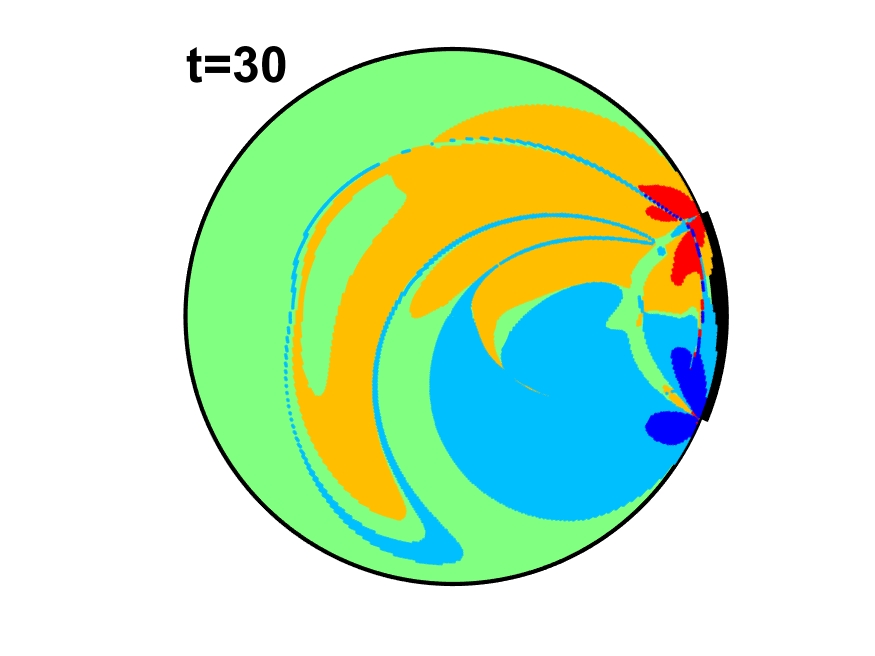}
}
\subfigure[Conventional scheme $\tau=10$]{
\includegraphics[trim=5cm 1cm 5cm 1cm,clip,width=.25\textwidth]{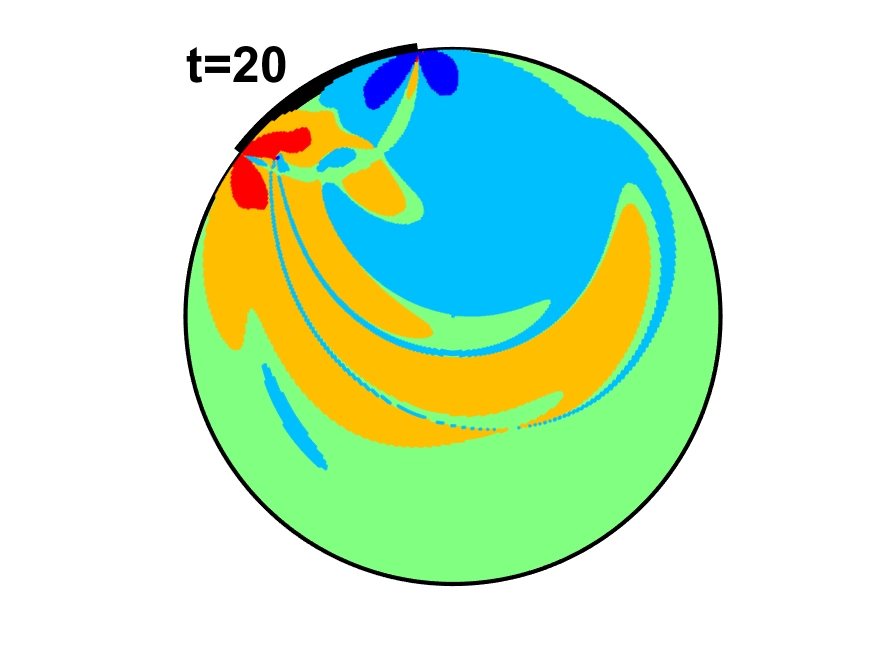}
\includegraphics[trim=5cm 1cm 5cm 1cm,clip,width=.25\textwidth]{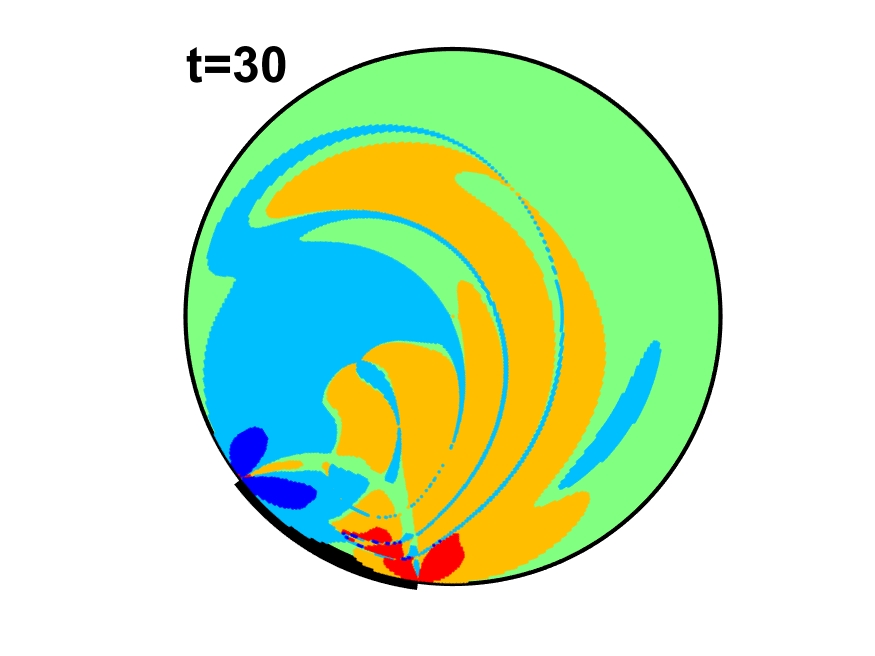}
}

\caption{Impact of fluid deformation on equilibration demonstrated by partition of relative change rate $\widetilde{\beta} = \widetilde{g}/\widetilde{f}$ via $\widetilde{\beta}_{\rm rel}\equiv \widetilde{\beta}/\max|\widetilde{\beta}|$ into regions with strong enhancement ($\widetilde{\beta}_{\rm rel}\geq\epsilon_2$; red) versus strong diminution ($\widetilde{\beta}_{\rm rel}\leq-\epsilon_2$; blue), moderate enhancement ($\epsilon_1\leq \widetilde{\beta}_{\rm rel}<\epsilon_2$; yellow) versus moderate diminution ($-\epsilon_2< \widetilde{\beta}_{\rm rel}\leq -\epsilon_1$; cyan) and insignificant impact ($|\widetilde{\beta}_{\rm rel}|<\epsilon_1$; green) using thresholds $(\epsilon_1,\epsilon_2)=(5\times 10^{-3},5\times 10^{-2})$ for $N=3$ and $Pe=1000$ at given $\tau$. Heavy boundary segments indicate active arc during step $[t-\tau,t]$ prior to given time level $t$.}
\label{HeatingAccelerationInterior}
\end{figure}

The step-wise impact of fluid deformation on equilibration for the above case is shown in Fig.~\ref{HeatingAccelerationInterior}
in terms of $\widetilde{\beta}$ and is also reminiscent of that of the base flow in Fig.~\ref{HeatingDynamics4}: emergence of localised high-impact regions (red/blue) near the arc edges and larger moderate-impact regions (yellow/cyan) in the interior. Moreover, these high-impact regions again vary only marginally in time and with flow reorientation; significant spatio-temporal variation occurs only in the moderate-impact regions. Thus the simplified $\widetilde{G}$ following \eqref{HeatingFluid11} to good approximation holds in general and in conjunction with the wall temperature gradients in Fig.~\ref{TemperatureGradBoundary} implies $\widetilde{G}>0$ and, inherently, acceleration of equilibration by the same reasoning as for the base flow in Sec.~\ref{GlobalHeatingDynamics2}.

The above behaviour is representative for the entire transient in arbitrary cases and demonstrates that fluid deformation generically indeed enhances energising and equilibration. However, as clearly demonstrated in Fig.~\ref{HeatingAccelerationInterior}, this primarily relies on the high-shear regions near the arc edges and their interaction with the temperature gradient; fluid deformation in the flow interior only plays a secondary role in this process due to its relative weakness compared to said regions. This has the major implication that the adaptive scheme attains its superior and consistent performance mainly from step-wise optimal orientation between arc edges and wall temperature gradient. The adaptive scheme thus creates peaks of comparable magnitude close to an upper limit set by the beforementioned thermo-physical conditions for all $\tau$, as e.g. demonstrated by the profiles at $t_{n+1}=20$ in Fig.~\ref{TemperatureGradBoundary}, yielding the nearly uniform $t_{\varepsilon,a}$ in Fig.~\ref{fig:L_slices}. The conventional scheme, on the other hand, sequentially creates peaks at the consecutive arcs that grow with $\tau$ (again demonstrated by the profiles at $t_{n+1}=20$ in Fig.~\ref{TemperatureGradBoundary}) due to the increasing fluid deformation associated with longer arc activation and approach said upper limit only for sufficiently large $\tau$. Reaching this limit is intimately related to the local breakdown of heat-transfer enhancement at $t\sim\mathcal{O}(\tau_\beta)$, with $\tau_\beta$ according to \eqref{RelaxationTime}, where $\tau_\beta\simeq\mathcal{O}(0.3-3)$ for the degree of fluid deformation in the arc region (Sec.~\ref{LocalAsymptotic}). The significant decline and subsequent flattening of $t_{\varepsilon,p}$ exactly at step durations $\tau$ of this magnitude in Fig.~\ref{fig:L_slices} support this scenario.

\begin{figure}[!hbt]
	\centering
		\includegraphics[width=.9\textwidth]{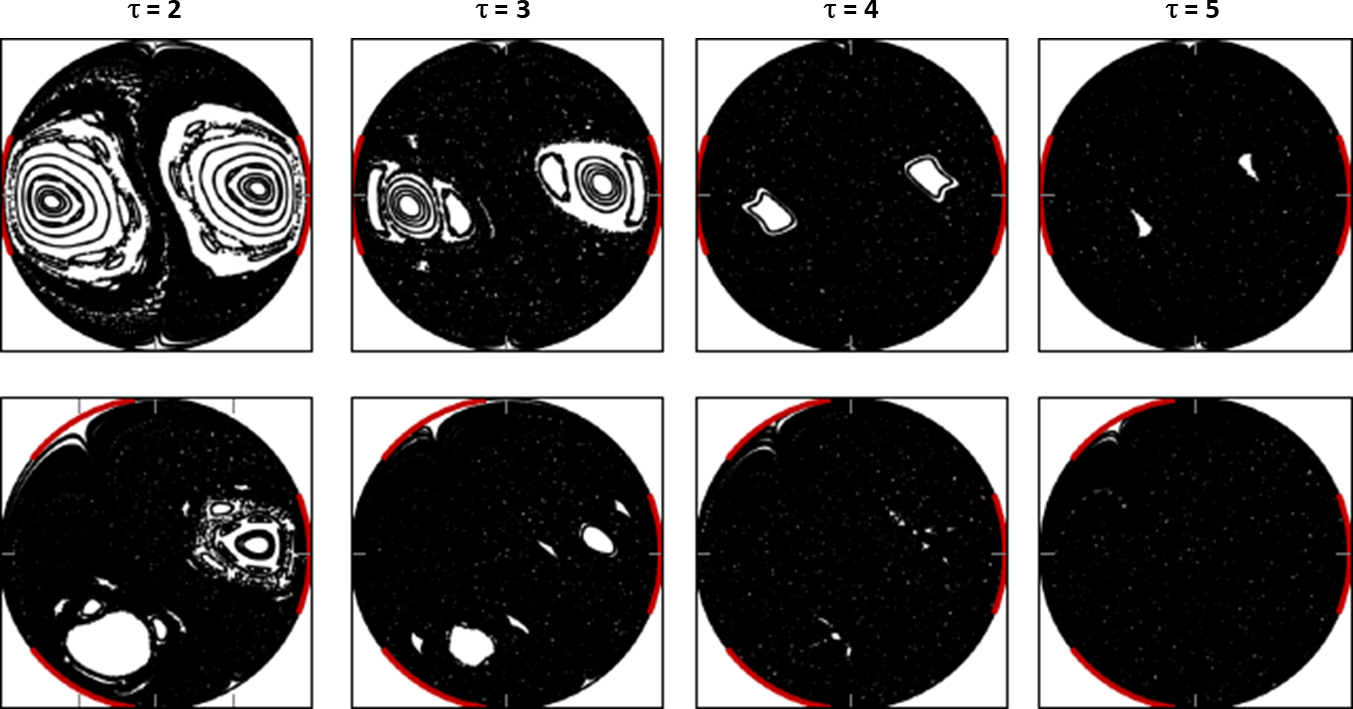}
	\caption{Lagrangian transport associated with periodic plume-forming schemes \eqref{PeriodicSchemes} versus $\tau$ demonstrated by stroboscopic map of 100 tracers released on $x$-axis for RAM
with $N=2$ (top) and $N=3$ (bottom) apertures.}
	\label{fig:poincare_sections}
\end{figure}

The relative weakness of fluid deformation in the domain interior suggests that global advection mainly serves as mechanism for exchanging (to be) heated fluid parcels between the arc regions and said interior and makes whether this fluid transport is chaotic or not of secondary importance. The following considerations substantiate this subordinate role of chaotic advection for the heating performance of the RAM. Lagrangian transport within the flow interior associated with the frequently-emerging periodic plume-forming schemes such as e.g. \eqref{PeriodicSchemes} transits from regular to chaotic with increasing $\tau$, as demonstrated in Fig.~\ref{fig:poincare_sections} by the stroboscopic map of 100 tracers released on the $x$-axis. However, the corresponding transient time $t_{\varepsilon,a}$ in Fig.~\ref{fig:L_slices} remains uniform, signifying a thermal performance of the adaptive scheme that is consistent for all $\tau$ and, in consequence, independent of the emergence of chaos.

The conventional scheme results in a similar transition from regular to chaotic advection in the ranges $1\lesssim\tau\lesssim 6$ ($0\lesssim\log_{10}\tau\lesssim 0.8$), $2\lesssim\tau\lesssim 5$ ($0.3\lesssim\log_{10}\tau\lesssim 0.7$) and $2\lesssim\tau\lesssim 5$ ($0.3\lesssim\log_{10}\tau\lesssim 0.7$) for $N=2$, $N=3$ and $N=4$, respectively, and yields stroboscopic maps as
shown in \cite{baskan2016}. (Stroboscopic maps in \cite{baskan2016} in fact concern companion schemes for $\Theta'=-\Theta$ and $\Omega') = -\Omega$ and on grounds of symmetry are identical
to maps of the current schemes.) The transition and fully-chaotic zones correlate with the regimes of declining and flat profiles of $t_{\varepsilon,p}$ in Fig.~\ref{fig:L_slices}, respectively, and this may
thereby suggest a significant role of the interior (chaotic) advection in the thermal performance. However, increasing $\tau$ and $N$ simultaneously changes the conditions near the active arc by two
effects: (i) circulation of more fluid parcels through the $N$ arc regions (thus promoting greater heat exchange between wall and interior) and (ii) larger fluid deformation (thus
promoting a steeper wall temperature gradient at the arc). This, together with the flattening of $t_{\varepsilon,p}$ at $\tau\sim\mathcal{O}(\tau_\beta)$ established above, advances the conditions at the active arc that drives the interior (chaotic) advection -- rather than the nature of this interior transport itself -- as the primary cause for the decline in $t_{\varepsilon,p}$ and corresponding enhancement in thermal performance.

\section{Conclusions}
\label{sec:conclusions_and_recommendations}

The present study aims to contribute to existing solutions for enhancement of scalar transport in laminar flows through flow reorientation. To this end a dedicated flow-control strategy is developed (i.e. ``adaptive flow reorientation'') that systematically determines the ``best'' flow reorientation for the fast heating of a cold fluid via a hot boundary in a representative case study.

The control strategy is founded on an in-depth analysis of the dynamics of heating in fluid flows. This exposes fluid deformation as the ``thermal actuator'' via which the flow affects the heat transfer. The link between former and latter is non-trivial, though. Fluid deformation may, depending on its orientation relative to the temperature gradient, both enhance and diminish local heat exchange between fluid parcels. Moreover, enhanced heat transfer promotes local thermal homogenisation and, by reducing temperature gradients, thus effectively counteracts itself. This fundamental ``conflict'' between local heat transfer and thermal homogenisation tends to restrict the beneficial impact
of flow to short-lived episodes. The impact of fluid deformation on the global fluid heating is primarily confined to the direct proximity of the moving boundary that drives the flow. Fluid deformation in the flow interior only plays a secondary role in this process due to its relative weakness compared to said regions. These insights imply that incorporation of the thermal behaviour is essential for effective flow-based enhancement strategies and efficient fluid mixing, the conventional approach adopted in industry for this purpose, is potentially sub-optimal.

Global heating encompasses two concurrent processes, i.e. increasing energy content (``energising'') and thermal homogenisation, and this fundamentally differentiates the current problem from the thermal homogenisation in adiabatic systems usually considered in related studies. Moreover, this notion yields the relevant metrics for the global dynamics and thus enables formulation of the control problem as the minimisation of a dedicated cost function that naturally emerges from the dynamic analyses and adequately incorporates both processes. This facilitates step-wise determination of the ``best'' flow reorientation from predicted future evolutions of actual intermediate states and, in tandem with an efficient predictor, paves the way to (real-time) regulation of scalar transport by flow control in practical applications. Key enablers for this predictor are (i) the property that flow reorientations carry over to the temperature field (ii) a compact reduced-order model for the Perron-Frobenius evolution operator that rapidly maps initial to final temperature fields for each step duration and flow reorientation.

Performance analyses reveal that adaptive flow reorientation significantly accelerates the fluid heating throughout the considered parameter space and thus is superior over conventional periodic schemes (designed for efficient fluid mixing) both in terms of consistency and effectiveness. Fluid heating is accelerated by at least $14\%$ everywhere and $24\%$ or more in large areas and process enhancement of this magnitude constitutes a dramatic reduction in energy and (potentially also) resource consumption in industries motivating the present study. The controller in fact breaks with conventions by, first, never selecting these periodic schemes and, second, achieving the same superior performance for all flow
conditions irrespective of whether said mixing occurs. The controller typically achieves this superiority by creating an essentially heterogeneous situation comprising of thermal plumes that extend from the hot wall into the cold(er) interior and are driven by two alternating and counter-rotating circulations. The performance analyses furthermore substantiate the primary and secondary roles of fluid deformation near the driving boundary segments and in the flow interior, respectively, in the heating enhancement by the flow.

Ongoing efforts concern, first, more advanced and precise regulation of the thermal process by continuous actuation of all moving boundary segments and, second, experimental validation and testing of the control strategy. Future studies aim at further paving the way to practical applications by development of observers for full state estimation from discrete sensor data and data-based construction of compact models as well as realisation of advanced control targets such as e.g. thermal fronts or heterogeneous temperature fields for promotion of chemical reactions.

\section*{Acknowledgments}
This research is supported by the Netherlands Organisation for Scientific Research (NWO) as part of the Computational Sciences for Energy Research Initiative (grant 15CSER15).

\bibliography{references}

\appendix

\section{Evolution of metric $\Ja$}
\label{EvolutionL2norm}

The evolution of metric $\Ja$ according to \eqref{HeatingNorms} is governed by
\begin{eqnarray}
\frac{d\Ja}{dt} = \int_{\mathcal{D}}\frac{\partial \widetilde{T}^2}{\partial t} d^2\xvec{x} = 2\int_{\mathcal{D}}\widetilde{T}\frac{\partial \widetilde{T}}{\partial t} d^2\xvec{x}
\stackrel{\eqref{eq:ade_pde_discrete}}{=} -2\int_{\mathcal{D}}\widetilde{T}\,\xvec{v}\cdot\pmb{\nabla}\widetilde{T}d^2\xvec{x} - 2\int_{\mathcal{D}}\widetilde{T}\,\pmb{\nabla}\cdot\widetilde{\xvec{q}}d^2\xvec{x}.
\label{HeatingFluid2aX}
\end{eqnarray}
The leading term on the RHS of \eqref{HeatingFluid2aX} admits reformulation as
\begin{eqnarray}
\int_{\mathcal{D}}\xvec{v}\cdot(\widetilde{T}\pmb{\nabla}\widetilde{T})d^2\xvec{x} = \int_{\mathcal{D}}\xvec{v}\cdot\pmb{\nabla}(\widetilde{T}^2/2)d^2\xvec{x}
\stackrel{\pmb{\nabla}\cdot\xvec{v}=0}{=} \int_{\mathcal{D}}\pmb{\nabla}\cdot\left(\xvec{v}T^2/2\right)d^2\xvec{x}
= \int_\Gamma\xvec{v}\cdot\xvec{n}\widetilde{T}^2/2 ds = 0,
\label{HeatingFluid2bX}
\end{eqnarray}
due to $\xvec{v}\cdot\xvec{n}=0$ and $\widetilde{T}=0$ on $\Gamma$. The trailing term on the RHS of \eqref{HeatingFluid2aX} admits reformulation as
\begin{eqnarray}
\int_{\mathcal{D}}\widetilde{T}\,\pmb{\nabla}\cdot\widetilde{\xvec{q}}d^2\xvec{x} &=&
\int_{\mathcal{D}}\pmb{\nabla}\cdot(\widetilde{T}\widetilde{\xvec{q}})d^2\xvec{x} - \int_{\mathcal{D}}\widetilde{\xvec{q}}\cdot\pmb{\nabla}\widetilde{T}d^2\xvec{x} =
\int_\Gamma\widetilde{\xvec{q}}\cdot\xvec{n}\widetilde{T}ds - \int_{\mathcal{D}}\widetilde{\xvec{q}}\cdot\pmb{\nabla}\widetilde{T}d^2\xvec{x}\nonumber\\
&=& -\int_{\mathcal{D}}\widetilde{\xvec{q}}\cdot\pmb{\nabla}\widetilde{T}d^2\xvec{x},
\label{HeatingFluid2cX}
\end{eqnarray}
due to $\widetilde{T}=0$ on $\Gamma$. Substitution of \eqref{HeatingFluid2bX} and \eqref{HeatingFluid2cX} into \eqref{HeatingFluid2aX} yields
\begin{eqnarray}
\frac{d\Ja}{dt} = 2\int_{\mathcal{D}}\widetilde{\xvec{q}}\cdot\pmb{\nabla}\widetilde{T}d^2\xvec{x}.
\label{HeatingFluid2dX}
\end{eqnarray}
as simplified evolution equation for metric $\Ja$.

\end{document}